\documentclass[twocolumn]{aastex63}

\usepackage{natbib}

\usepackage[maxfloats=256]{morefloats}
\usepackage{color}
\usepackage{mathrsfs}
\usepackage{comment}

\newcommand{\HI}{\mbox{H\,{\sc i}}}

\received{June 1, 2019}
\revised{January 10, 2019}
\accepted{\today}
\submitjournal{ApJ}

\shorttitle{Cold gas in Virgo galaxies}
\shortauthors{Morokuma-Matsui et al.}

\begin{document}

\title{A phase-space view of cold-gas properties of Virgo-cluster galaxies: multiple quenching processes at work?}

\correspondingauthor{Kana Morokuma-Matsui}
\email{kanamoro@ioa.s.u-tokyo.ac.jp}

\author[0000-0003-3932-0952]{Kana Morokuma-Matsui}
\altaffiliation{JSPS Fellow}
\affiliation{Institute of Astronomy, Graduate School of Science, The University of Tokyo, 2-21-1 Osawa, Mitaka, Tokyo 181-0015, Japan}
\affiliation{Institute of Space and Astronautical Science, Japan Aerospace Exploration Agency, 3-1-1 Yoshinodai, Chuo-ku, Sagamihara, Kanagawa 252-5210, Japan}
\affiliation{Chile Observatory, National Astronomical Observatory of Japan,
    2-21-1 Osawa, Mitaka-shi, Tokyo 181-8588, Japan}

\author[0000-0002-2993-1576]{Tadayuki Kodama}
\affiliation{Graduate School of Science, Tohoku University, 6-3 Aramaki Aza-Aoba, Sendai, Miyagi 980-8578, Japan}


\author[0000-0001-7449-4814]{Tomoki Morokuma}
\affiliation{Institute of Astronomy, Graduate School of Science, The University of Tokyo, 2-21-1 Osawa, Mitaka, Tokyo 181-0015, Japan}

\author[0000-0002-6939-0372]{Kouichiro Nakanishi}
\affiliation{National Astronomical Observatory of Japan, 2-21-1 Osawa, Mitaka, Tokyo 181-8588, Japan}
\affiliation{Department of Astronomy, School of Science, The Graduate University for Advanced Studies, SOKENDAI, Mitaka, Tokyo 181-8588, Japan}


\author[0000-0002-0479-3699]{Yusei Koyama}
\affiliation{Subaru Telescope, National Astronomical Observatory of Japan, 650 North A’ohoku Place, Hilo, HI 96720, U.S.A.}

\author[0000-0002-4999-9965]{Takuji Yamashita}
\affiliation{National Astronomical Observatory of Japan, 2-21-1 Osawa, Mitaka, Tokyo 181-8588, Japan}
\affiliation{Research Center for Space and Cosmic Evolution, Ehime University, 2-5 Bunkyo-cho, Matsuyama, Ehime 790-8577, Japan}

\author[0000-0002-0100-1238]{Shuhei Koyama}
\affiliation{National Astronomical Observatory of Japan, 2-21-1 Osawa, Mitaka, Tokyo 181-8588, Japan}
\affiliation{Research Center for Space and Cosmic Evolution, Ehime University, 2-5 Bunkyo-cho, Matsuyama, Ehime 790-8577, Japan}

\author[0000-0003-0137-2490]{Takashi Okamoto}
\affiliation{Department of Cosmosciences, Graduate School of Science, Hokkaido University, N10 W8, Kitaku, Sapporo, 060-0810, Japan}




\begin{abstract}
We investigate the cold-gas properties of massive Virgo galaxies ($>10^9$ M$_\odot$) at $<3R_{200}$ ($R_{200}$ is the radius where the mean interior density is 200 times the critical density) on the projected phase-space diagram (PSD) with the largest archival dataset to date to understand the environmental effect on galaxy evolution in the Virgo cluster.
We find:
lower \HI~and H$_2$ mass fractions and higher star-formation efficiencies (SFEs) from \HI~and H$_2$ in the Virgo galaxies than the field galaxies for matched stellar masses;
the Virgo galaxies generally follow the field relationships between the offset from the main sequence of the star-forming galaxies [$\Delta$(MS)] with gas fractions and SFEs but slightly offset to lower gas fractions or higher SFEs than field galaxies at \replaced{$\Delta({\rm MS})\lesssim 0$}{$\Delta({\rm MS})< 0$};
lower gas fractions in galaxies with smaller clustocentric distance and velocity;
lower gas fractions in the galaxies in the W cloud, a substructure of the Virgo cluster.
Our results suggest the cold-gas properties of some Virgo galaxies are affected by their environment at least at $3 R_{200}$ maybe via strangulation and/or pre-processes and \HI~and H$_2$ in some galaxies are removed by ram pressure at $<1.5 R_{200}$.
Our data cannot rule the possibility of the other processes such as strangulation and galaxy harassment accounting for the gas reduction in some galaxies at $<1.5 R_{200}$.
Future dedicated observations of a mass-limited complete sample are required for definitive conclusions.
\end{abstract}



\section{Introduction} \label{sec:intro}

As the large-scale structure of the universe develops with time, the environmental effects on galaxies, especially quenching of star formation activity, are getting more and more important \citep{Peng:2010eq,Darvish:2016gj}.
It is shown that galaxies in dense regions in the local universe tend to be red and passive \citep{Dressler:1980yl,Gomez:2003hb,Balogh:2004vf,Kauffmann:2004rj,Hogg:2004oj,Baldry:2006kd} and the red and passive galaxy fraction in cluster environments increases with time \citep{Butcher:1984xt,Couch:1987yw,Rakos:1995xt,Margoniner:2001xy,Ellingson:2001ms,Kodama:2001zx,Saintonge:2008tj,Webb:2013gr}.

In the cluster environment, several processes that are not effective for galaxies in low-density regions (hereafter, field galaxies) are considered to become relatively important in galaxy evolution, such as ``ram-pressure stripping'' \citep[e.g.,][]{Gunn:1972kc} by intra-cluster medium (ICM), ``galaxy harassment'' \citep[e.g.,][]{Moore:1996mv}, and ``strangulation'' \citep[e.g.,][]{Larson:1980ok}.
It is also claimed that the star formation in galaxies had been already suppressed before entering the cluster potential when they were members of a galaxy group that would be eventually merged into the cluster \citep[``pre-process'', e.g.,][]{Fujita:2004fa}.

To understand the quenching processes working on cluster galaxies, it is essential to investigate the physical properties of their cold-gas (atomic and molecular gas) reservoir, which is the raw material for star formation. Recent detailed observations in \HI~and CO of so-called ``Jellyfish galaxies'' have provided us an important insight into the effect of ram-pressure stripping on the cold gas in galaxies.
The Jellyfish galaxies are characterized by their fascinating ionized-gas tails extended from the galaxy and considered to be in the peak phase of the ram-pressure stripping \citep{Ebeling:2014ez}.
They are found to be molecular gas rich in tail, and the length of the tail and the velocity gradient along the tail suggest that most molecular gas is formed in situ from atomic gas or diffuse molecular gas rather than stripped from the galaxy as molecular gas \citep{Jachym:2014oc,Jachym:2017vd,Jachym:2019hx,Moretti:2018hp,Moretti:2020we,Moretti:2020td}.
Given the depletion time of cold gas by star formation in the tail, which is longer than the Hubble time, the cold gas is considered to become ICM before being consumed by star formation \citep{Jachym:2014oc,Jachym:2017vd,Verdugo:2015is,Moretti:2018hp,Moretti:2020we}.
The main body of the Jellyfish galaxies is reported to be atomic-gas poor \citep{Ramatsoku:2019kt,Ramatsoku:2020ut,Deb:2020rl}, molecular-gas rich \citep{Moretti:2020td}, and actively forming stars, suggesting that the ram-pressure stripping promote the molecular gas formation from atomic gas and consequently star formation in galaxies. 
After this phase, the star-formation activity in galaxies is expected to be quenched.

For more general view of the cold gas properties in cluster galaxies, the atomic gas deficit is reported to be ubiquitous in the cluster environment in the various observational studies of nearby clusters, such as the Virgo cluster \citep{Davies:1973jj,Chamaraux:1980rs,Cayatte:1990qy,Solanes:2001nq,Gavazzi:2005kb}, the Coma cluster \citep{Gavazzi:2006xu,Healy:2020bz}, the Fornax cluster \citep{Schroder:2001rc,Waugh:2002rr}, and a combination of multiple clusters \citep{Solanes:2001nq,Brown:2017bu}.

On the other hand, no definitive conclusion has been reached for whether or not the cluster environment affects the molecular gas content of galaxies.
In the local universe, the contradicting results have been reported in different clusters and even in the same clusters.
For example, the CO-deficit in cluster galaxies is reported for the Virgo cluster \citep{Rengarajan:1992kx,Fumagalli:2009bt,Corbelli:2012wv,Boselli:2014qs}, the Coma cluster \citep{Fumagalli:2009bt}, the Fornax cluster \citep{Zabel:2019ne}, and the Abell 262 cluster \citep{Bertram:2006kj}.
In contrast, no significant difference in CO contents between field and cluster galaxies is reported for the Virgo cluster \citep{Kenney:1986eu,Kenney:1989vp,Stark:1986js,Boselli:1994gj,Perea:1997wd,Nakanishi:2006zv,Chung:2017kj}, the Coma cluster \citep{Casoli:1991jr,Boselli:1997jk}, and the Antlia cluster \citep{Cairns:2019ju}.
\cite{Mok:2017ey} found that molecular gas is even more abundant in the Virgo galaxies than the field counterparts.
However, based on detailed mapping observations toward individual galaxies in the Virgo cluster, it is shown that there certainly exist galaxies whose molecular gas is being stripped by the ram pressure \citep[e.g., NGC~4330, NGC~4402, NGC~4438, NGC~4522,][]{Combes:1988jm,Vollmer:2008dk,Vollmer:2009zg,Vollmer:2012jz,Lee:2017ut,Cramer:2020nt}.

It is not obvious that all the galaxies in a cluster with different masses and evolutionary stages are experiencing the quenching processes in the same way.
In addition, there should be various galaxies under the different kinds and levels of quenching processes even within a cluster.
Molecular gas is unlikely to be as easily stripped as atomic gas \citep[e.g.,][]{Kenney:1986eu,Cortese:2016yh,Jachym:2017vd}, considering that atomic gas disk typically extends out to a few (or more) times the size of optical disk \citep[e.g.,][]{Warren:2004hx,Koribalski:2009zi,Koribalski:2018la} whereas molecular gas disk is confined within $50$~\% of the optical disk \citep{Young:1995jq} where the anchoring force per area is strong against the ram pressure.
Simply comparing field and cluster galaxies does not give us a clear conclusion on the molecular deficiency of cluster galaxies.
 
The clustocentric distance is one of the measures of the time since the first infall to the cluster (accretion phase) and has been used to investigate galaxy properties \citep[e.g.,][]{Andersen:1996vm,Coenda:2009ff,Coenda:2009kz,Sheen:2012nt,Gu:2013xa,Gu:2016td}, although the projected distance alone is not enough since there could be galaxies located at various clustocentric distances along the line of sight \citep{Smith:2015dv}.
If the clustocentric distance and clustocentric velocity are considered simultaneously, i.e., so-called ``phase-space diagram'' (PSD), it is claimed to be possible to classify galaxies according to their accretion phase \citep[e.g.,][]{Mahajan:2011qg,Oman:2013tf,Rhee:2017kl} and the degree of tidal mass loss \citep[e.g.,][]{Smith:2015dv,Rhee:2017kl} based on numerical simulations.
Many observational studies find that galaxies in different locations on the PSD have different properties, such as morphology, stellar mass, star formation, active galactic nuclei activity, and dust temperature \citep{Biviano:2002df,Mahajan:2011qg,Noble:2013he,Pimbblet:2013wp,Noble:2016wl,Hernandez-Fernandez:2014zx,Muzzin:2014ev}.

The cold gas properties of galaxies on the PSD have been also investigated.
\HI~gas observations toward cluster galaxies showed that galaxies that are expected to be affected by ram pressure on the PSD tend to have lower levels of atomic gas than field galaxies \citep{Jaffe:2015pq,Wang:2020av}.
\cite{Yoon:2017jl} investigated a relationship between \HI~morphology, as a measure of the ram-pressure stripping, and the location on the PSD of Virgo galaxies and showed that the galaxies with severely stripped \HI~disks are found deep inside the cluster.
For molecular gas properties on the PSD, on the other hand, there is no systematic study in nearby clusters but in distant clusters at $z\sim1.5-2.5$ \citep[e.g.,][]{Hayashi:2017dn,Wang:2018rz}.
\cite{Wang:2018rz} clearly showed that molecular gas properties are correlated with accretion phase in the cluster, where those closer to the cluster center and those with a smaller relative velocity to the cluster tend to be increasingly gas poor.

In this work, we investigate both the atomic- and molecular-gas and star-formation properties of the Virgo galaxies in the PSD with the largest CO and \HI~data so far by combining several datasets from the literature in order to understand the environmental effects on these galaxy properties.
The Virgo cluster is the nearest massive galaxy cluster to the Milky Way and therefore provides abundant multi-wavelength data.
The cluster is considered to be a dynamically young system \citep{Arnaboldi:2004ti,Aguerri:2005fo}, contains a relatively high fraction of blue spiral galaxies \citep{Cappellari:2013rg}, and shows an asymmetric structure of hot ICM traced in X-ray \citep{Bohringer:1994td}.

The structure of the paper is as follows:
we first introduce our sample and the data used in this study in Section~\ref{sec:sampledata}.
We present our results on cold-gas and star-formation properties of the Virgo galaxies in Section~\ref{sec:coldgassf} and briefly summarize the results in Section~\ref{sec:summaryresults}.
The environmental effects on cold gas and star formation properties of Virgo galaxies are discussed in Section~\ref{sec:discussions}.
Finally, we present out conclusions in Section~\ref{sec:summary}.
We adopt the coordinate of M~87, a brightest cluster galaxy in the Virgo cluster, of $(\alpha, \delta)=(12^{h}30^{m}49.4^{s}, +12^{\circ}23'28'')$ as the cluster center, a distance of $16.5$~Mpc \citep{Mei:2007wt}, $R_{200}$, the radius where the mean interior density is 200 times the critical density, of 5.38~deg \citep{McLaughlin:1999gx}, a velocity dispersion of 800 km s$^{-1}$ \citep{Binggeli:1993qt}, and the initial mass function (IMF) of \cite{Chabrier:2003oe} throughout this paper.

\section{Sample and Data} \label{sec:sampledata}

\begin{table*}
\begin{center}
\caption{Crossmatch summary. Numbers of secure/possible Virgo members.\label{tab:xmatch}}
\begin{tabular}{lcccccc}
\tableline
\tableline
& \HI~or H$_2$ & Telescope$^{\rm a}$ & Original & Matched$^{\rm b}$ & Detected$^{\rm b}$ & Reference$^{\rm c}$\\
\tableline
EVCC & -- & -- & 1,028/561 & -- & -- & 1\\
z0MGS & -- & -- & 15,738 & 266/455 & -- & 2\\
\tableline
Boselli+1995 & H$_2$ & S15 \& O20 & 24$^{\rm d}$ & 19/0 (16/0) & 15/0 (12/0) & 3\\
Chung+2009 & H$_2$ & F14 & 28$^{\rm e}$ & 14/4 (9/4) & 14/4 (9/4) & 4\\
Atlas3D & H$_2$ & I30 & 264 & 60/26 (57/25) & 12/5 (12/4) & 5\\
AMIGA & H$_2$ & F14 \& I30 & 285 & 0/6 (0/6) & 0/3 (0/3) & 6\\
HRS & H$_2$ & K12 & 344 & 50/20 (50/20) & 50/20 (50/20) & 7\\
COMING & H$_2$ & N45 & 147 & 10/7 (9/6) & 10/7 (9/6) & 8\\
ALFALFA & \HI & A305 & 5118 & 292/212 (117/97) & 292/212 (117/97) & 9\\
VIVA & \HI & VLA & 53$^{\rm f}$ & 45/7 (37/7) & 45/7 (37/7) & 10\\
Atlas3D & \HI & W & 260 & 39/2 (38/2) & 4/0 (3/0) & 11\\
\tableline
\end{tabular}
\end{center}
\vspace{-4mm}
\tablecomments{
$^{\rm a}$ S15: SEST 15~m; O20: Onsala 20~m; F14: FCRAO 14~m; I30: IRAM 30~m; K12: NRAO KP 12~m; N45: NRO 45~m; A305: Arecibo 305~m; W: WSRT; VLA: Very Large Array.\\
$^{\rm b}$ Secure members/possible members classified in the EVCC project and those values in parentheses for galaxies with $M_{\rm star}$ and SFR measurements in the z0MGS project.\\
$^{\rm c}$ 1: \cite{Kim:2014nb}; 2: \cite{Leroy:2019cu}; 3: \cite{Boselli:1995sf}; 4: \cite{Chung:2009xq}; 5: \cite{Young:2011sq}; 6: \cite{Lisenfeld:2011sy}; 7: \cite{Boselli:2014yq}; 8: \cite{Sorai:2019hs}; 9: \cite{Giovanelli:2005ua}; 10: \cite{Chung:2009ys}; 11: \cite{Serra:2012oc}.\\
$^{\rm d}$ Among the Virgo sample in \cite{Boselli:1995sf}, five (NGC~3258, NGC~4413, NGC~3414, NGC~3476, and NGC~3521) are not listed in EVCC.\\
$^{\rm e}$ \cite{Chung:2009xq} have observed 28 galaxies while $M_{\rm mol}$ values are presented only for 19 galaxies (see Table~3 in \cite{Chung:2009xq}). Additionally, we did not use $M_{\rm mol}$ of NGC~4576 since it is the total value for NGC~4567 and NGC~4568 as a pair.\\
$^{\rm f}$ Among the Virgo sample in \cite{Chung:2009ys}, IC~3418 is not included in EVCC.
}
\end{table*}

\begin{table}
\begin{center}
\caption{Numbers of secure/possible Virgo members for each subsample.\label{tab:data}}
\begin{tabular}{lcc}
\tableline
\tableline
Subsample & Matched & Detected\\
\tableline
H$_2$ & 132/54 & 75/26 \\
\HI  & 353/217 & 318/215 \\
H$_2$ \& \HI & 94/24 & 49/14 \\
\tableline
H$_2$ \& z0MGS  & 122/53 & 68/25 \\
\HI~\& z0MGS  & 169/102 & 134/100 \\
H$_2$ \& \HI~\& z0MGS & 90/24 & 46/14 \\
\tableline
H$_{2, M_{\rm star9}}$$^{\rm a}$  & 121/52 & 68/25\\
\HI$_{,M_{\rm star9}}$ & 124/70 & 89/68 \\
H$_{2, M_{\rm star9}}$ \& \HI$_{,M_{\rm star9}}$ & 90/23$^{\rm b}$ & 46/14$^{\rm c}$\\
\tableline
\end{tabular}
\end{center}
\vspace{-4mm}
\tablecomments{
$^{\rm a}$ Galaxies with $M_{\rm star}>10^9$~M$_\odot$, i.e., with $\Delta$(Field) measurements,\\
$^{\rm b}$ ``{\it CO+\HI-obs. sample}'',\\
$^{\rm c}$ ``{\it CO+\HI-det. sample}''.
}
\end{table}

\begin{figure*}[]
\begin{center}
\includegraphics[width=.49\textwidth, bb=0 0 593 596]{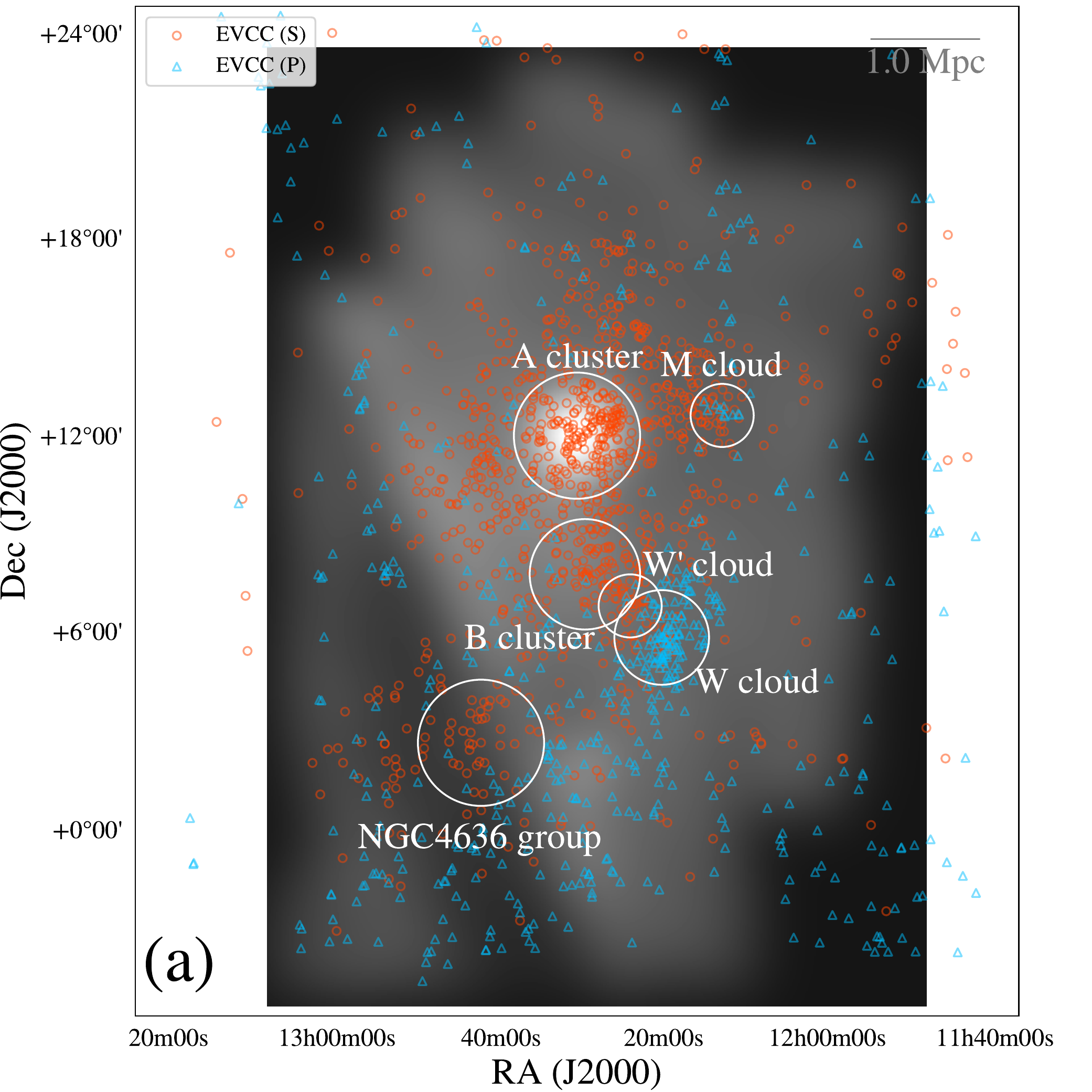}
\includegraphics[width=.49\textwidth, bb=0 0 593 596]{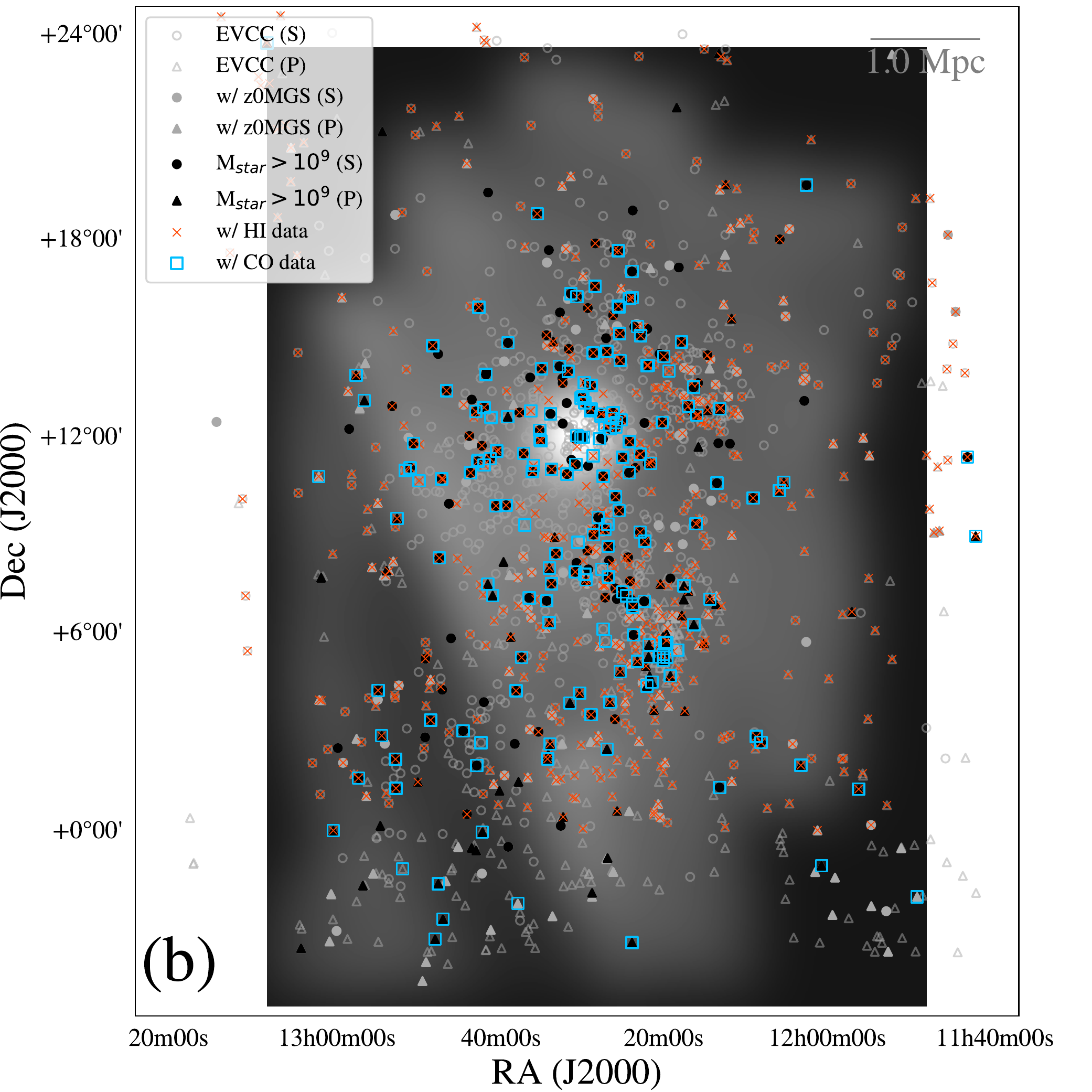}
\end{center}
\caption{
Sky distributions of (a) the secure (orange open circle) and possible (blue open triangle) member galaxies in the EVCC,
(b) those with $M_{\rm star}$ and SFR measurements (grey filled circle/triangle), those with $M_{\rm star}>10^9~M_\odot$ (black filled circle/triangle), those with CO measurements (blue open square), and those with \HI~measurements (orange X marks).
The grey open circle and triangle respectively indicate the secure and possible member galaxies in the EVCC without the measurements of $M_{\rm star}$, SFR, CO, or~\HI.
The substructures in the Virgo cluster are indicated with white open circles in the left panel.
Background is the ROSAT X-ray image \citep{Bohringer:1994td}.
}
\label{fig:radec}
\end{figure*}

\begin{figure*}[]
\begin{center}
\includegraphics[width=0.49\textwidth, bb=0 0 471 539]{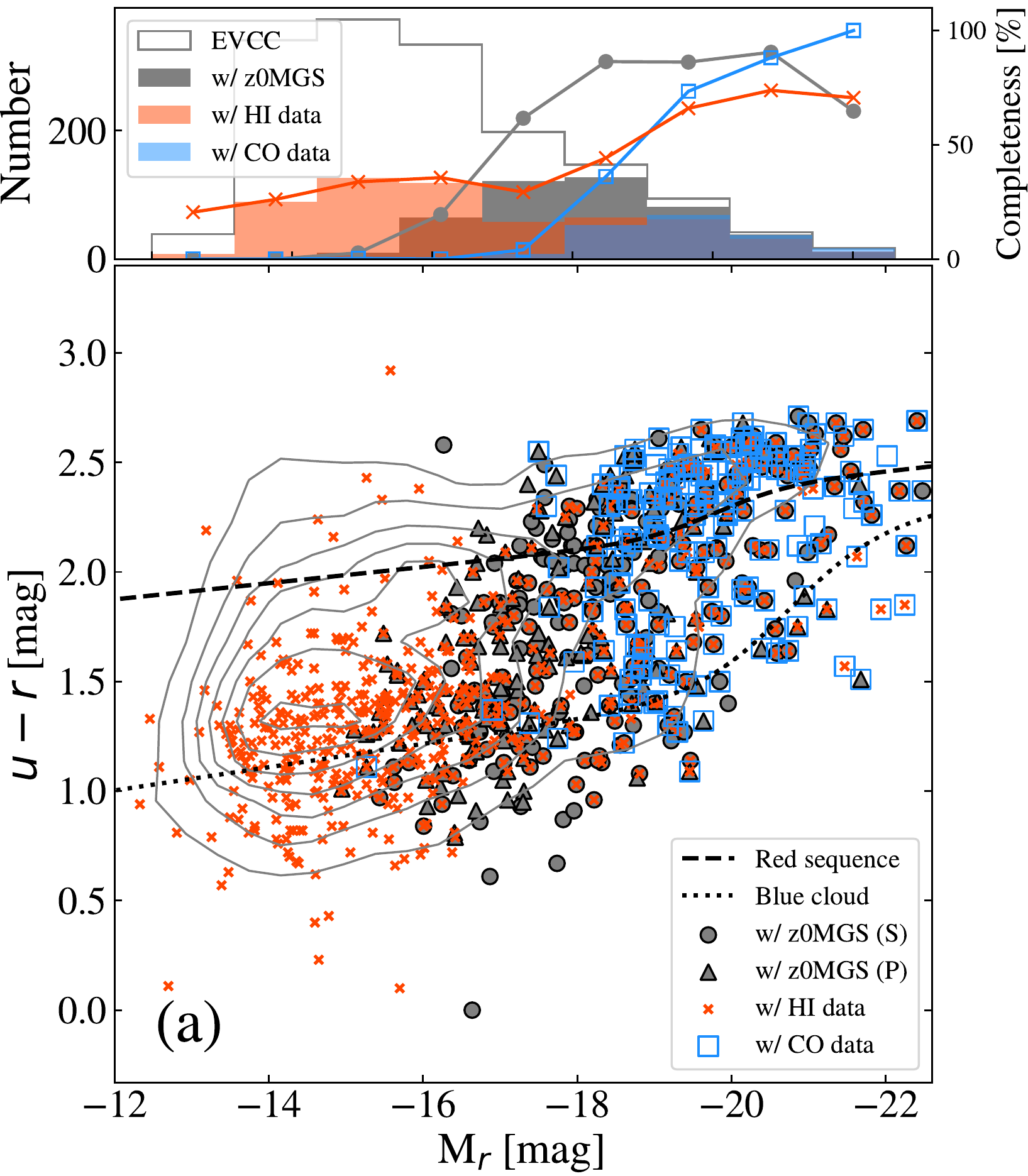}
\includegraphics[width=0.49\textwidth, bb=0 0 471 539]{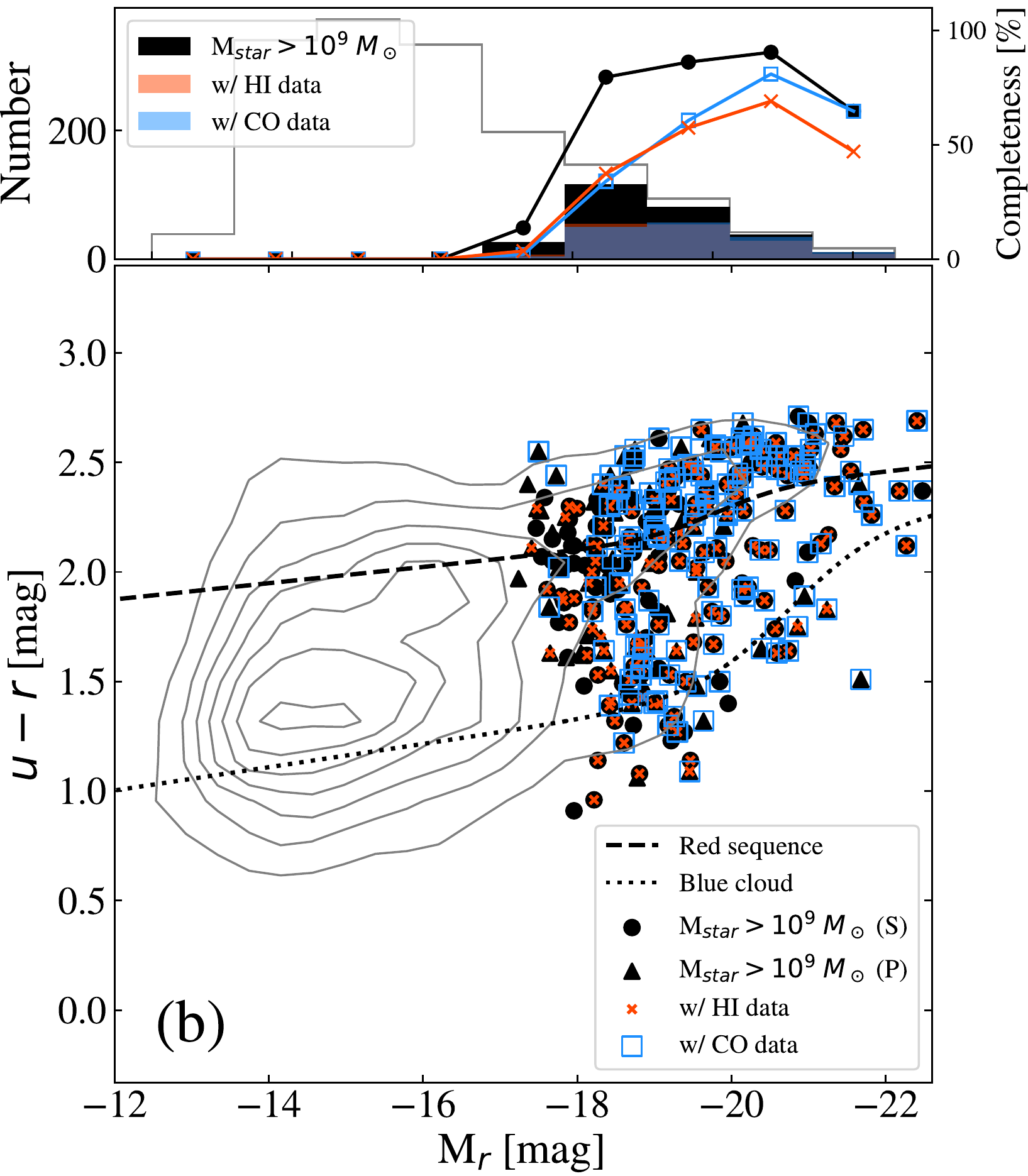}
\end{center}
\caption{
Crossmatch result between EVCC and literatures: color magnitude relations and the completeness of each subsample as a function of $r$-band magnitude for
(a) the galaxies without a $M_{\rm star}$ cut and (b) those with $M_{\rm star}>10^9$~$M_\odot$.
The grey contour in the both panel indicate the entire EVCC galaxies.
The galaxies with $M_{\rm star}$ and SFR, \HI, and CO measurements including upper limits are indicated as filled grey circles/triangles for the secure/possible members, orange X marks, and blue open square, respectively.
Black filled symbols indicate galaxies with $M_{\rm star}>10^9$~M$_\odot$, which are the main targets in this study.
The dashed and dotted lines indicate the red sequence and the blue cloud of local SDSS galaxies \citep{Baldry:2004tw}.
The fractions of galaxies listed in the z0MGS (grey), those with $M_{\rm star}>10^9$~M$_\odot$ (black), those with \HI~measurements (orange), and those with CO measurements (blue) among the EVCC are also shown in the upper panels of each plot.
}
\label{fig:z0mgs_evcc}
\end{figure*}

We adopted the Extended Virgo Cluster Catalog \citep[EVCC,][]{Kim:2014nb}, containing SDSS optical data of 1,589 galaxies, as a reference list of Virgo galaxies, then searched for their stellar mass ($M_{\rm star}$), star-formation rate (SFR), \HI~and CO data.
In this section, we first mention about the reference galaxies for the crossmatch especially in terms of the membership to the Virgo cluster in Section~\ref{sec:member}.
Then we describe the crossmatchs with literatures and derivation of each physical quantity in Section~\ref{sec:mstarsfr} for $M_{\rm star}$ and SFR,
Section~\ref{sec:mmol} for molecular gas mass ($M_{\rm H_2}$),
and Section~\ref{sec:matom} for atomic gas mass ($M_{\rm HI}$).
The crossmatch summary is presented in Section~\ref{sec:xmatch},
and we introduce key quantities adopted in this study to investigate star-formation and cold-gas properties of galaxies in Section~\ref{sec:key}.
In Section~\ref{sec:subsample}, we describe several subsamples treated in this study.
For galaxies with multiple CO or \HI~measurements, we prioritize data with detections rather than upper limits, and calculate error-weighted means in case of multiple detections.

\subsection{Membership to the Virgo cluster and substructures} \label{sec:member}

In the EVCC, the listed galaxies are categorized into two categories in terms of the membership to the Virgo cluster, ``secure'' and ``possible'' members, based on the redshift comparison with a cluster infall model \citep{Kim:2014nb}.
It is known that there are several substructures in the Virgo cluster \citep[Figure~\ref{fig:radec} (a),][]{Binggeli:1985os,Binggeli:1987ur,Binggeli:1993qt,Sandage:1985bw}: A (or M~87) cluster, B (or M~49) cluster, NGC~4636 group, M cloud \citep{Ftaclas:1984ae}, W' cloud \citep{de-Vaucouleurs:1961ln}, and W cloud \citep{de-Vaucouleurs:1961ln}.
Additionally, the galaxies at declination less than $5^\circ$ are called the ``southern extension''.
According to the EVCC classification, galaxies in the W and M clouds are partially and largely categorized to possible members, respectively.
The mean velocity of these W and M cloud galaxies is $\sim2000$~km~s$^{-1}$ ($\sim1000$~km~s$^{-1}$ for the main body of the Virgo cluster) and they are considered to be located at twice the distance of the Virgo cluster \citep{Binggeli:1993qt}.
The M and W clouds galaxies are also categorized as possible members in the classical Virgo Cluster Catalogue \citep[VCC,][]{Binggeli:1985os}.

On the other hand, \cite{Yoon:2012pj,Yoon:2017fi} found that Ly$\alpha$-absorbing warm-hot gas ($T=10^{4-6}$~K) is associated with the substructures in the Virgo cluster including both the M and W clouds, and suggested that the observed warm gas is part of large-scale gas flows into the Virgo cluster.
Such warm-hot gas is preferentially found in the outskirts rather than the central region of galaxy clusters \citep{Yoon:2012pj,Yoon:2017fi,Burchett:2018jv}, which is also predicted in the hydrodynamic simulations of the galaxy clusters \citep{Emerick:2015dm}, although the mass contribution of the warm-hot gas is not so large compared to the X-ray emitting hot gas \citep[$\sim3$~\%,][]{Burchett:2018jv}.
Thus, we use both the secure and possible members of the Virgo cluster in the following analysis.
The effect of the inclusion of the possible members on our results is discussed in Section~\ref{sec:membership}.

\subsection{Stellar mass and star-formation rate}\label{sec:mstarsfr}

For the $M_{\rm star}$ and SFR data, the EVCC is crossmatched with the sample galaxies in the project of ``z = 0 Multiwavelength Galaxy Synthesis'' \citep[z0MGS,][]{Leroy:2019cu}, resulting in 721 galaxies.
\cite{Leroy:2019cu} measured $M_{\rm star}$ and SFR of $\sim15,750$ galaxies located at $<50$~Mpc which are observed with the Galaxy Evolution Explorer \citep[GALEX,][]{Martin:2005wd} and the Wide-field Infrared Survey Explorer \citep[WISE,][]{Wright:2010oi}.
The z0MGS project derived empirical relationship between $M_{\rm star}$ and WISE photometry at 3.4~$\mu$m, and SFR and WISE (22~$\mu$m) and GALEX (far-ultraviolet or near ultraviolet) photometry using data obtained in the GALEX-SDSS-WISE Legacy Catalog \citep[GSWLC,][]{Salim:2016wi}.
\cite{Leroy:2019cu} prioritizes a ``hybrid'' SFR estimation for galaxies which are observed and detected both with GALEX and WISE, i.e., SFR estimation with ultraviolet and mid-infrared data to account for both the obscured and unobscured components.
The GSWLC project combined GALEX and WISE photometry with SDSS observations and conducted population synthesis modeling using the CIGALE code \citep{Boquien:2019ex}.
\cite{Salim:2016wi} assumed a Chabrier IMF to derive $M_{\rm star}$ and SFR \citep{Chabrier:2003oe}.

\subsection{Molecular gas mass}\label{sec:mmol}

For the CO data, we used the literature data of \cite{Boselli:1995sf}, \cite{Chung:2009xq}, Atlas 3D \citep{Young:2011sq}, the Analysis of the interstellar Medium in Isolated GAlaxies \citep[AMIGA,][]{Lisenfeld:2011sy}, Herschel Reference Survey \citep[HRS,][]{Boselli:2014yq}, and CO Multi-line Imaging of Nearby Galaxies \citep[COMING,][]{Sorai:2019hs} and found 186 galaxies (175 with $M_{\rm star}$ and SFR measurements) without double counting.
$M_{\rm H_2}$ is calculated from CO line luminosity ($L_{\rm CO}'$) with a CO-to-H$_2$ conversion factor for the Milky Way ($\alpha_{\rm CO}=3.21$~M$_\odot$ [K km s$^{-1}$ pc$^2$]$^{-1}$, which corresponds to $X_{\rm CO}=2.0\times10^{20}$~cm$^{-2}$~[K km s$^{-1}$]$^{-1}$) as
\begin{equation}
M_{\rm H_2} = 1.36 \times \alpha_{\rm CO} L_{\rm CO}',
\end{equation}
where a factor of 1.36 accounts for the Helium contribution to mass \citep[e.g.,][]{Bolatto:2013vn}.
Here we do not consider the metallicity dependence of $\alpha_{\rm CO}$ since our sample galaxies are massive galaxies whose $M_{\rm star}$ are larger than $\sim10^9$ M$_\odot$, which is near the turnover mass for the mass-metallicity relation of local galaxies \citep{Andrews:2013nx} and roughly corresponds to the Solar metallicity \citep[$12+\log({\rm O/H})=8.69$,][]{Asplund:2009fh}.
At metallicities larger than the Solar value, the variation in $\alpha_{\rm CO}$ is small \citep[Figure~9 of][]{Bolatto:2013vn}.
Additionally, it is claimed that the mass-metallicity relation of galaxies does not strongly depend on the galaxy environments and the metallicity of cluster galaxies is found to be slightly higher than that of field galaxies \citep[$\sim0.05$ dex, e.g.,][]{Mouhcine:2007dd,Cooper:2008td,Ellison:2009dd}.

\subsection{Atomic gas mass}\label{sec:matom}

For the \HI~data, we crossmatched the EVCC galaxies with literature data obtained in The Arecibo Legacy Fast ALFA survey \citep[ALFALFA,][]{Giovanelli:2005ua}, Atlas 3D \citep{Serra:2012oc}, and VLA Imaging of Virgo Spirals in Atomic Gas \citep[VIVA,][]{Chung:2009ys}, resulting in 570 galaxies (271 with $M_{\rm star}$ and SFR measurements) without double counting.
$M_{\rm atom}$ is calculated based on the optically thin assumption as
\begin{equation}
M_{\rm HI} = 1.36 \times 2.36 \times 10^5 D^2 \times \Sigma_i S_i \Delta v,
\end{equation}
where the factor of 1.36 again accounts for the Helium contribution to mass, $D$ is the distance to the galaxies in Mpc (16.5~Mpc) and $\Sigma_i S_i \Delta v$ is the summation over the total emission in each channel in Jy km s$^{-1}$ \citep[e.g.,][]{Wild:1952bt}.

\subsection{Crossmatch summary}\label{sec:xmatch}
The sky distribution of the secure/possible member galaxies in the EVCC is shown in Figure~\ref{fig:radec} and the crossmatch summary is presented in Table~\ref{tab:xmatch}.
Table~\ref{tab:data} summarizes the subsamples in this study.
Figure~\ref{fig:z0mgs_evcc} shows the color-magnitude relation of target galaxies in this study and the EVCC galaxies with data of the Sloan Digital Sky Survey (SDSS) Data Release 7 \citep[DR7,][]{Abazajian:2009cz}.
We can see that our sample galaxies are biased to bright objects, or massive galaxies that are expected to be less affected by their environments than less massive galaxies \citep{Peng:2010eq}.
Our main target is limited to massive galaxies with $M_{\rm star}>10^9$~M$_\odot$ that is the lowest $M_{\rm star}$ value of field samples for a comparison (sed Section~\ref{sec:field}).
The completeness of our sample with $M_{\rm star}>10^9$~M$_\odot$ is $\sim65-90$~\% at absolute $r$-band magnitudes of $\lesssim-16.7$ mag (see the histograms in Figure~\ref{fig:z0mgs_evcc}).

\subsection{Key quantities}\label{sec:key}

\begin{figure*}[]
\begin{center}
\includegraphics[width=\textwidth, bb=0 0 1852 943]{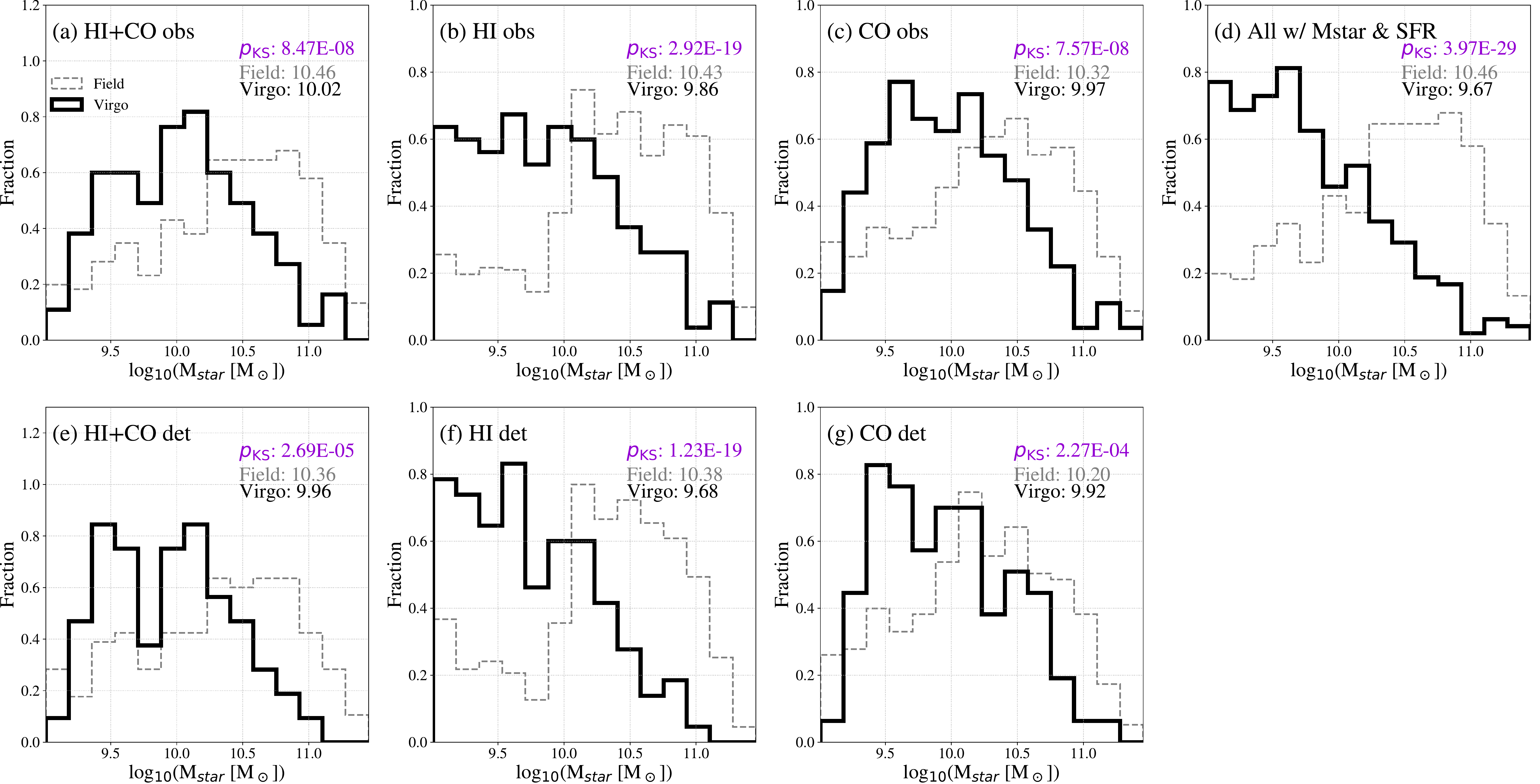}
\end{center}
\caption{
The histograms of the stellar mass of seven subsamples of the Virgo galaxies with $M_{\rm star}>10^9$~M$_\odot$ in this study (thick black line).
The histograms for field galaxies in the same classification are also shown as grey dashed lines.
The medians of $\log{(M_{\rm star}~{\rm [M_\odot]})}$ for each subsample and the $p$-value of the KS test for a comparison between the Virgo and field galaxies are also shown on the upper left corner.
}
\label{fig:hist_mstar}
\end{figure*}

With the physical values estimated above, we calculate the following ``key quantities'' of galaxies:
specific SFR (${\rm sSFR}={\rm SFR}/M_{\rm star}$), 
atomic-gas mass fraction ($\mu_{\rm HI}=M_{\rm HI}/M_{\rm star}$),
molecular-gas mass fraction ($\mu_{\rm H_2}=M_{\rm H_2}/M_{\rm star}$),
(total) gas mass fraction [$\mu_{\rm gas}=(M_{\rm HI}+M_{\rm H_2})/M_{\rm star}=M_{\rm gas}/M_{\rm star}$],
molecular gas to atomic gas mass ratio ($R_{\rm H_2}=M_{\rm H_2}/M_{\rm HI}$),
star-formation efficiency (SFE) from atomic gas (SFE$_{\rm HI}={\rm SFR}/M_{\rm HI}$),
SFE from molecular gas (SFE$_{\rm H_2}={\rm SFR}/M_{\rm H_2}$), and
SFE from total gas (SFE$_{\rm gas}={\rm SFR}/M_{\rm gas}$).

\subsection{Subsamples}\label{sec:subsample}
In the following sections, we compare the key quantities of Virgo galaxies on the best effort basis, i.e., we use all the galaxies with measurements including both detections and upper limits, of the numerators and denominators of the key quantities (hereafter, ``{\it best-effort sample}''), and thus the number of galaxies in each plot is different depending on the quantities.
To assess the sample bias due to this treatment, we also compute these values for galaxies with all the measurements of $M_{\rm star}$, SFR, \HI, and CO (``{\it CO+\HI-obs. sample}'', 113 galaxies) or galaxies with detections (``{\it CO+\HI-det. sample}'', 70 galaxies) (see Table~\ref{tab:data}).

Figure~\ref{fig:hist_mstar} shows the $M_{\rm star}$ distribution of our seven subsample galaxies with $M_{\rm star}>10^9$~M$_\odot$, such as
(a) galaxies with \HI~and CO measurements (``{\it CO+\HI-obs. sample}''),
(b) galaxies with \HI~measurements,
(c) galaxies with CO measurements,
(d) galaxies with $M_{\rm star}$ and SFR measurements,
(e) galaxies with \HI~and CO detections (``{\it CO+\HI-det. sample}''),
(f) galaxies with \HI~detections, and
(g) galaxies with CO detections.
The ``{\it best-effort sample}'' consists of (a) for $\mu_{\rm gas}$, SFE$_{\rm gas}$, and $R_{\rm H_2}$, (b) for $\mu_{\rm HI}$ and SFE$_{\rm HI}$, (c) for $\mu_{\rm H_2}$ and SFE$_{\rm H_2}$, and (d) for SFR and sSFR.

The $\log{(M_{\rm star} [{\rm M}_\odot])}$ medians of the field and Virgo galaxies and the $p$-value of the KS test are also presented in each panel (see Section~\ref{sec:field} for the field sample).
The $M_{\rm star}$ medians of the field galaxies are larger than those of the Virgo galaxies and the KS test shows that the difference of the $M_{\rm star}$ distribution between the field and the Virgo galaxies is statistically significant.
Additionally, the KS test among the different Virgo subsamples show that the $M_{\rm star}$ distribution is different for (a) versus (d), (a) versus (f), (c) versus (d), \deleted{and} (c) versus (f)\added{, (d) versus (e), (d) versus (g), and (f) versus (g)}.
The $M_{\rm star}$ medians of these subsample suggest that the galaxies with CO measurements are more massive (\replaced{$\sim10^{10.1}$~M$_\odot$}{$\sim10^{10.0}$~M$_\odot$}) than those only with $M_{\rm star}$ and SFR measurements and those with \HI~detections (\replaced{$\sim10^{9.9}$~M$_\odot$}{$\sim10^{9.7}$~M$_\odot$}).
Therefore, the ``{\it best-effort sample}'' in plots for SFR and sSFR is slightly biased to lower mass systems compared to the other subsamples.

\section{Cold-gas and star-formation properties} \label{sec:coldgassf}

In this section, we focus on the key quantities (sSFR, $\mu_{\rm HI}$, $\mu_{\rm H_2}$, $\mu_{\rm gas}$, $R_{\rm H_2}$, SFE$_{\rm HI}$, SFE$_{\rm H_2}$, and SFE$_{\rm gas}$) to investigate the relationship between the cold-gas and star-formation properties of galaxies and locations in the Virgo cluster.
In this study, we consider that two samples are different when the $p$-value of the Kolmogorov-Smirnov (KS) test becomes less than 0.05.
The correlation between two values is assessed based on the Spearman's rank-order correlation coefficient and the reliability on the correlation coefficient is assessed with the $p$-value, i.e., we consider that two quantities are correlated when the $p$-value is less than 0.05.

We first compare the Virgo galaxies and the field galaxies in Section~\ref{sec:field}, and then investigate the dependence of the key quantities on the clustocentric distance in Section~\ref{sec:distance}, on the local galaxy density in Section~\ref{sec:density}, and on the accretion phase in the cluster in Section~\ref{sec:accretion}.

\subsection{Comparison with field galaxies}\label{sec:field}

\begin{figure*}[]
\begin{center}
\includegraphics[width=150mm, bb=0 0 1398 1384]{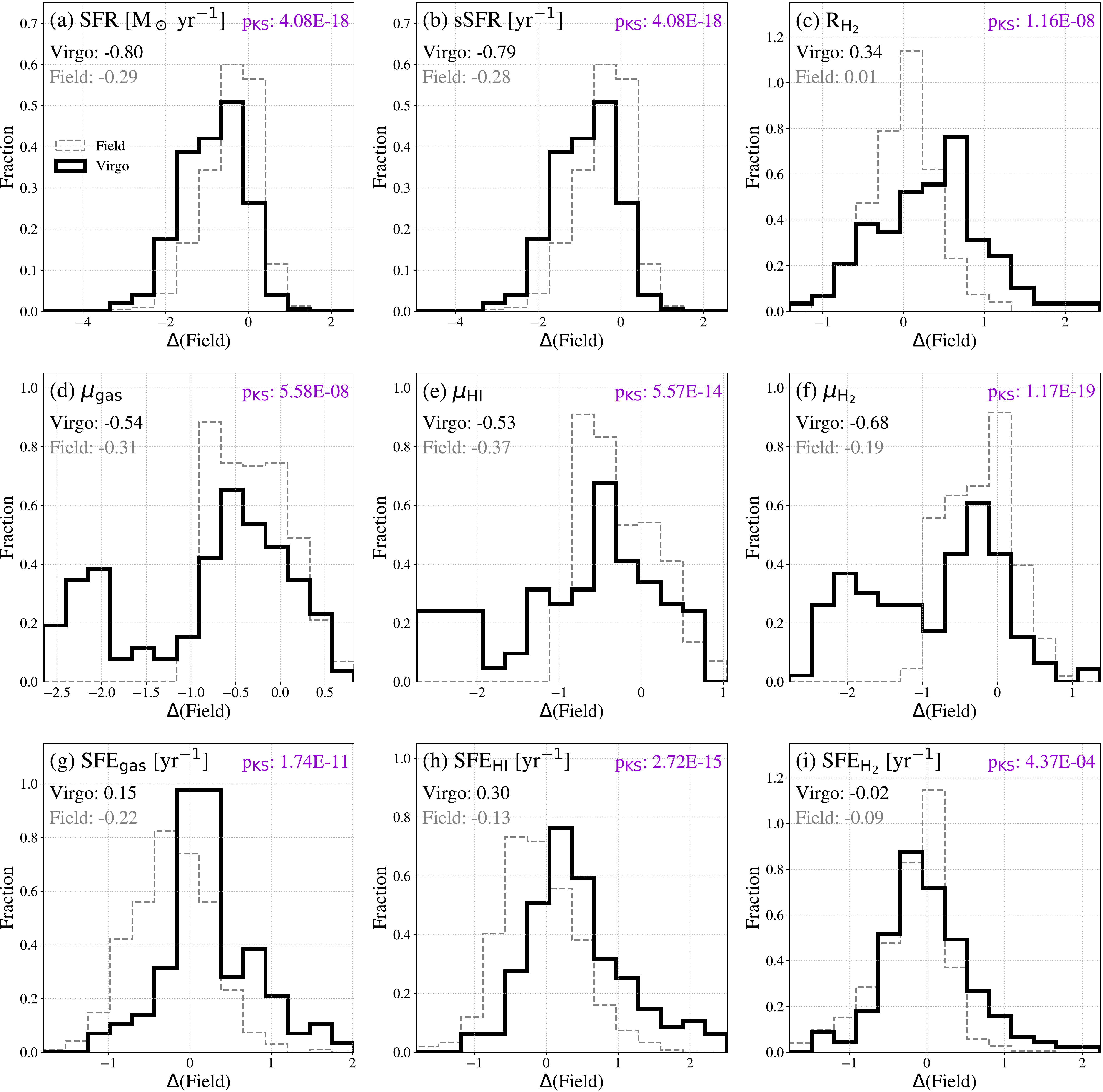}
\end{center}
\caption{
Comparison between the Virgo galaxies (solid line) and field galaxies (dashed line) with $M_{\rm star}>10^9$~M$_\odot$: histograms for the offsets from the field relations, $\Delta({\rm Field})$.
Note that $\Delta({\rm Field})$s of SFR and sSFR are calculated based on the SFMS definition in \cite{Speagle:2014by}.
The medians for the Virgo and field galaxies are indicated at the upper left corner.
The $p$-value for the KS test comparing the Virgo and field galaxies is indicated at the upper right corner.
The left peak of the bimodal distributions of the gas fractions mainly consists of galaxies with upper limits.
Figure~\ref{fig:comp_field2b} shows histograms for the galaxies with CO/\HI~detection.
}
\label{fig:comp_field2}
\end{figure*}

We compare the Virgo galaxies and field galaxies in Figure~\ref{fig:comp_field2}.
Note that the left peaks for the gas fractions consist of the galaxies with the upper limits.
In Appendix~\ref{sec:fieldcomparison_plots}, we also show the comparison of the key quantities of the Virgo and field galaxies without the upper limits (Figure~\ref{fig:comp_field2b}).
To derive empirical relationships of key quantities as a function of stellar mass in the field galaxies, we make use of data of the extended GALEX Arecibo SDSS Survey \citep[xGASS,][]{Catinella:2018ib} for \HI-related relations, and the extended CO Legacy Database for GASS \citep[xCOLD GASS,][]{Saintonge:2017ve} for H$_2$-related relations.
We adopted the star-forming main sequence (SFMS) galaxies defined in \cite{Speagle:2014by}.
The difference between the Virgo and field galaxies [$\Delta({\rm Field})$] in Figure~\ref{fig:comp_field2} is calculated as
$\Delta({\rm Field})=\log_{10}{\rm (X_{\rm Virgo}(M_{\rm star})/X_{\rm Field}(M_{\rm star}))}$,
where $X_{\rm Virgo}(M_{\rm star})$ and $X_{\rm Field}(M_{\rm star})$ are key quantities for the Virgo and field galaxies at stellar mass of $M_{\rm star}$, which are derived based on polynomial regression with an order of three except for SFR and sSFR.
Speagle's SFMS is adopted for $X_{\rm Field}(M_{\rm star})$ for SFR and sSFR.

We find that the Virgo galaxies have lower medians of SFR, sSFR, gas fractions, and higher medians of $R_{\rm H_2}$ and SFEs than the field galaxies in both cases that the sample includes those with CO or \HI~upper limits (``{\it best-effort sample}'', Figure~\ref{fig:comp_field2}) and that the sample is limited to those with CO or \HI~detections (Figure~\ref{fig:comp_field2b} in Appendix~\ref{sec:fieldcomparison_plots}).
These differences are confirmed to be statistically significant according to the KS test.
Even in case we limit the sample to galaxies with CO and \HI~measurements, i.e., the ``{\it CO+\HI-obs. sample}'', the conclusions do not change.

\begin{figure*}[]
\begin{center}
\includegraphics[width=\textwidth, bb=0 0 1867 939]{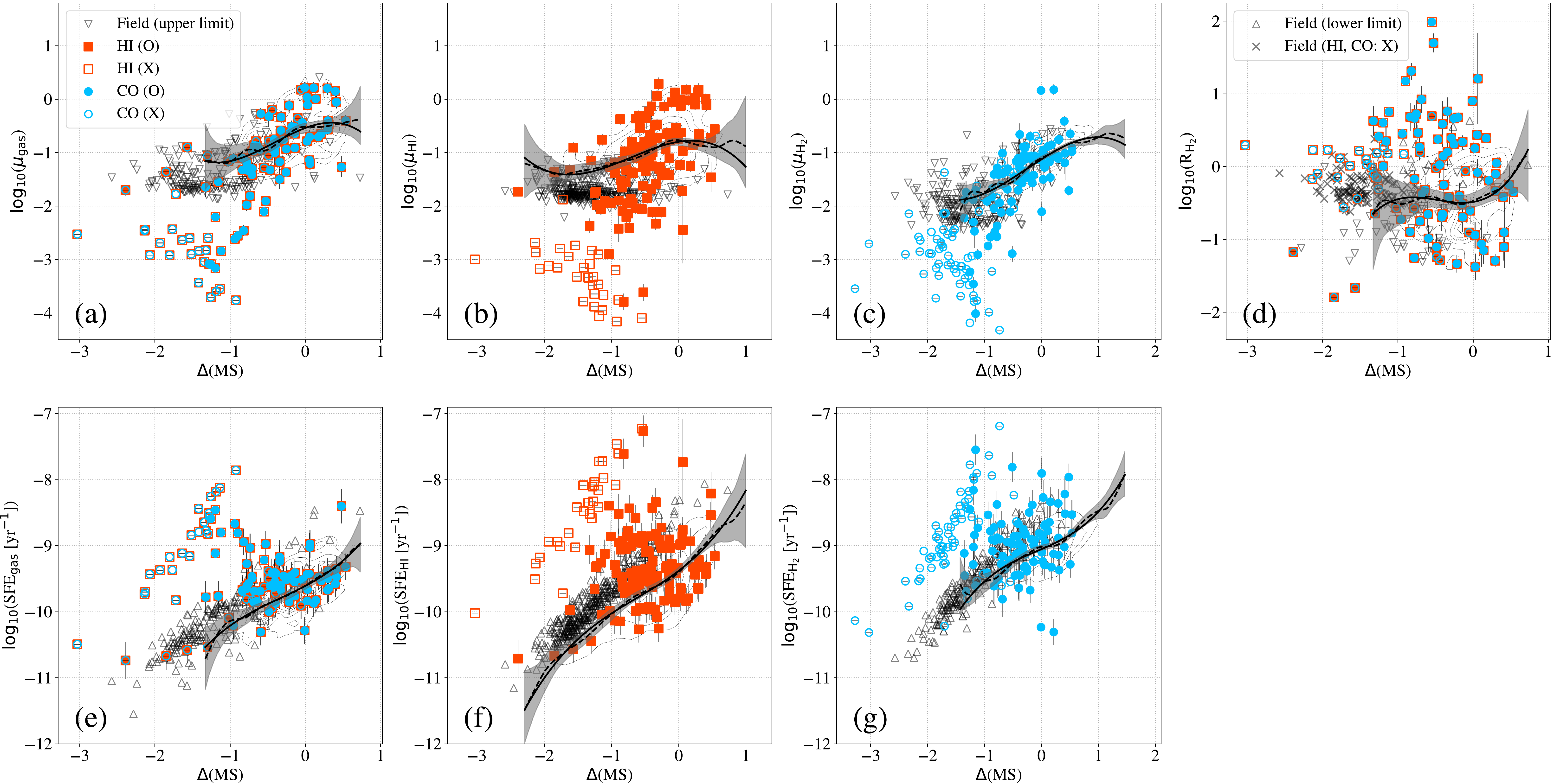}
\end{center}
\caption{
Comparison between the Virgo galaxies and field galaxies with $M_{\rm star}>10^9$~M$_\odot$: relationships between the offsets from the main sequence of star-forming galaxies, $\Delta({\rm MS})$, with gas fractions, $R_{\rm H_2}$, and SFEs.
The Virgo galaxies with an \HI~detection, an \HI~upper limit, a CO detection and a CO upper limit are indicated as filled orange square, open orange square, filled blue circle, and open blue circle, respectively.
The field relations are estimated with CO- or \HI-detected galaxies, i.e., star-forming galaxies, and indicated as dashed (with non-parametric fitting) and (with 3rd-order polynomial fitting) solid lines in the panels (c)-(i).
Grey-shaded regions are the errors of the non-parametric fitting and estimated by the bootstrap method with a 95~\% confidence interval.
The contours indicate field galaxies used to derive the field relations.
The lower/upper limits of field galaxies are indicated as open triangles/inverted triangles.
For the $R_{\rm H_2}$ panel, the field galaxies with \HI~and CO upper limits are indicated as X marks.
}
\label{fig:comp_field3}
\end{figure*}

It has been claimed that both the $\mu_{\rm H_2}$ and SFE$_{\rm H_2}$ values of galaxies are related to the offset from the SFMS of the field galaxies, $\Delta$(MS)$=\log_{10}{\rm (SFR/SFR_{MS})}$, where ${\rm SFR_{MS}}$ is SFR of the SFMS field galaxies for a fixed stellar mass and redshift \citep[e.g.,][]{Saintonge:2012nj,Genzel:2015gn,Scoville:2017jw,Ellison:2020kf}.
These relations are independent of the local galaxy number density at least for field and group galaxies \citep{Koyama:2017lf}.
Figure~\ref{fig:comp_field3} compares the relationships of $\Delta$(MS) with cold-gas mass fractions, $R_{\rm H_2}$, and SFEs for the Virgo and field galaxies.
Although most Virgo galaxies follow the same relation as field galaxies, the Virgo galaxies tend to be distributed at the lower $\Delta$(MS) regime in these plots.
Interestingly, some Virgo galaxies, especially below the main sequence ($\Delta$(MS)$\sim-1$), seem to have decreased $\mu_{\rm H_2}$ and elevated $R_{\rm H_2}$ and SFEs compared to field galaxies for fixed $\Delta$(MS).
We compare the medians of gas fractions, $R_{\rm H_2}$, and SFEs of the field and the Virgo galaxies with CO or \HI~detection depending on the quantities on the SFMS ($-0.5<\Delta$(MS)$<0.5$) as well as those below the SFMS ($-1.5<\Delta$(MS)$<-0.5$).
We find that the Virgo galaxies have lower gas fractions and higher $R_{\rm H2}$ and SFEs than the field galaxies \deleted{both on and} below the SFMS.
The KS test shows that the differences between the field and the Virgo galaxies are statistically significant for \deleted{both} the subsamples \deleted{on and} below the SFMS\deleted{ except for the $\mu_{\rm gas}$ values for the SFMS galaxies}.
\added{For the SFMS galaxies, only the differences in $\mu_{\rm H_2}$ and $R_{\rm H_2}$ are statistically significant where $\mu_{\rm H_2}$ and $R_{\rm H_2}$ of the Virgo galaxies are lower and higher than those of the field galaxies, respectively.}
It should be also noted that \HI-related values of the Virgo galaxies show a wider variation than the field galaxies at a fixed $\Delta$(MS).

In the following sections, we use the $\Delta({\rm Field})$ values, i.e., the values from which $M_{\rm star}$-dependences of field galaxies have been subtracted, in order to investigate the environmental effect without the $M_{\rm star}$ effect.

\subsection{Dependence on the clustocentric distance}\label{sec:distance}

\begin{figure*}[]
\begin{center}
\includegraphics[width=150mm, bb=0 0 1442 1833]{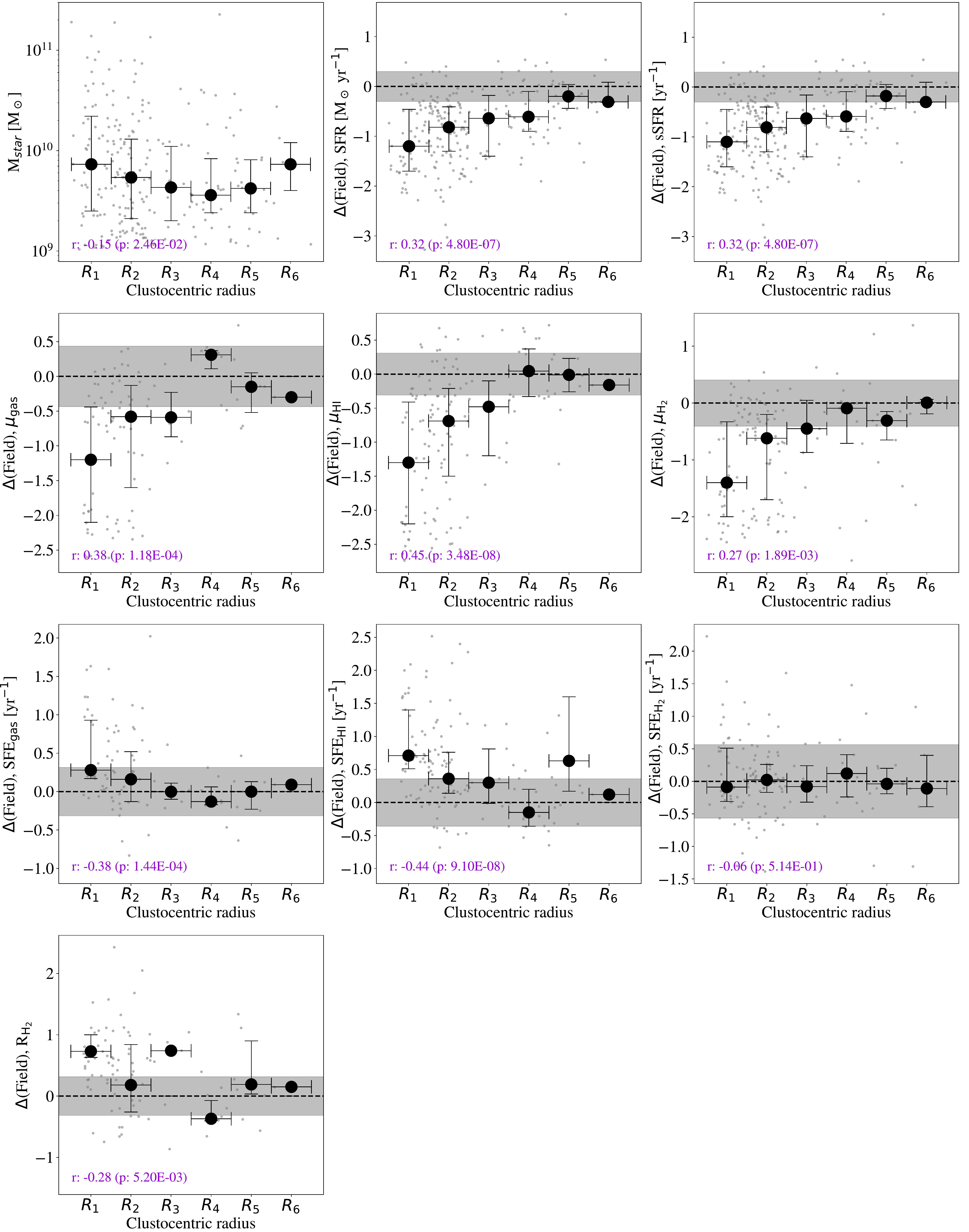}
\end{center}
\caption{
Radial variation of key quantities (median, and, the 1st and 3rd quartiles) whose $M_{\rm star}$ dependence have been subtracted, i.e., $\Delta({\rm Field})$ values.
The individual data is indicated as grey filled circles.
Zero of each key quantity indicates the values of the field galaxies (dashed line) and the grey-shaded regions indicate the 1-$\sigma$ of the $\Delta({\rm Field})$ values for the field galaxies.
Note that the numbers of galaxies used to plot in each panel are different.
The Spearman's rank-order correlation coefficient $r$ and the $p$-value are shown on the lower left corner of each panel in purple.
The W-cloud galaxies that are mostly in the $R_{3}$ bin are excluded from this plot.
}
\label{fig:radial_wul}
\end{figure*}

The galaxy properties as a function of clustocentric distance have been investigated in many previous studies showing that galaxies nearer to the cluster center tend to be more passive and \HI-deficient \citep[e.g.][]{Giovanelli:1985kt,Solanes:2001nq}.
To investigate the radial variation of the galaxy properties, we divided our Virgo galaxies into six radial bins where each bin spans $0.5 R_{200}$ as
$R_{\rm proj}/R_{200}<0.5$ ($R_1$),
$0.5 \leq R_{\rm proj}/R_{200}<1.0$ ($R_2$),
$1.0 \leq R_{\rm proj}/R_{200}<1.5$ ($R_3$),
$1.5 \leq R_{\rm proj}/R_{200}<2.0$ ($R_4$),
$2.0 \leq R_{\rm proj}/R_{200}<2.5$ ($R_5$), and
$R_{\rm proj}/R_{200} \geq 2.5$ ($R_6$),
where $R_{\rm proj}$ is the projected distance from a galaxy to M~87.
The most outskirt galaxy in our ``{\it best-effort sample}'' locates at $\sim3 R_{200}$.

Figure~\ref{fig:radial_wul} shows the medians, 1st and 3rd quartiles of the key quantities for galaxies in each category.
There is a radial tendency in SFR and cold gas fractions in galaxies where these values are lower for galaxies nearer to the cluster center than their counterparts in the outskirts.
The Spearman's rank-order correlation coefficient and the $p$-value confirm that the clustocentric distance positively correlates to SFR, sSFR, and gas fractions with correlation coefficients of \replaced{$0.2-0.4$}{$0.3-0.5$}, and negatively correlates to SFE$_{\rm gas}$, SFE$_{\rm HI}$, and $R_{\rm H_2}$ with correlation coefficients ranging from \replaced{$-0.3$}{$-0.4$} to \replaced{$-0.2$}{$-0.3$}.
\replaced{SFEs and $R_{\rm H_2}$ do}{SFE$_{\rm H_2}$ does} not show a clear dependence on the clustocentric distance\deleted{, whereas SFE$_{\rm HI}$ at $R_4$ is lower than those at the other radii ($p$-values of the $R_3$ vs $R_4$ and $R_4$ vs $R_5$ comparisons are $0.0042$ and $0.0019$, respectively)}.

Additionally, the radial trends of the gas fractions are not completely {\it continuous} but there seems to be a {\it bend} between $R_3$ and $R_4$.
Note that the galaxies in the W-cloud galaxies are excluded in Figure~\ref{fig:radial_wul} since the cold-gas and star-formation properties are different from the surrounding galaxies.
The inclusion of the W-cloud galaxies causes a {\it jump} rather than a {\it bend} between $R_3$ and $R_4$ in sSFR and $\mu_{\rm H_2}$.
Here, the W-cloud galaxies are defined on the PSD by eye, and the definition is presented in Section~\ref{sec:accretion}.
The properties of the W-cloud galaxies are discussed more in Sections~\ref{sec:accretion} and \ref{sec:preprocess}.

For a comparison with the field galaxies (zero corresponds to the values of field galaxies), we can see that \replaced{SFR}{gas fractions, $R_{\rm H_2}$, SFE$_{\rm gas}$, and SFE$_{\rm H_2}$} are already comparable to field galaxies at $>R_5$ whereas the medians of \replaced{sSFR and the gas fractions}{SFR and sSFR} are slightly lower than those of the field galaxies.
On the other hand, \deleted{SFEs, especially} SFE$_{\rm HI}$\deleted{,} \replaced{tend}{tends} to be higher than the field galaxies even at \replaced{$R_6$}{$>R_5$}.

Note that the same tendency is observed even when we consider only the sample with CO and \HI~measurements, i.e., the ``{\it CO+\HI-obs. sample}''.
If we limit the sample to galaxies with CO and \HI~detection (``{\it CO+\HI-det. sample}''), the radial trends seen in SFR, sSFR, and gas fractions disappear.

\subsection{Dependence on the galaxy number density}\label{sec:density}

\begin{figure*}[]
\begin{center}
\includegraphics[width=150mm, bb=0 0 1422 1390]{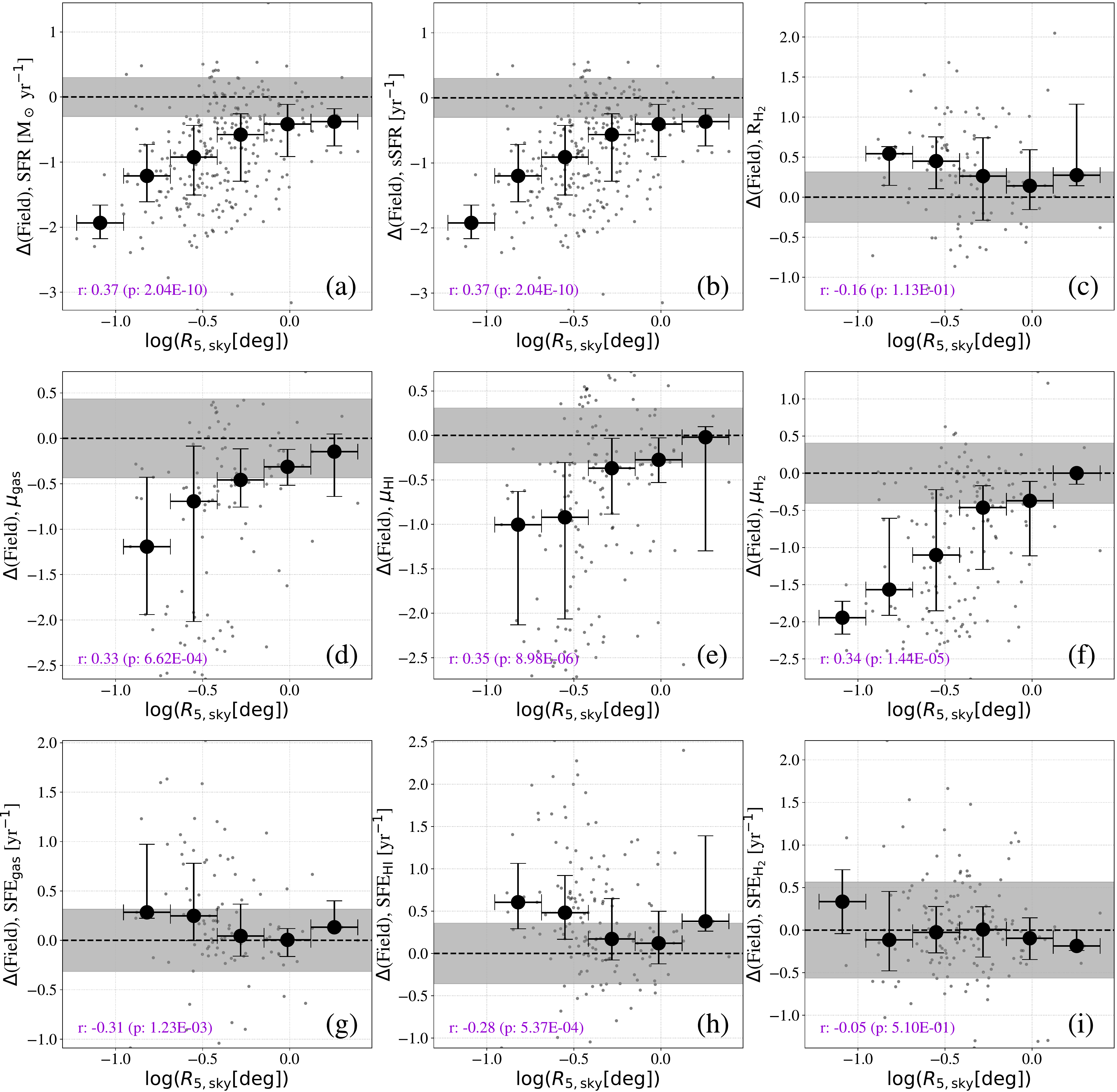}
\end{center}
\caption{
The relationships between $R_{\rm 5, sky}$ with the key quantities [$\Delta({\rm Field})$ values].
Symbols are same as in Figure~\ref{fig:radial_wul}.
}
\label{fig:numberdensity_radec}
\end{figure*}

Galaxy number density is one of the important parameters to understand the dominant quenching mechanism in cluster galaxies.
If the dominant quenching mechanism is related to the local number density rather than the cluster ICM or potential, galaxy-galaxy interactions are expected to be important.

We adopt two distances to see the dependences of the key quantities on the local galaxy density:
the distances to the 5-th nearest galaxy (1) on the sky ($R_{\rm 5, sky}$) and (2) on the PSD ($R_{\rm 5, PSD}$) where the vertical and horizontal axes are respectively $R_{\rm proj}/R_{200}$ and $\Delta v_{\rm gal}/\sigma_{\rm Virgo}$ and both values are dimensionless quantities ($\Delta v_{\rm gal}$ is the velocity difference between a galaxy to M~87).
We make use of these two values complementarily to evaluate the effect of galaxy-galaxy interactions including the galaxy harassment (multiple high-speed galaxy-galaxy encounters) since some galaxy pairs can have a smaller $R_{\rm 5, sky}$ by chance even though the actual distance between them along the line of sight would be much larger than the projected distance, which can dilute the true relations between $R_{\rm 5, sky}$ and the key quantities.
On the other hand, the comparison with $R_{\rm 5, PSD}$ would miss the effect of the galaxy harassment.

First, we compare $R_{\rm 5, sky}$ or $R_{\rm 5, PSD}$ with the projected clustocenteric distance (plots are shown in Figure~\ref{fig:dist_density} of the Appendix~\ref{sec:localdensity_plots}).
Overall, there are moderate and weak positive correlations between $R_{\rm 5, sky}$ and $R_{\rm 5, PSD}$ with the clustocenteric distance, respectively.
The obtained positive correlations suggest that the dependence of the key quantities on clustocentric distance seen in Figure~\ref{fig:radial_wul} might just reflect the dependence on the local densities, and vice versa.
The exclusion of the W-cloud galaxies increases $R_{\rm 5, sky}$ and $R_{\rm 5, PSD}$ at the $R_3$ bin, but does not drastically change the overall trend.

Next, we compare $R_{\rm 5, sky}$ or $R_{\rm 5, PSD}$ with the key quantities.
Plots for $R_{\rm 5, sky}$ and $R_{\rm 5, PSD}$ are shown in Figure~\ref{fig:numberdensity_radec} and Figure~\ref{fig:numberdensity} in Appendix~\ref{sec:localdensity_plots}, respectively.
We observe positive correlations between $R_{\rm 5, sky}$ with SFR, sSFR, and gas fractions with correlation coefficients of $\sim0.3-0.4$, and \deleted{a} negative \replaced{correlation}{correlations} with \added{SFE$_{\rm gas}$ and} SFE$_{\rm HI}$ with a correlation coefficient of \replaced{$-0.16$}{$\sim -0.3$}.
In case of the $R_{\rm 5, PSD}$, we find positive correlations with SFR, sSFR, and $\mu_{\rm HI}$ with correlation coefficients of $\sim0.2$.
The exclusion of the W-cloud galaxies basically does not change the results for $R_{\rm 5, sky}$ \replaced{but results in an additional positive correlation with $\mu_{\rm gas}$}{and $R_{\rm 5, PSD}$}.
\deleted{On the other hand for the  $R_{\rm 5, PSD}$, the correlation with $\mu_{\rm HI}$ becomes insignificant when excluding the W-cloud galaxies.}
If we limit sample to those with both the CO and \HI~measurements (``{\it CO+\HI-obs. sample}''), the results do not change but the \replaced{correlations}{correlation} between \deleted{$R_{\rm 5, sky}$ with $\mu_{\rm HI}$ and} $R_{\rm 5, PSD}$ with $\mu_{\rm HI}$ \replaced{become}{becomes} insignificant.
Next if we limit sample to those with both the CO and \HI~detections (``{\it CO+\HI-det. sample}''), all the correlations for $R_{\rm 5, sky}$ \added{and $R_{\rm 5, PSD}$} become insignificant\deleted{ but $R_{\rm 5, PSD}$ positively correlates to SFR, SFE$_{\rm gas}$, and SFE$_{\rm HI}$}.

Therefore the radial tendency observed in SFR, sSFR, and gas fraction can be partially attributed to the radial dependence of the local galaxy density where the local environment of galaxies becomes denser at nearer to the cluster center.

\subsection{Dependence on the accretion phase to the cluster}\label{sec:accretion}

\begin{figure*}[]
\begin{center}
\includegraphics[width=\textwidth, bb=0 0 1061 495]{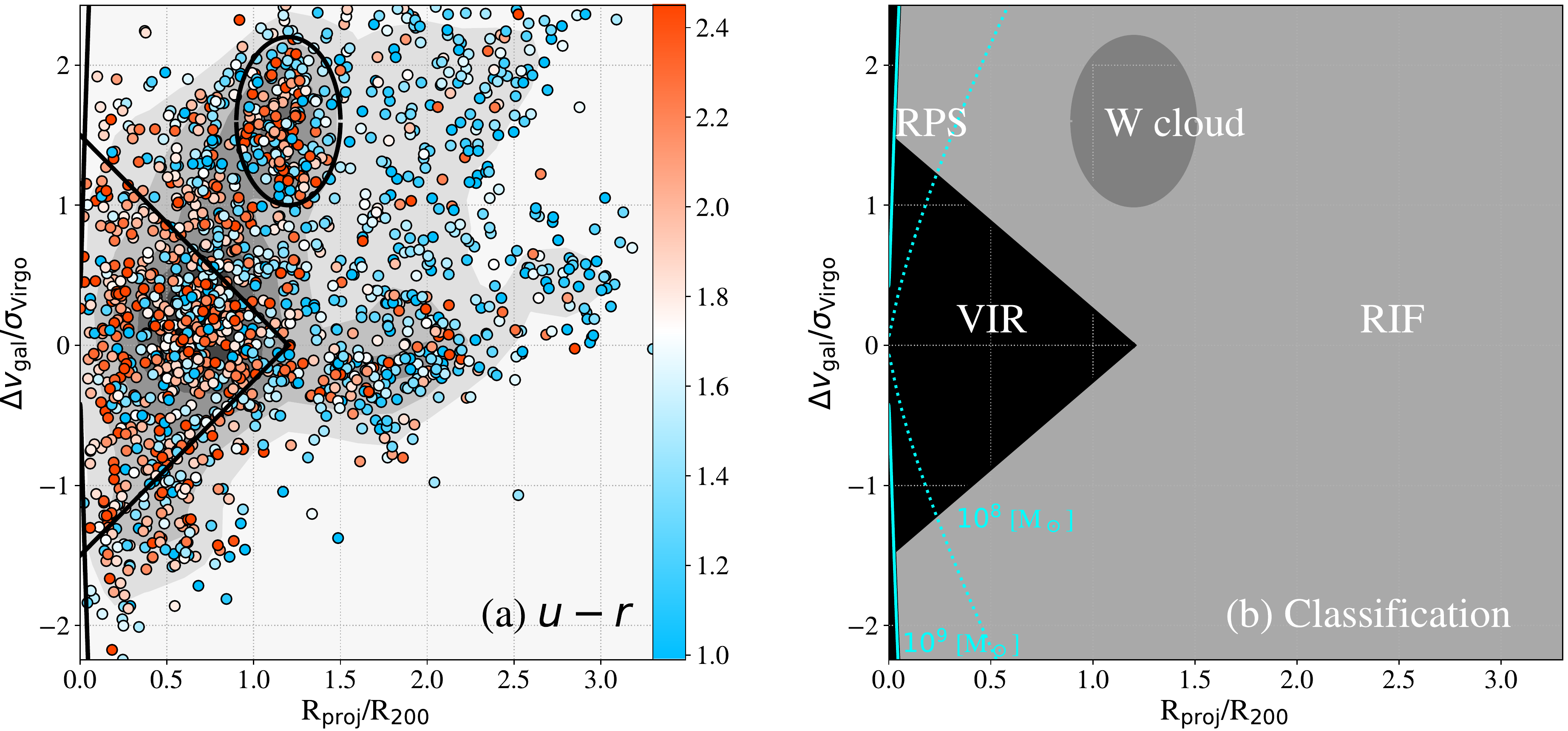}
\end{center}
\caption{
(a) SDSS $u-r$ color, and (b) galaxy classification on the PSD of the whole EVCC galaxies.
The light blues lines in the panel (b) indicate the boundary where the whole cold gas (both atomic and molecular gas) in galaxies with $M_{\rm star}$ of $10^8$ (doted) and $10^9$~M$_\odot$ (solid) is removed by the ram pressure.
The assumed $\mu_{\rm gas}$ values are 2.3 and 1.0 for galaxies with $M_{\rm star}$ of $10^8$ and $10^9$~M$_\odot$, respectively \citep{Hunt:2015yd}.
}
\label{fig:psd_sdss}
\end{figure*}

\begin{figure*}[]
\begin{center}
\includegraphics[width=150mm, bb=0 0 1402 1824]{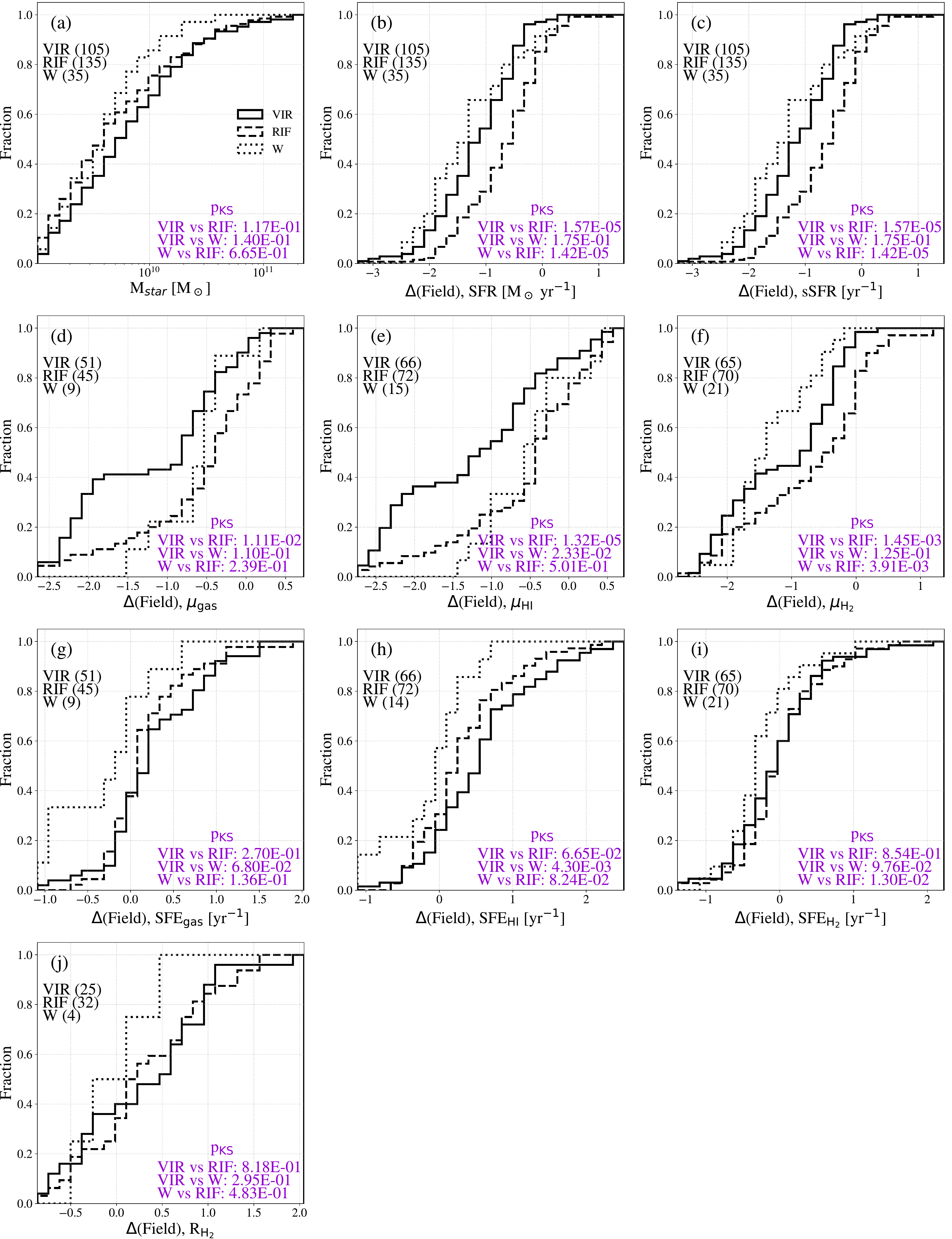}
\end{center}
\caption{
Cumulative histograms of stellar mass and the key quantities of galaxies in the VIR (solid line), RIF (dashed line), and W cloud regions (dotted line).
The numbers of galaxies in the VIR and RIF regions are indicated at the upper left corner in each panel.
The $p$-values of the KS test is indicated in purple at the lower right corner.
}
\label{fig:hist_c2_inclp}
\end{figure*}

\begin{figure*}[]
\begin{center}
\includegraphics[width=150mm, bb=0 0 1422 1390]{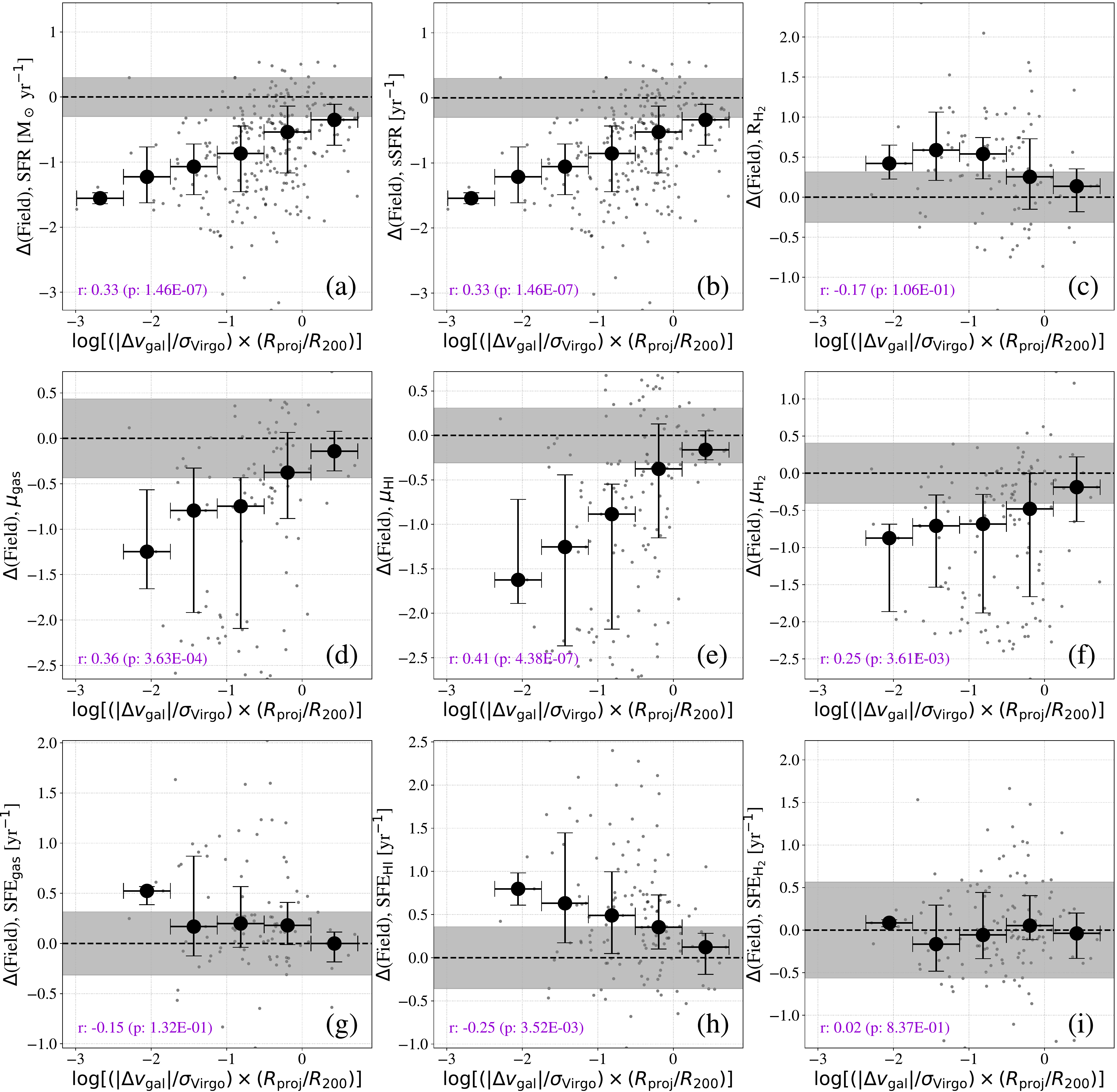}
\end{center}
\caption{
Relationship between the accretion phase, $(|\Delta v_{\rm gal}|/\sigma_{\rm Virgo})\times(R_{\rm proj}/R_{200})$, with the key quantities.
The W-cloud galaxies are not included in this plots.
Symbols are same as in Figure~\ref{fig:radial_wul}.
}
\label{fig:accretionphase}
\end{figure*}

\begin{table}
\begin{center}
\caption{ICM and galaxy models for a boundary line of the ram-pressure stripping (RPS) galaxies \label{tab:str_model}}
\begin{tabular}{lcc}
\tableline\tableline
Parameters & Values & References\\
\tableline
&--- ICM model$^{a}$ ---&\\
$\rho_0$ & $4.0\times10^{-2}$ cm$^{-3}$ & \cite{Vollmer:2001ir}\\
$R_{\rm C}$ & 13.4 kpc & \cite{Vollmer:2001ir}\\
$\beta$ & 0.5 & \cite{Vollmer:2001ir}\\
\tableline
&--- Galaxy model$^{b}$ ---&\\
$M_{\rm d,star}$ & $10^{9}$ M$_\odot$ & this study\\
$M_{\rm d,gas}$ & $1.0 \times M_{\rm d,star}$ & \cite{Hunt:2015yd}\\
$R_{\rm d,star}$ & $0.9$ kpc & this study\\
$R_{\rm d,gas}$ & $1.7 \times R_{\rm d,star}$ & \cite{Cayatte:1994hr}\\
\tableline
\end{tabular}
\end{center}
\tablecomments{
$^{\rm a}$ Standard $\beta$-model of ICM gas density distribution \citep{Cavaliere:1976mb}:
$\rho_{\rm ICM}(r_{\rm 3D})=\rho_0\left[1+\left(\frac{r_{\rm 3D}}{R_{\rm C}}\right)^2\right]^{-3\beta/2}$, where $r_{\rm 3D}=(\pi/2)R_{\rm proj}$, $\rho_0$, $R_{\rm C}$, and $\beta$ are 3D distance from the cluster center, the number density at the center of the cluster, core radius, and the power of the distribution.\\
$^{\rm b}$ Exponential surface density distribution:
$\Sigma_{\rm d, star~or~gas}(r) = \frac{M_{\rm d, star~or~gas}}{2 \pi R_{\rm d, star~or~gas}^2} \exp{\left(\frac{-r}{R_{\rm d, star~or~gas}}\right)}$, where $M_{\rm d,star}$ and $M_{\rm d,gas}$ are the stellar and gas masses, and $R_{\rm d,star}$ and $R_{\rm d,gas}$ are the scale lengths of stellar and gas disk.
We adopted a RPS boundary for the $M_{\rm star}=10^9$~M$_\odot$ galaxies where their entire gas (both atomic and molecular gas) is removed.
The $R50$ for the $M_{\rm star}\sim10^9$~M$_\odot$ galaxies listed in EVCC is converted to the scale length with an empirical relation between $R50$ and scale length of $R50=1.69 \times R_{\rm d,star}$.
}
\end{table}

In order to see the effect of ram pressure, we divide the galaxies into two groups in terms of accretion phase on the projected PSD of the cluster:
(1) ``virialized'' (VIR) and ``stripping'' (RPS),
(2) ``recent infall'' (RIF).
The RPS and VIR galaxies are respectively expected to be currently affected and to have been affected by the ram pressure but it turned out that there is no RPS galaxy in our sample.
\cite{Jaffe:2015pq} shows that on the projected PSD, galaxies oscillates in position and velocity drawing a ``wedge'' and fall into a cluster core in their simulations \citep[galaxy trajectories on the PSD shown in Figure~4 of][]{Jaffe:2015pq}.
However, it should be noted that not all galaxies in the VIR region have necessarily experienced the ram-pressure stripping.
The position on the PSD provides a statistical probability that each galaxy is in a different accretion phase and is not univocally associated to the quenching mechanisms.
The probability for Milky Way-type galaxies being stripped by ram pressure is estimated to be $\gtrsim60$~\% in the VIR region of a $z\sim0.2$ cluster \citep[Figure~6 of][]{Jaffe:2015pq}.
The W-cloud galaxies are treated separately from the other RIF galaxies.
Figure~\ref{fig:psd_sdss} shows the galaxy distribution on the PSD, color-coded according to the SDSS $(u-r)$ color and the galaxy classification on the PSD.

The classification is basically done following the procedures presented in \cite{Jaffe:2015pq} and \cite{Yoon:2017jl}.
The VIR region is enclosed by the line connecting the points of $(R_{\rm proj}/R_{200}$, $|\Delta v_{\rm gal}|/\sigma_{\rm Virgo})=(1.2, 0.0)$ and $(0.0,1.5)$.
Although there is no RPS galaxy in our sample, the RPS region is determined with the balance between the ram pressure, $P_{\rm ram}=\rho_{\rm ICM} v_{\rm 3D,gal}^2$, and the galaxy's anchoring force per area, $\Pi_{\rm gal}=2\pi G \Sigma_{\rm star} \Sigma_{\rm gas}$, where $\rho_{\rm ICM}$, $v_{\rm 3D,gal}$, $G$, $\Sigma_{\rm d,star}$, and $\Sigma_{\rm d,gas}$ are the ICM density distribution, 3D velocity of the galaxy ($v_{\rm 3D,gal}=\sqrt{3} v_{\rm obs,gal}$), gravitational constant, surface mass density of stellar and gas disks, respectively \citep{Gunn:1972kc}.
The ram pressure felt by a member galaxy is a function of the clustocentric velocity of the galaxy and the density distribution of the ICM.
On the other hand, the galaxy's anchoring force per area is a function of surface densities of stellar and gas components at a galactic radius where stripping is active.
The adopted boundary line in this study represents a galaxy with a stellar mass of $10^{9}$ M$_\odot$ and a scale length of 0.9~kpc (the median value for EVCC galaxies with $M_{\rm star}\sim10^9$~M$_\odot$).
This RPS boundary is a measure to assess whether or not the entire gas of a $10^{9}$ M$_\odot$ galaxy is stripped by ram pressure, i.e., even the gas at the galaxy center where the anchoring force is the strongest is removed.
Generally, cold gas in less massive galaxies tends to be more easily stripped by ram pressure (see Figure~\ref{fig:psd_sdss}).
The model parameters to calculate a boundary line for the RPS galaxies are summarized in Table~\ref{tab:str_model}.

The W-cloud galaxies are defined on the PSD by eye as mentioned in Section~\ref{sec:distance}.
We consider that galaxies in an ellipse on the PSD with a central coordinate and semi-major/semi-minor axes of $(R_{\rm proj}/R_{200}, |\Delta v_{\rm gal}|/\sigma_{\rm Virgo})=(1.2, 1.6)$ and 0.6/0.3, respectively, are the W-cloud galaxies.

Figure~\ref{fig:hist_c2_inclp} shows the comparison of cumulative histogram of the key quantities for the VIR, RIF, and W-cloud galaxies (the distributions of the key quantities on the PSD are presented in Appendix~\ref{sec:psd_plots}).
The VIR and W-cloud galaxies tend to have lower medians of SFR, sSFR, and gas fractions than the RIF galaxies.
Furthermore, the W-cloud galaxies have lower medians of \added{SFR,} sSFR\replaced{ and}{,} $\mu_{\rm H_2}$\added{, and SFEs} than the VIR galaxies.
The W-cloud galaxies have higher \replaced{SFE$_{\rm H_2}$}{$\mu_{\rm gas}$ and $\mu_{\rm HI}$} \deleted{and lower $R_{\rm H_2}$} medians than the \deleted{VIR and} RIF galaxies.
Based on the KS test, the following differences are significant:
(1) SFR, sSFR, and all the gas fractions between the VIR and RIF galaxies,
(2) \replaced{$\mu_{\rm H_2}$ and SFE$_{\rm H_2}$}{$\mu_{\rm HI}$ and SFE$_{\rm HI}$} between the VIR and W-cloud galaxies, and
(3) \added{SFR,} sSFR, \replaced{all the gas fractions}{$\mu_{\rm H_2}$}, and SFE$_{\rm H_2}$ between the W-cloud and RIF galaxies.

If we limit the sample to galaxies with \HI~or CO measurements (``{\it CO+\HI-obs. sample}''), all the the differences in \deleted{(2) and the differences in} (3) \deleted{except for $\mu_{\rm H_2}$ and SFE$_{\rm H_2}$} disappear \added{while the difference of SFE$_{\rm HI}$ becomes significant}.
If we limit the sample to galaxies with \HI~or CO detections (``{\it CO+\HI-det. sample}''), on the other hand, \added{only} the differences of SFR and sSFR between the VIR and RIF galaxies\deleted{, the differences of SFR, SFE$_{\rm gas}$, and SFE$_{\rm H_2}$ between the VIR and W-cloud galaxies, and the difference of SFE$_{\rm H_2}$ between the W-cloud and RIF galaxies} are statistically significant.

When we do not distinguish the W-cloud galaxies from the RIF galaxies, the VIR galaxies tend to have lower SFR, \replaced{$\mu_{\rm gas}$, $\mu_{\rm HI}$}{sSFR, gas fractions}, and higher $R_{\rm H_2}$\added{, SFE$_{\rm gas}$, SFE$_{\rm HI}$} medians than the RIF galaxies, while no significant difference is seen in \replaced{$\mu_{\rm H_2}$ or SFEs}{SFE$_{\rm H_2}$}.
SFR, sSFR, \replaced{$\mu_{\rm gas}$, and $\mu_{\rm HI}$}{gas fractions, and SFE$_{\rm HI}$} are statistically different between VIR and RIF galaxies, according to the KS test.

The product of $(|\Delta v_{\rm gal}|/\sigma_{\rm Virgo})\times(R_{\rm proj}/R_{200})$ is claimed to be a measure of the accretion epoch where the galaxies with the lower value were accreted earlier time \citep{Noble:2013he}.
Figure~\ref{fig:accretionphase} shows the relationship between $(|\Delta v_{\rm gal}|/\sigma_{\rm Virgo})\times(R_{\rm proj}/R_{200})$ with the key quantities of galaxies without those in the W cloud.
There are positive correlations between the accretion epoch and SFR, sSFR, and gas fractions with the correlation coefficients of \replaced{$\sim0.2-0.5$}{$\sim0.3-0.4$}, i.e., galaxies with smaller clustocentric distance and velocity tend to have lower star-formation activity and cold-gas contents.
\added{Additionally, SFE$_{\rm HI}$ negatively correlates to the product of $(|\Delta v_{\rm gal}|/\sigma_{\rm Virgo})\times(R_{\rm proj}/R_{200})$ with a correlation coefficient of $-0.25$.}
The inclusion of the W-cloud galaxies diminishes the significance of the correlation with $\mu_{\rm H_2}$ ($r$ and $p$-value of \replaced{$-0.01$ and $0.9$}{$0.12$ and $0.1$}, respectively) while the \replaced{correlation}{negative correlations} with \replaced{SFE$_{\rm H_2}$ becomes}{SFE$_{\rm gas}$ and $R_{\rm H_2}$ become} significant ($r$ and $p$-value of \replaced{$0.21$ and $0.007$}{$-0.23$ and $0.02$ for SFE$_{\rm gas}$} \added{and those of $-0.26$ and $0.007$ for $R_{\rm H_2}$}, respectively) by decreasing $\mu_{\rm H_2}$\replaced{ and increasing SFE$_{\rm H_2}$}{, SFE$_{\rm gas}$ and $R_{\rm H_2}$} at the largest bin of $(|\Delta v_{\rm gal}|/\sigma_{\rm Virgo})\times(R_{\rm proj}/R_{200})$.

The results do not drastically change \deleted{(the correlation coefficients slightly reduce to $\sim0.2-0.4$)} even if we limit the sample to galaxies with \HI~and CO measurements (``{\it CO+\HI-obs. sample}'').
On the other hand, if we limit the sample to galaxies with \HI~and CO detections (``{\it CO+\HI-det. sample}''), \replaced{only the correlation with SFR remains to be significant}{all the correlations disappear}.

\section{Summary of the results} \label{sec:summaryresults}

In the previous sections, we obtained the following results:
\begin{itemize}
\item The 1,589 EVCC galaxies are crossmatched with 10 literatures' sample and reduced to 132/54 (secure/possible members) galaxies with CO measurement, 353/217 galaxies with \HI~measurement, and 94/24 galaxies with both the CO and \HI~measurements (Figures~\ref{fig:radec} and \ref{fig:z0mgs_evcc} in Section~\ref{sec:sampledata}).

\item The galaxy numbers are reduced to 121/52 (CO), 124/70 (\HI), and 90/23 (CO+\HI) when we limit the sample to $M_{\rm star}>10^9$~M$_\odot$ galaxies, for which we can subtract the stellar mass dependence of the field galaxies from the key quantities (Figures~\ref{fig:radec} and \ref{fig:z0mgs_evcc} in Section~\ref{sec:sampledata}).

\item Compared to field galaxies, our Virgo sample have lower SFR, sSFR, cold gas fractions ($\mu_{\rm gas}$, $\mu_{\rm HI}$, $\mu_{\rm H_2}$) and higher SFEs (SFE$_{\rm gas}$, SFE$_{\rm HI}$, SFE$_{\rm H_2}$) \added{and $R_{\rm H_2}$}.
Most our Virgo galaxies follow the same relations of $\Delta$(MS) with the gas fractions and SFEs as the field galaxies, while the Virgo galaxies with low $\Delta$(MS) tend to have low $\mu_{\rm H_2}$ and high SFE$_{\rm H_2}$ compared to field galaxies.
Dispersions of the \HI-related quantities are larger for the Virgo galaxies than field galaxies at fixed $\Delta$(MS) (Figures~\ref{fig:comp_field2} and \ref{fig:comp_field3} in Section~\ref{sec:field} and Figures~\ref{fig:comp_field} and \ref{fig:comp_field2b} in Appendix~\ref{sec:fieldcomparison_plots}).

\item For the dependence of the $\Delta({\rm Field})$ values on the clustocentric distance of the key quantities, there are positive correlations with SFR, sSFR, and gas fractions with correlation coeffecients of \replaced{$0.2-0.4$}{$0.3-0.5$}, and negative correlations with SFE$_{\rm gas}$, SFE$_{\rm HI}$ and $R_{\rm H_2}$ with correlation coefficients ranging from \replaced{$-0.3$ to $-0.2$}{$-0.4$ to $-0.3$}.
There seems to be a bend in these values between $1.5 R_{200}$ and $2.0 R_{200}$.
The inclusion of W-cloud galaxies decreases $\mu_{\rm H_2}$ at $\sim(1.0-1.5) R_{200}$.
No significant dependence on the clustocentric distance is found for SFE$_{\rm H_2}$ (Figure~\ref{fig:radial_wul} in Section~\ref{sec:distance}).

\item For the dependence of the $\Delta({\rm Field})$ values on the projected distance to the 5-th nearest neighbour on the sky, there are positive correlations with SFR, sSFR and gas fractions with correlation coefficient of \replaced{$\sim0.2-0.4$}{$\sim0.3-0.4$} and \deleted{a} weak negative \replaced{correlation}{correlations} with \added{SFE$_{\rm gas}$ and} SFE$_{\rm HI}$ with a correlation coefficient of \replaced{$-0.16$}{$\sim -0.3$}.
In case of the distance to the 5-th nearest neighbour on the PSD, there are correlations with SFR, sSFR, and $\mu_{\rm HI}$ with correlation coefficients of $\sim0.2$.
There also exist positive correlations between these distances with the clustocentric distance (Figure~\ref{fig:numberdensity_radec} in Section~\ref{sec:density} and Figures~\ref{fig:dist_density} and \ref{fig:numberdensity} in Appendix~\ref{sec:localdensity_plots}).

\item For the dependence of the $\Delta({\rm Field})$ values on the accretion phase of galaxies in the cluster, SFR, sSFR, and all the gas fractions in the VIR galaxies are lower than those of the RIF galaxies when the W-cloud galaxies are distinguished from the other RIF galaxies.
There are positive correlations \added{between the product of $(|\Delta v_{\rm gal}|/\sigma_{\rm Virgo})\times(R_{\rm proj}/R_{200})$} with SFR, sSFR, and all the gas fractions with correlation coefficients of \replaced{$\sim0.2-0.5$}{$\sim0.3-0.4$} \added{and a negative correlation with SFE$_{\rm HI}$ with a correlation coefficient of $-0.25$}.
When the W-cloud galaxies are included, the correlation with $\mu_{\rm H_2}$ becomes insignificant while the \replaced{correlation with SFE$_{\rm H_2}$ becomes}{correlations with SFE$_{\rm gas}$ and $R_{\rm H_2}$ become} significant
(Figures~\ref{fig:hist_c2_inclp} and \ref{fig:accretionphase} in Section~\ref{sec:accretion} and Figure~\ref{fig:psd_inclp} in Appendix~\ref{sec:psd_plots}).

\item Comparing three subsamples, the dependence of the key quantities on the clustocentric distance, local galaxy density and accretion phase is basically observed for ``{\it best-effort sample}'' and ``{\it CO+\HI-obs. sample}'' but not for ``{\it CO+\HI-det. sample}''.
This is because ``{\it CO+\HI-det. sample}'' is biased to somehow similar systems which are less affected by the environments.
The information of upper limits are inevitable to understand the cold-gas properties of the Virgo galaxies.

\end{itemize}

\section{Discussions}\label{sec:discussions}


Our data show that our Virgo galaxies have lower SFR, sSFR and gas fractions and higher SFEs for their stellar masses compared to field galaxies as previous studies reported \citep[e.g.,][]{Rengarajan:1992kx,Fumagalli:2008gf,Fumagalli:2009bt,Boselli:2014qs}, even though they are massive galaxies ($M_{\rm star}>10^9$~M$_\odot$), whose star-formation activities are expected to be less affected by their environment \citep{Peng:2010eq}.
In the following sections, we discuss the effects of the inclusion of the ``possible'' members to our sample on the obtained results (Section~\ref{sec:membership}) and possible quenching mechanisms in the Virgo cluster (Section~\ref{sec:quenching}).

\subsection{Membership of the Virgo cluster}\label{sec:membership}

In this section, we investigate the effect of the inclusion of the possible members on the obtained results by removing them from the analysis.
As we mentioned in Section~\ref{sec:member}, we use data of both the secure and possible members in the EVCC.
The possible members mainly consist of the W-cloud galaxies and the southern-extension galaxies.
We find that the galaxies in the W cloud \deleted{seem to} have lower SFR\added{, sSFR,} \deleted{and} $\mu_{\rm H_2}$\added{, and SFE$_{\rm H_2}$} compared to the surrounding galaxies (Section~\ref{sec:accretion}).
On the other hand, the southern-extension galaxies (galaxies with higher $R_{\rm proj}/R_{200}$ and higher $\Delta v_{\rm gal}/\sigma_{\rm Virgo}$ values on the PSD) tend to have a bluer SDSS color as seen in Figure~\ref{fig:psd_sdss}, suggesting a higher SFR and abundant cold gas. 

For the dependences on the clustocentric distance, overall trends are basically the same as what we find with the sample including both the secure and possible members.
More specifically, we find the bend of the key quantities at $R_{3}$.
For the dependence on the local galaxy densities, the results for $R_{\rm 5, sky}$ do not significantly change\deleted{(there is an additional negative correlation with $\mu_{\rm gas}$)}, while all the correlations disappear for $R_{\rm 5, PSD}$.
This difference can be attributed for the disappearance of the galaxies with high SFR, sSFR, and $\mu_{\rm HI}$ at the sparser region on the PSD, i.e., the southern-extension galaxies.
For the dependence on the accretion phase, we find that the median values of SFR and all the gas fractions \deleted{(SFEs and $R_{\rm H_2}$)} are lower \deleted{(higher)} in the VIR galaxies than the RIF galaxies.
According to the KS test, the differences in SFR and all the gas fractions are confirmed to be statistically significant.
The correlations between $(|\Delta_{\rm gal}|/\sigma_{\rm Virgo})\times(R_{\rm proj}/R_{200})$ with SFR, sSFR, and gas fractions are still statistically significant even though the correlation coefficients decreases to $0.2-0.3$.
Therefore the exclusion of the possible member galaxies of the EVCC does not drastically change our results obtained in the Sections~\ref{sec:coldgassf} and \ref{sec:summaryresults} except for the $R_{\rm 5, PSD}$ dependence of the key quantities.
Hereafter, we discuss possible quenching mechanisms at work in the Virgo galaxies based on the obtained results with both the secure and possible members in the following sections.

\subsection{Quenching mechanism in the Virgo cluster}\label{sec:quenching}

Our results show that SFR, sSFR, and gas fractions of galaxies are different in samples chosen at $R_{\rm proj}<1.5 R_{200}$ and those at $R_{\rm proj}>1.5 R_{200}$ (Figure~\ref{fig:radial_wul}).
Considering that there is the boundary between the VIR and RIF regions around $R_{\rm proj}\sim 1.5 R_{200}$, our results suggest that ram-pressure and/or long-period quenching mechanisms such as the strangulation and the tidal interaction with the cluster potential play a role in affecting cold-gas and star-formation properties of the Virgo galaxies.
Additionally, the correlations between the projected local densities of galaxies on the sky with SFR, sSFR, and gas fractions suggest that galaxy-galaxy interaction also plays some roles in decreasing the star-formation activity in these galaxies (Figure~\ref{fig:numberdensity_radec}).
The weaker dependence of these key values on the local density on the PSD suggests that the galaxy harassment is especially important for decreasing gas reservoirs and star-formation activities of these galaxies (Figure~\ref{fig:numberdensity}).

Furthermore, we find that our Virgo sample tends to have a lower gas fraction and a higher SFE than the field galaxies.
The field galaxies are known to follow ``universal relations'' between $\Delta({\rm MS})$ and $\mu_{\rm H_2}$ and between $\Delta({\rm MS})$ and SFE$_{\rm H_2}$, where galaxies with lower $\Delta({\rm MS})$ tend to have a lower $\mu_{\rm H_2}$ and SFE$_{\rm H_2}$ \citep[e.g.,][]{Saintonge:2012nj}.
However, our Virgo sample with lower $\Delta({\rm MS})$ is found to have lower gas fractions and higher SFEs than the field galaxies.
This is why our Virgo sample has higher SFEs than the field galaxies even though they are less actively forming stars.
This may suggest that the gas reduction process in those galaxies with lower $\Delta({\rm MS})$ takes place over a shorter timescale than the gas-depletion timescale by star formation, i.e., $M_{\rm gas}/{\rm SFR}\sim3$~Gyr.
Considering that $\Delta({\rm MS})$ is equal to $\Delta({\rm Field})$ of SFR, the positive correlations between the $\Delta({\rm Field})$ value of SFR with the clustocentric distance, the distance to the 5-th nearest neighbour, and $(|\Delta v_{\rm gal}|/\sigma_{\rm Virgo})\times(R_{\rm proj}/R_{200})$ indicate that the galaxies with lower $\Delta({\rm MS})$ values are located at nearer to the cluster center, have smaller clustocentric velocity, and reside in denser regions.

We discuss the environmental effects on cold-gas and star-formation properties in the following sections by focusing on ram-pressure and tidal stripping (Section~\ref{sec:ram}), galaxy harassment (Section~\ref{sec:harassment}), preprocessing in a group environment (Section~\ref{sec:preprocess}), and strangulation (Section~\ref{sec:strangulation}).

\subsubsection{Ram-pressure or tidal stripping}
\label{sec:ram}

Over recent decades, ram-pressure stripping of cold gas in galaxies has been considered as one of the main candidate mechanisms of star-formation quenching in clusters \citep{Gunn:1972kc,Balsara:1994nn,Fujita:1999iv,Abadi:1999id,Quilis:2000gz,Mori:2000om,Vollmer:2001ir,Okamoto:2003xb}, and there is observational evidence that ram-pressure stripping modifies atomic gas \citep{Chung:2009ys} and ionized gas \citep[e.g.,][]{Yoshida:2004vc,Koopmann:2006hj,Cortese:2007ty,Ebeling:2014ez,Poggianti:2016bc}.
It is also shown that the poststarburst galaxies are preferentially found at small clustocentric radii with high clustocentric velocities, suggesting that the ram-pressure stripping is the most plausible candidate for the physical mechanism to quench their star formation \citep{Muzzin:2014ev,Paccagnella:2017tx}.
At the same time, ram pressure is claimed to enhance star formation activity in galaxies \citep{Fujita:1999iv,Bekki:2003tu,Kronberger:2008li,Bekki:2014ip,Ruggiero:2017qq} and there are supportive observations \citep[e.g.,][]{Vulcani:2018oy,Roberts:2020he,Wang:2020av}.
Additionally, ram pressure is found to induces a gas funneling to the central supermassive black hole of galaxies \citep{Poggianti:2017qg}.
However, the effect of ram pressure on molecular gas in cluster galaxies is still an open question.

Comparing the VIR galaxies with the RIF galaxies without the W-cloud galaxies, we find that star-formation activity and cold-gas reservoirs are significantly smaller in the former than the latter.
The atomic gas in the VIR and RPS galaxies has been known to be deficient compared to the RIF galaxies in previous studies \citep{Jaffe:2015pq,Wang:2020av}, but this is the first time one has found a similar tendency in molecular gas in the Virgo galaxies on the PSD.
The tendency is also confirmed by the positive correlations between $(|\Delta v_{\rm gal}|/\sigma_{\rm Virgo})\times(R_{\rm proj}/R_{200})$ with SFR, sSFR, and gas fractions as seen in the $z\sim2.5$ cluster \citep{Wang:2018rz}.
These results suggest the importance of not only the long-period quenching processes but also the ram-pressure stripping for star-formation quenching since the galaxies that were affected by the ram pressure also exist in the VIR region \citep{Jaffe:2015pq,Wang:2020av}.

In addition, we find that the gas fractions and SFEs of the Virgo galaxies are lower and higher than the field galaxies for fixed $\Delta({\rm MS})$ preferring a gas removal process whose timescale is shorter than the gas depletion time.
These results suggest that ram pressure plays an important role in the star-formation quenching in the Virgo galaxies and affects not only atomic gas but also molecular gas in some cluster galaxies as previous studies claimed \citep[e.g.,][]{Rengarajan:1992kx,Fumagalli:2008gf,Fumagalli:2009bt,Boselli:2014qs}.
However, there is no pure RPS galaxies with CO or \HI~measurements in our sample, so our data cannot completely rule out the possibility that the other long-period processes such as the strangulation and the tidal interaction with the cluster potential cause these changes.

In this study, SFEs are not significantly different between the VIR and RIF galaxies.
Previous observational studies showed a wide variety in the spatial distribution of SFE \citep{Fumagalli:2008gf,Fumagalli:2009bt,Vollmer:2012jz,Mok:2017ey}.
\cite{Vollmer:2012jz} investigated the SFE spatial distribution of 12~Virgo spiral galaxies and found that not all but some galaxies experiencing stripping have an increased SFE$_{\rm gas}$ (but not SFE$_{\rm H_2}$) on the windward side of the galactic disks where $R_{\rm H_2}$ is also high.
They also showed that the extraplanar region of such galaxies tends to have reduced SFE$_{\rm gas}$ (also SFE$_{\rm H_2}$ for one galaxy).
Therefore, the spatial variation may hamper the difference in the integrated galactic values of SFEs as explored in this study.

On the PSD, galaxies that are affected by the tidal force of the cluster potential are predicted to be distributed similarly to the galaxies that are affected by ram pressure from the ICM according to the numerical simulations \citep[Figures 8 and 9 of][]{Rhee:2017kl}.
In our data, when the sample is limited to ``secure'' members with both the CO and \HI~measurements, the difference in $\mu_{\rm HI}$ between the VIR and RIF galaxies becomes insignificant ($p$-value of \replaced{0.07}{$0.1$}) whereas SFR and $\mu_{\rm H_2}$ are still lower in VIR galaxies than RIF galaxies.
This is rather strange, given that the ram pressure is expected to affect atomic gas more than molecular gas.
This may suggest that the tidal force is more important for star-formation quenching in cluster galaxies than ram pressure.
Spatially resolved data of both the gaseous and stellar components of galaxies would help us to distinguish which process plays a major role in affecting star-formation properties, since the ram pressure should interact only with gas whereas the tidal force is expected to affect both gas and stellar components. 

\subsubsection{Galaxy harassment}
\label{sec:harassment}
The so-called density-morphology relation \citep[i.e.,][]{Dressler:1980yl} suggests that the galaxy environment plays a role in changing the galaxy morphology.
Galaxy harassment is one of the mechanisms leading to morphological changes due to the combined effect of tidal interactions between an infalling galaxy and the cluster potential and of the tidal interaction with close high speed encounters with other cluster members \citep{Moore:1996mv,Moore:1998ps}.
The change is claimed to depend on various parameters such as the orbit of an infalling galaxy within a galaxy cluster \citep{Mastropietro:2005gf,Smith:2010xy}, the surface density of the infalling galaxy \citep{Moore:1999hw,Chang:2013qv}, and the inclination between the orbital plane and the disk of the infalling galaxy \citep{Villalobos:2012mp}.
\cite{Haynes:2007wy} shows that the observed properties of a strongly asymmetric Virgo galaxy NGC~4254 are consistent with what are expected for galaxies in the process of the galaxy harassment.
However, the galaxy harassment is also claimed to be only significant on specific orbits close to the cluster core and does not induce significant structural or dynamical transformation on most typical cluster orbits \citep{Bialas:2015ty,Smith:2015dv}.

We find a stronger dependence of the star-formation activity and the cold-gas contents of galaxies on $R_{\rm 5, sky}$ than $R_{\rm 5, PSD}$, suggesting the importance of galaxy harassment.
At the same time, we also find positive correlations between $R_{\rm 5, sky}$ and $R_{\rm 5, PSD}$ with the clustocentiric distance, i.e., the galaxies at nearer to the cluster center tend to reside in denser region.
Therefore, the dependence of the star-formation activity and the cold-gas contents of galaxies on the local density could be just reflecting the dependence of these properties on the clustocentric distance, and vice versa.
It is difficult to conclude which are more essential correlations with our data given that the strengths of the correlations are comparable ($r$ of \replaced{$0.2-0.4$}{$\sim0.3-0.5$}).
Again, spatially resolved properties of both the stellar and cold-gas components in these galaxy would provide us a clue to assess the importance of the galaxy harassment for them.

\subsubsection{Pre-process in group environments}
\label{sec:preprocess}

There are many studies suggesting that suppression of star formation in galaxies has already started before the galaxies fall deep into the cluster \citep[``pre-process'', e.g.,][]{Kodama:2001zg,Lewis:2002pw,Gomez:2003hb,Tanaka:2005vn,Koyama:2008dn,Lu:2012zy,Rasmussen:2012ej,Matsuki:2017nl,Kleiner:2021un}.
More than half of the galaxies in the local universe reside in group and cluster environments \citep{Zwicky:1938fz,Abell:1958hs,Huchra:1982gk,Geller:1983mw,Eke:2004mt,Eke:2005pz}.
It is possible that the most quenched galaxies in the cluster observed today may have been pre-processed before falling into the cluster.

The star-formation quenching of group galaxies has been investigated \citep[e.g.,][]{Zabludoff:1998ew,Balogh:2004vf,Balogh:2009cd,Wilman:2005vm,Weinmann:2006ve,McGee:2011wd,Rasmussen:2012ej,Hou:2013aq,Tal:2014mx} and it is generally considered that galaxy harassment and tidal interactions are the most common quenching processes in the group environment \citep{Wilman:2009to}.
X-ray emitting hot gas has been observed in some groups \citep{Mulchaey:2000zn} and the interaction with the intra-group medium (IGM) is also one of the possible quenching processes for such groups \citep{Bekki:2002qa,Rasmussen:2008kb}.
It is also claimed that not one but several quenching processes are necessary to account for the observations \citep{Smethurst:2017ws,Vulcani:2018ku}.

In Section~\ref{sec:accretion}, we find that the W-cloud galaxies have lower $\mu_{\rm H_2}$ and \deleted{higher} SFE$_{\rm H_2}$ than the VIR and the RIF galaxies.
Since most of the W-cloud galaxies with CO upper limits, i.e., low $\mu_{\rm H_2}$ galaxies, do not have \HI~measurements, it is difficult to assess whether their \HI~gas is also deficient.
However, considering the non-detections of these sources in the blind \HI~survey of the ALFALFA, \HI~gas is not so abundant compared with the other Virgo galaxies which do show \HI~detection in the ALFALFA \citep[$5\sigma$ sensitivity of 0.72~Jy~km~s$^{-1}$, corresponding to the \HI~mass of $4.6\times10^7$~M$_\odot$  at 16.5~Mpc distance,][]{Haynes:2018rt}.

In order to reduce a significant amount of molecular gas, either very strong gas stripping by IGM or galaxy-galaxy interactions are required.
According to X-ray observations of the Virgo cluster, the hot gas does not seem to be strong at the location of the W cloud (see Figure~\ref{fig:radec}), suggesting that the galaxy-galaxy interactions are a more likely scenario for the CO deficiency in the W-cloud galaxies.
If the W cloud is falling into the Virgo cluster, pre-processing in the group environment is one of the important paths to increase the fraction of passive galaxies in the cluster environment.

Recently, \cite{Castignani:2021tr} claimed that the star-formation quenching is already at play in the filaments, by investigating CO and \HI~properties of 245 galaxies in the filaments around the Virgo cluster.
They found a clear tendency from field to filament and to cluster galaxies for decreasing SFR and gas fraction.
Furthermore, it is shown that $\mu_{\rm HI}$ and $\mu_{\rm H_2}$ decrease with increasing the local galaxy density in the filament, suggesting an importance of tidal interactions.

There are few CO or \HI~observations toward group galaxies.
Low \HI~gas fractions have been observed in galaxies in the Hickson Compact Groups \citep[HCG,][]{Hickson:1982pe,Verdes-Montenegro:2001dh,Borthakur:2010uk}, although HCGs are smaller galaxy groups than the W cloud of the Virgo cluster.
As distinct from the W cloud in the Virgo cluster, no significant deficiency in molecular gas has been reported in the HCG galaxies \citep{Verdes-Montenegro:1998xt,Leon:1998qq,Martinez-Badenes:2012dj}.
However, detailed CO mapping observations of HCG galaxies with interferometers have revealed disturbed molecular gas and star formation quenching \citep{Alatalo:2014lf,Alatalo:2015fg}.
More samples with both CO and \HI~measurements are required to understand the pre-processes in the group environments.

\subsubsection{Strangulation}
\label{sec:strangulation}

Many observational studies with a significant number of galaxy groups and clusters have shown that strangulation, not ram-pressure stripping, is the dominant process for quenching star formation in galaxies \cite[e.g.,][]{Tanaka:2004uh,Weinmann:2006ve,Peng:2015ar,Paccagnella:2016nw}.
\cite{Peng:2015ar} showed that the difference in stellar metallicity between star-forming and passive galaxies is explained by long-timescale quenching with a typical timescale of 4~Gyr, which is longer than that expected for ram-pressure stripping \citep[$\sim200$~Myr,][]{Steinhauser:2016dk}.

In our sample in the Virgo cluster, \replaced{both the \HI-gas and H$_2$-gas masses}{the H$_2$-gas mass} in some RIF galaxies have been already reduced compared to field galaxies with matched $M_{\rm star}$ at least at $R_{\rm proj}\sim3 R_{200}$ (Figures~\ref{fig:radial_wul} and \ref{fig:psd_inclp}).
Thus hot-gas, as well as cold-gas, stripping by the ICM and pre-processes in group or filament environments are expected to play some role in the reduction of cold gas reservoirs in those galaxies.
Near the cluster core, the \HI~gas and perhaps H$_2$ gas are likely to be stripped by ram pressure, even for the relatively massive galaxies with $M_{\rm star}>10^9$~M$_\odot$ in our sample.
Galaxies whose cold gas is not completely removed by ram pressure would evolve similarly to those affected only by strangulation, but the ram pressure would somehow accelerate the quenching of star-formation.
In this case, the quenching timescale can be estimated by their gas depletion timescale of $M_{\rm gas}$/SFR~$\sim3$~Gyr (Figure~\ref{fig:comp_field}), which is shorter than but still comparable to the value obtained in the previous study \citep[$\sim4$~Gyr,][]{Peng:2015ar}.
This is consistent with previous studies, suggesting that ram pressure stripping ``finishes the job'' in quenching star formation in some cluster galaxies \citep{van-der-Burg:2018tf,Roberts:2019bp}.

\section{Summary}
\label{sec:summary}

We investigated atomic- and molecular-gas-related properties of massive Virgo galaxies ($M_{\rm star}>10^9$~M$_\odot$) on the phase-space diagram (PSD) with the world's largest current dataset of \HI~and CO (194 for \HI, 173 for CO, and 113 for both).
For a comparison, we utilize the medians of the key quantities of galaxy subsamples and the $p$-value of the KS test to assess the magnitude relationship and the statistical significance of the difference in the distribution of the key quantities, respectively.
Our main results are as follows:

\begin{description}
\item[Virgo vs field (Section~\ref{sec:field})]
Our sample galaxies in the Virgo cluster have lower SFR, sSFR, and gas fractions ($\mu_{\rm HI}$, $\mu_{\rm H_2}$, and $\mu_{\rm gas}$) and higher SFEs (SFE$_{\rm HI}$, SFE$_{\rm H_2}$ and SFE$_{\rm gas}$) compared to field galaxies with matched $M_{\rm star}$.
The Virgo galaxies basically follow the field relations between $\Delta({\rm MS})$ with the gas fractions and SFEs, but deviate to lower gas fractions and higher SFEs regimes at \replaced{$-1.0 \lesssim \Delta({\rm MS}) \lesssim 0.0$}{$\Delta({\rm MS})<0.0$}, suggesting the importance of shorter timescale gas stripping processes than the gas-depletion timescale by star formation.
The Virgo galaxies with lower $\Delta({\rm MS})$ values tend to be located at closer to the cluster center, have a lower clustocentric velocity, and reside in denser regions.

\item[Clustocentric distance (Section~\ref{sec:distance})] 
Overall, the galaxies nearer to the cluster center have low SFR, sSFR, and gas fractions (the Spearman's rank-order correlation coefficients $r$ of \replaced{$0.2-0.4$}{$0.3-0.5$}), and high SFE$_{\rm gas}$, SFE$_{\rm HI}$, and $R_{\rm H_2}$ ($r$ ranging from \replaced{$-0.3$ to $-0.2$}{$-0.4$ to $-0.3$}).
The radial trends seem to change at $R_{\rm proj}\sim1.5 R_{200}$, near the location of the boundary between the ``virialized'' and the ``recent infallers'' regions.
Even in the outskirt galaxies, the medians of the gas fractions are slightly lower than the those of the field galaxies.
These observations suggest the importance of ram-pressure gas stripping and long-period quenching processes in the Virgo cluster.

\item[Local galaxy density (Section~\ref{sec:density})]
The distance to the 5-th nearest neighbours on the sky ($R_{\rm 5, sky}$) positively correlates to the clustocentric distance, SFR, sSFR, and gas fractions (\replaced{$r\sim0.2-0.4$}{$r\sim0.3-0.4$}) and negatively correlates to \added{SFE$_{\rm gas}$ ($r=-0.31$) and} SFE$_{\rm HI}$ (\replaced{$r=-0.16$}{$r=-0.28$}).
On the other hand, the dependence of these values on the distance to the 5-th nearest neighbours on PSD becomes weak, suggesting the importance of the galaxy harassment.
However, the positive correlation between $R_{\rm 5, sky}$ with the clustocentric distance indicates that the dependence of the key quantities on the local density just reflect the dependence on the clustocentric distance, and vice versa.

\item[Accretion phase (Section~\ref{sec:accretion})]
The product of $(|\Delta_{\rm gal}|/\sigma_{\rm Virgo})\times(R_{\rm proj}/R_{200})$ positively correlates to SFR, sSFR, and gas fractions ($r$ of \replaced{$\sim0.2-0.5$}{$\sim0.3-0.4$}), suggesting that galaxies that were accreted to the Virgo cluster earlier epoch tend to have lower star-formation activity and cold-gas contents.
Provided that the galaxies that were affected by ram pressure in the past can also have lower $(|\Delta_{\rm gal}|/\sigma_{\rm Virgo})\times(R_{\rm proj}/R_{200})$, ram-pressure stripping as well as long-period quenching processes such as strangulation and tidal interaction with the cluster potential plays a role in the star-formation quenching in those galaxies.

\item[SF quenching processes in Virgo (Section~\ref{sec:quenching})] 
Our results support the ``hybrid'' quenching scenario where multiple processes account for decreasing star-formation activity in galaxies in the Virgo cluster.
Combining with the recent findings by \cite{Castignani:2021tr}, our results indicate that the pre-processes in the filament and group environments as well as strangulation have already affected cold-gas contents in some outskirt galaxies at $R_{\rm proj}\sim3 R_{200}$.
In addition, ram-pressure stripping is likely to reduce both the atomic and molecular gas contents and consequently decrease the star-formation activity in some galaxies at $R_{\rm proj}\sim1.5 R_{200}$.
However, our data cannot rule out the possibility that the tidal interaction with the cluster potential, strangulation and/or galaxy harassment cause these changes in some galaxies depending on the accretion parameters such as orbits within the cluster, velocity, and the inclinations of accreting galaxies with respect to the cluster.

\end{description}

However, these results are based on the incomplete and inhomogeneous data set.
In order to determine definitively the effect of ram pressure on molecular gas, CO and \HI~observations of a $M_{\rm star}$-limited complete sample of the Virgo cluster are required.

\acknowledgments
We thank the anonymous referee for his/her insightful comments that strengthen the results presented in this paper.
This work was supported by JSPS KAKENHI Grant Numbers 16H02158, 19J40004, 19H01931, and 20H05861.
KMM is grateful to Dr.~Hyein Yoon and Dr.~Aeree Chung for helpful comments on PSD plot of the Virgo cluster.
We also appreciate the kind help of Prof. Michael W. Richmond in improving the English grammar of the manuscript.

%

\facilities{GALEX, WISE, SEST, OSO:20m, FCRAO, IRAM:30m, NRAO:12m, No:45m, Arecibo, WSRT, VLA}


\software{astropy \citep{Astropy-Collaboration:2013uf},
	APLpy \citep{Robitaille:2012lz}
	}




\vspace{5mm}

\appendix

\section{Comparison of key quantities between the field and Virgo galaxies}\label{sec:fieldcomparison_plots}

\begin{figure*}[]
\begin{center}
\includegraphics[width=\textwidth, bb=0 0 1412 1384]{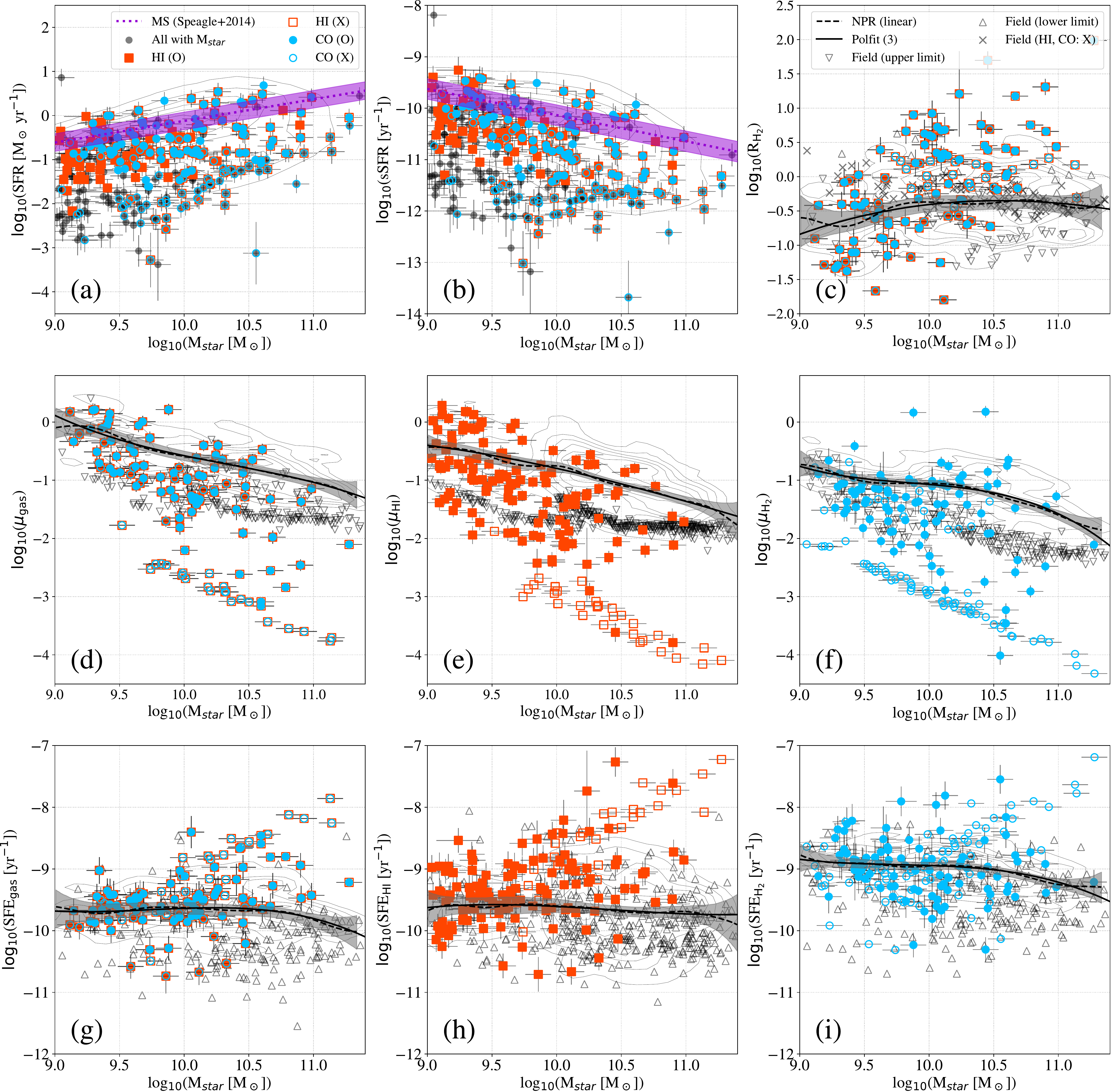}
\end{center}
\caption{
Comparison between the Virgo galaxies and field galaxies from the z0MGS \citep{Leroy:2019cu}, xGASS \citep{Catinella:2018ib} and xCOLDGASS \citep{Saintonge:2017ve} projects: $M_{\rm star}$ dependences of
(a) SFR,
(b) sSFR,
(c) $M_{\rm H_2}/M_{\rm HI}$,
(d) $M_{\rm gas}/M_{\rm star}$,
(e) $M_{\rm HI}/M_{\rm star}$,
(f) $M_{\rm H_2}/M_{\rm star}$,
(g) SFR/$M_{\rm gas}$,
(h) SFR/$M_{\rm HI}$,
(i) SFR/$M_{\rm H_2}$.
The Virgo galaxies with $M_{\rm star}$ and SFR measurements are indicated as filled grey circles in the panels (a) and (b).
The Virgo galaxies with an \HI~detection, an \HI~upper limit, a CO detection and a CO upper limit are indicated as filled orange square, open orange square, filled blue circle, and open blue circle, respectively.
The main-sequence of star-forming galaxies defined in \cite{Speagle:2014by} is indicated as purple dotted lines and the 0.2 dex scatter range is indicated as purple shading in the panels (a) and (b).
The field relations are estimated with CO- or \HI-detected galaxies, i.e., star-forming galaxies, and indicated as dashed (with non-parametric fitting) and (with 3rd-order polynomial fitting) solid lines in the panels (c)-(i).
Grey-shaded regions are the errors of the non-parametric fitting and estimated by the bootstrap method with a 95~\% confidence interval.
The contours indicate field galaxies used to derive the field relations.
The lower/upper limits of field galaxies are indicated as open triangles/inverted triangles.
For the $R_{\rm H_2}$ panel, the field galaxies with \HI~and CO upper limits are indicated as X marks.
}
\label{fig:comp_field}
\end{figure*}

\begin{figure*}[]
\begin{center}
\includegraphics[width=150mm, bb=0 0 1398 1384]{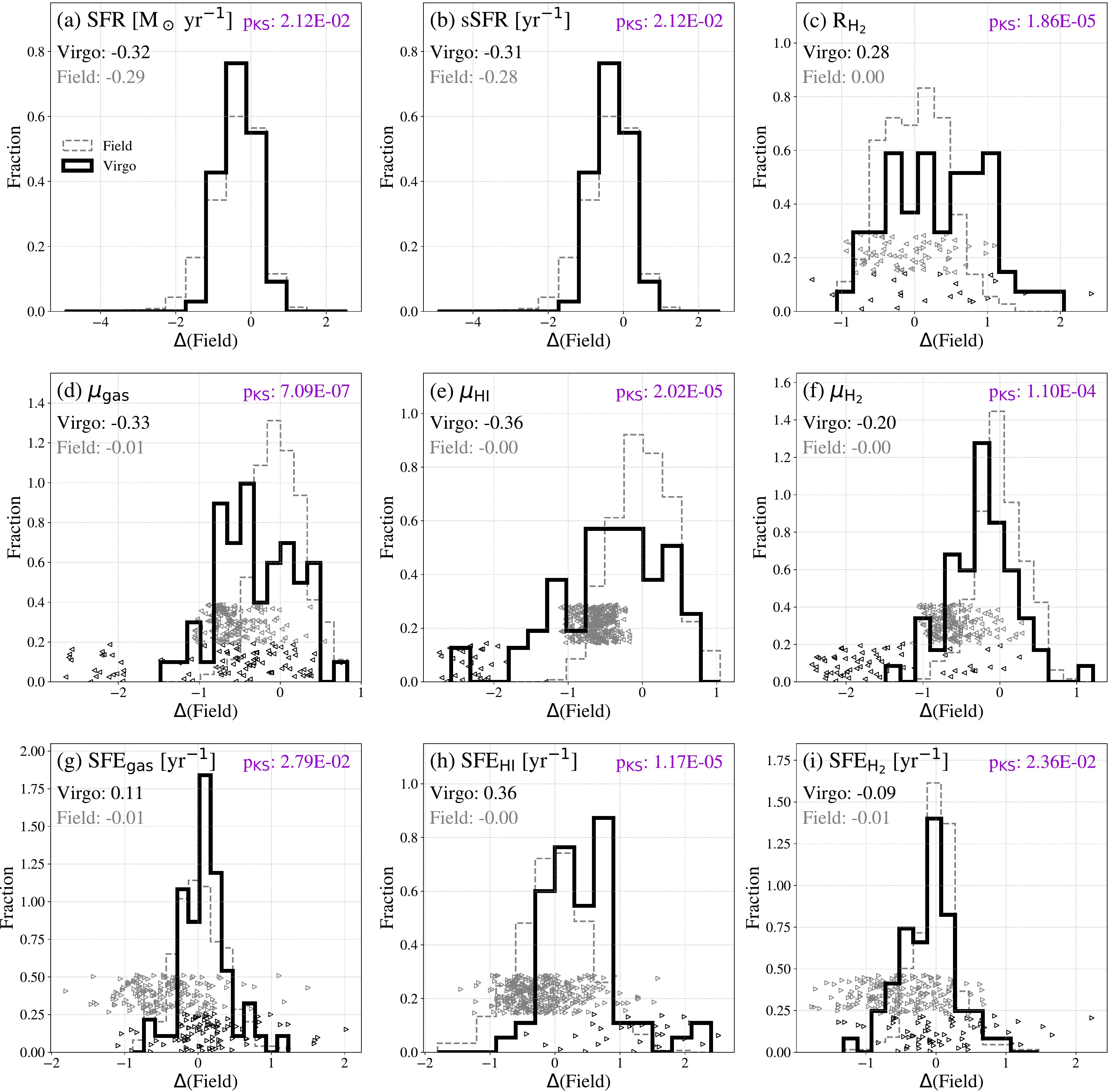}
\end{center}
\caption{
Same as Figure~\ref{fig:comp_field2} but the galaxies with CO/\HI~detections and the galaxies with CO/\HI~upper limits are presented separately.
The upper limits for the field and the Virgo galaxies are indicated as open grey and blue triangles, respectively.
The values of the y-axis for the upper limits are assigned randomly within certain ranges.
}
\label{fig:comp_field2b}
\end{figure*}

In this section, we present supplementary figures for the comparison of the key quantities between the field and the Virgo galaxies (Section~\ref{sec:field}).
Figure~\ref{fig:comp_field} shows the stellar mass dependence of the key quantities.
We can see that the fraction of galaxies below the SFMS is larger for the Virgo galaxies than the field galaxies and the Virgo galaxies tend to have lower gas fractions, a higher $R_{\rm H_2}$, and higher SFEs compared to the field galaxies.
Figure~\ref{fig:comp_field2b} compares the $\Delta({\rm Field})$ values of the field and the Virgo galaxies when limiting to those with CO or \HI~detections.
Note that the SFR and sSFR histograms of field galaxies includes all the sample in the z0MGS project regardless of the CO/\HI~measurements.
The upper/lower limits are indicated as left-pointing/right-pointing triangles.

\section{Local density dependence of the key quantities}\label{sec:localdensity_plots}

\begin{figure*}[]
\begin{center}
\includegraphics[width=.49\textwidth, bb=0 0 519 495]{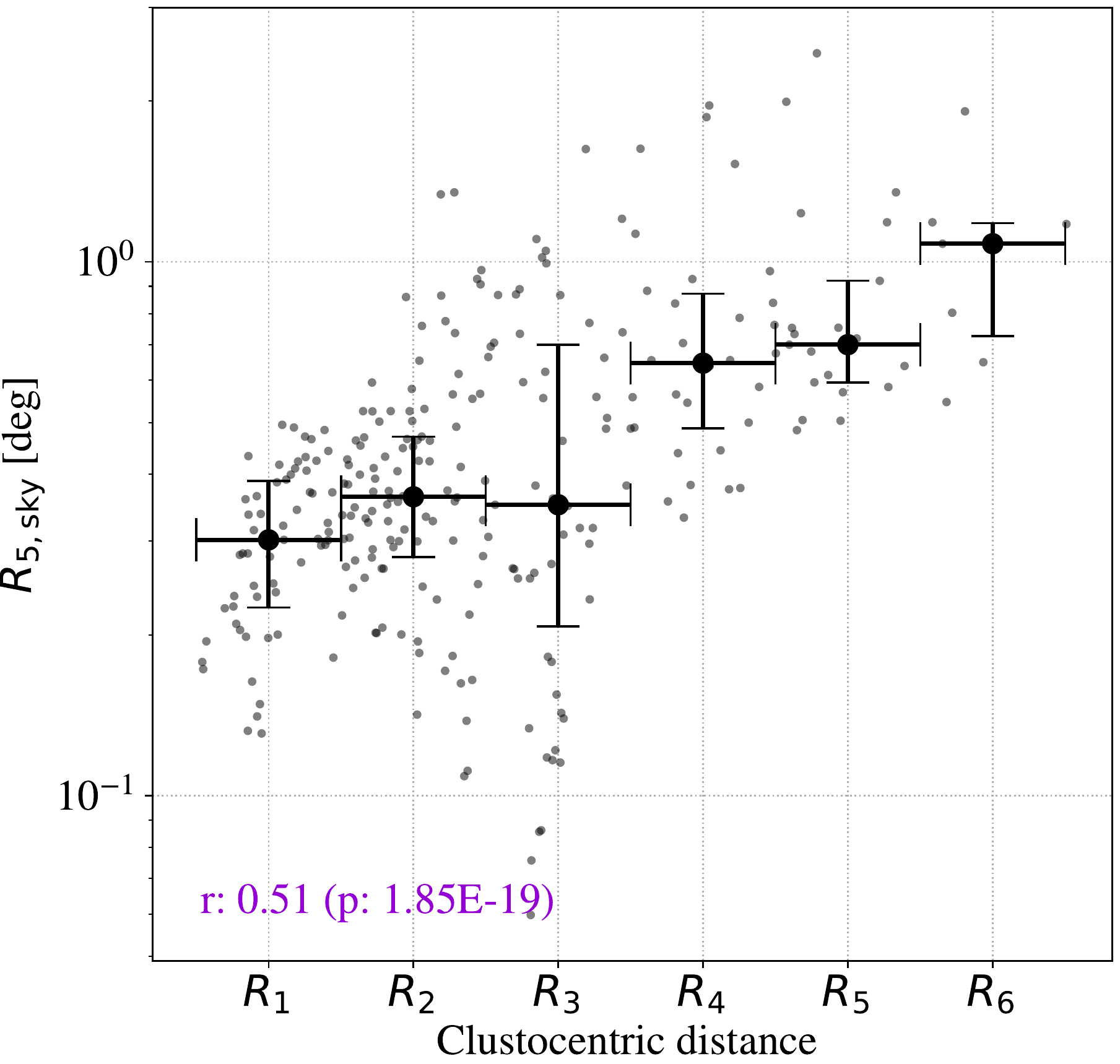}
\includegraphics[width=.49\textwidth, bb=0 0 519 495]{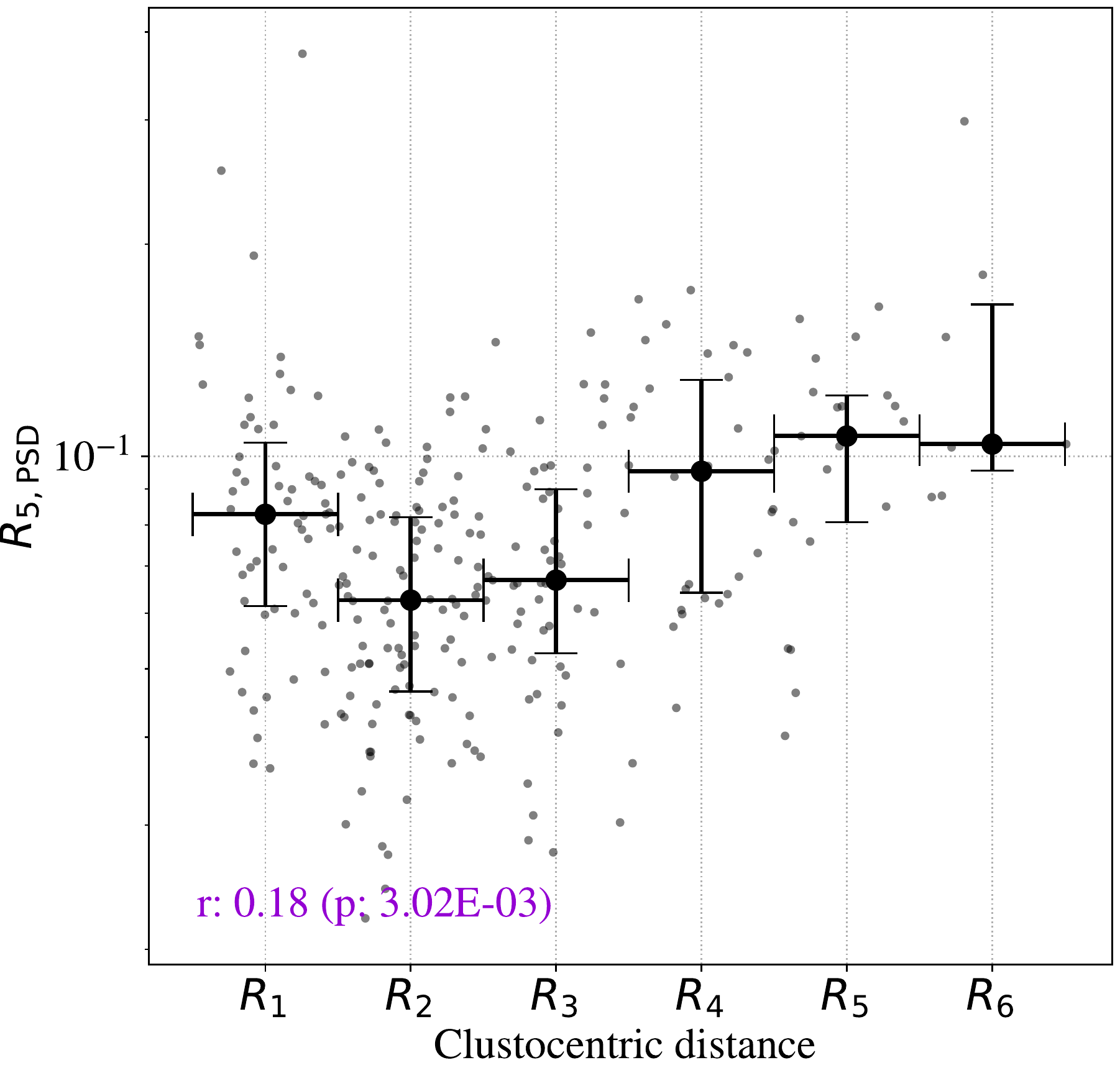}
\end{center}
\caption{
Relationship between the clustocentric distance and the distance to the 5-th nearest galaxy on the sky $R_{\rm 5, sky}$ (left) and on the PSD $R_{\rm 5, PSD}$ (right) as measures of local galaxy density.
The individual data is indicated as grey filled circles and the median values for each radius bin are indicated as black filled circles.
The error bars indicate the 1st and 3rd quartiles.
The correlation coefficient is indicated at the lower-left corner in purple.
}
\label{fig:dist_density}
\end{figure*}

\begin{figure*}[]
\begin{center}
\includegraphics[width=150mm, bb=0 0 1422 1388]{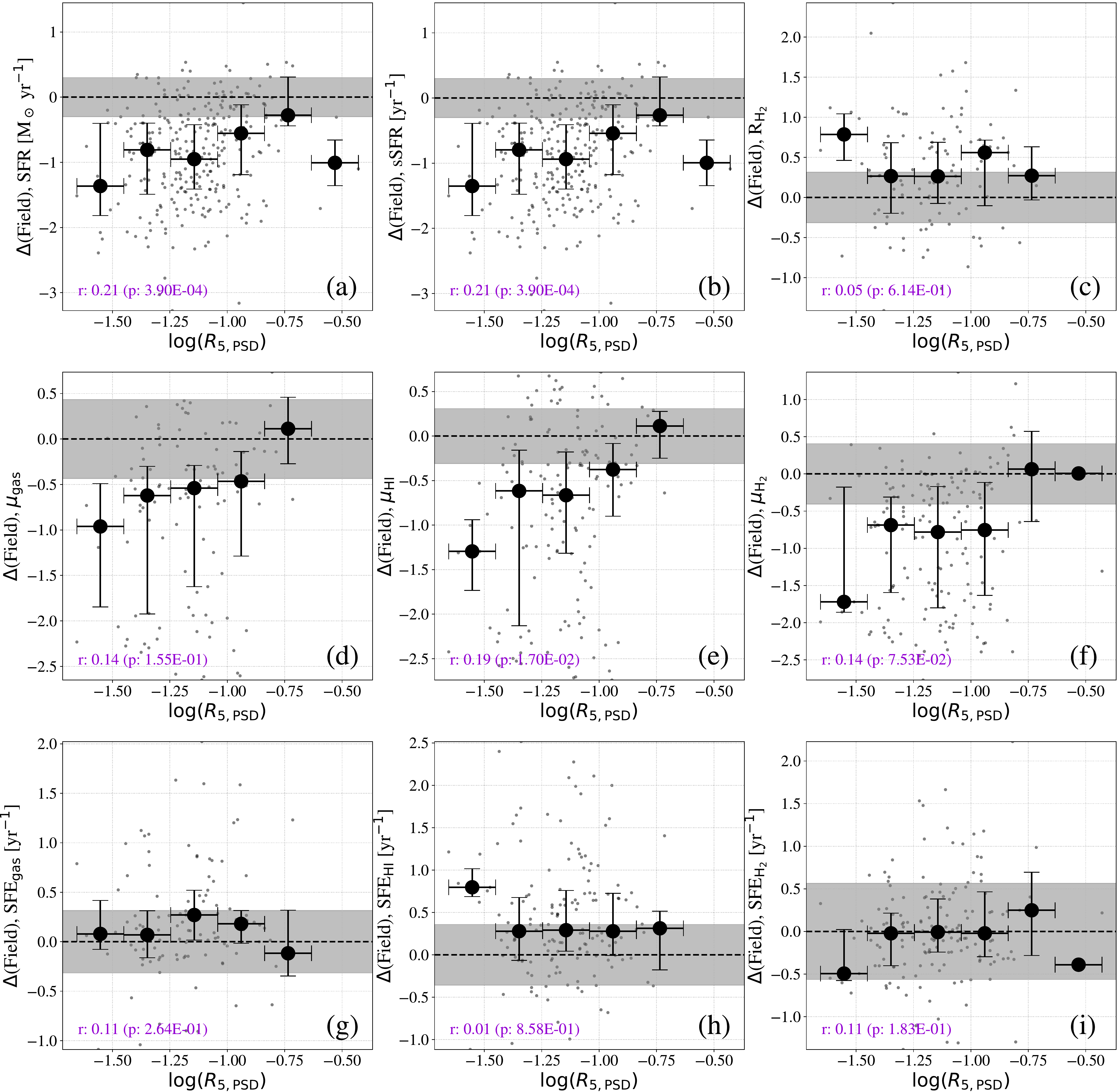}
\end{center}
\caption{
Same as Figure~\ref{fig:numberdensity_radec}, but for $R_{\rm 5, PSD}$.
}
\label{fig:numberdensity}
\end{figure*}

In this section, we present supplementary figures for the dependence of the key quantities on the local galaxy density (Section~\ref{sec:density}).
Figure~\ref{fig:dist_density} shows the relations between $R_{\rm 5, sky}$ and $R_{\rm 5, PSD}$ with clustocentric distance.
There are moderate and weak correlations between $R_{\rm 5, sky}$ and $R_{\rm 5, PSD}$ with clustocentric distance, respectively.
Figure~\ref{fig:numberdensity} shows the relations between $R_{\rm 5, PSD}$ with the key quantities.
As mentioned in Section~\ref{sec:density}, overall, there are weaker dependence of the key quantities on $R_{\rm 5, PSD}$ than $R_{\rm 5, sky}$.
According to the $p$-values, the correlations between $R_{\rm 5, PSD}$ with SFR, sSFR, and $\mu_{\rm HI}$ are statistically significant.

\section{Key quantities on the PSD}\label{sec:psd_plots}

\begin{figure*}[]
\begin{center}
\includegraphics[width=\textwidth, bb=0 0 1571 1826]{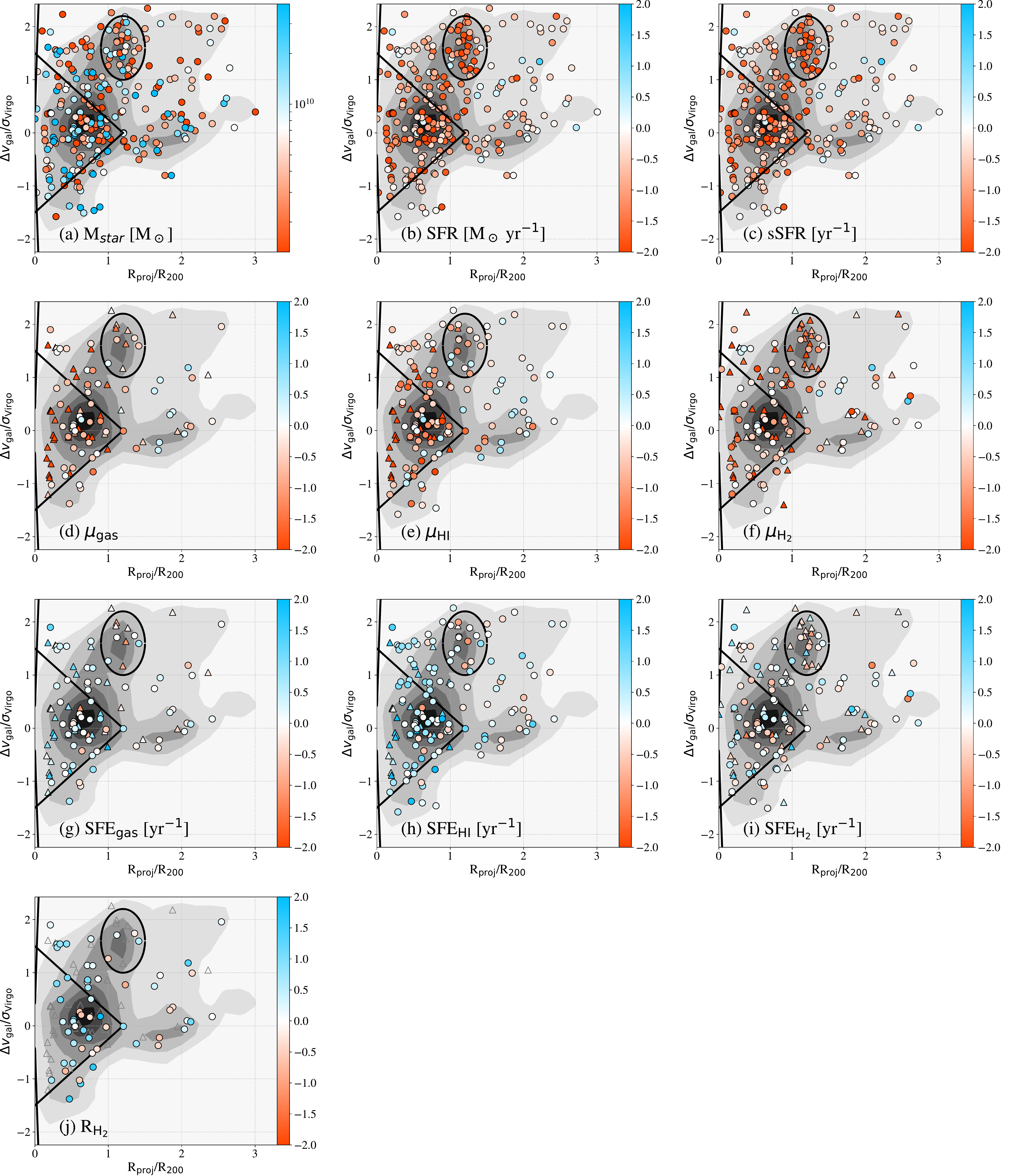}
\end{center}
\caption{
Stellar mass and the key quantities [$\Delta({\rm Field})$] varieties on PSD.
The colors of the symbols indicate the values of the key quantities.
Zero of each key quantity indicates the values of the field galaxies except for the stellar mass (white), i.e., the blue-ish and red-ish colors indicate the values are higher and lower than field galaxies, respectively.
}
\label{fig:psd_inclp}
\end{figure*}

\begin{figure*}[]
\begin{center}
\includegraphics[width=\textwidth, bb=0 0 1585 1826]{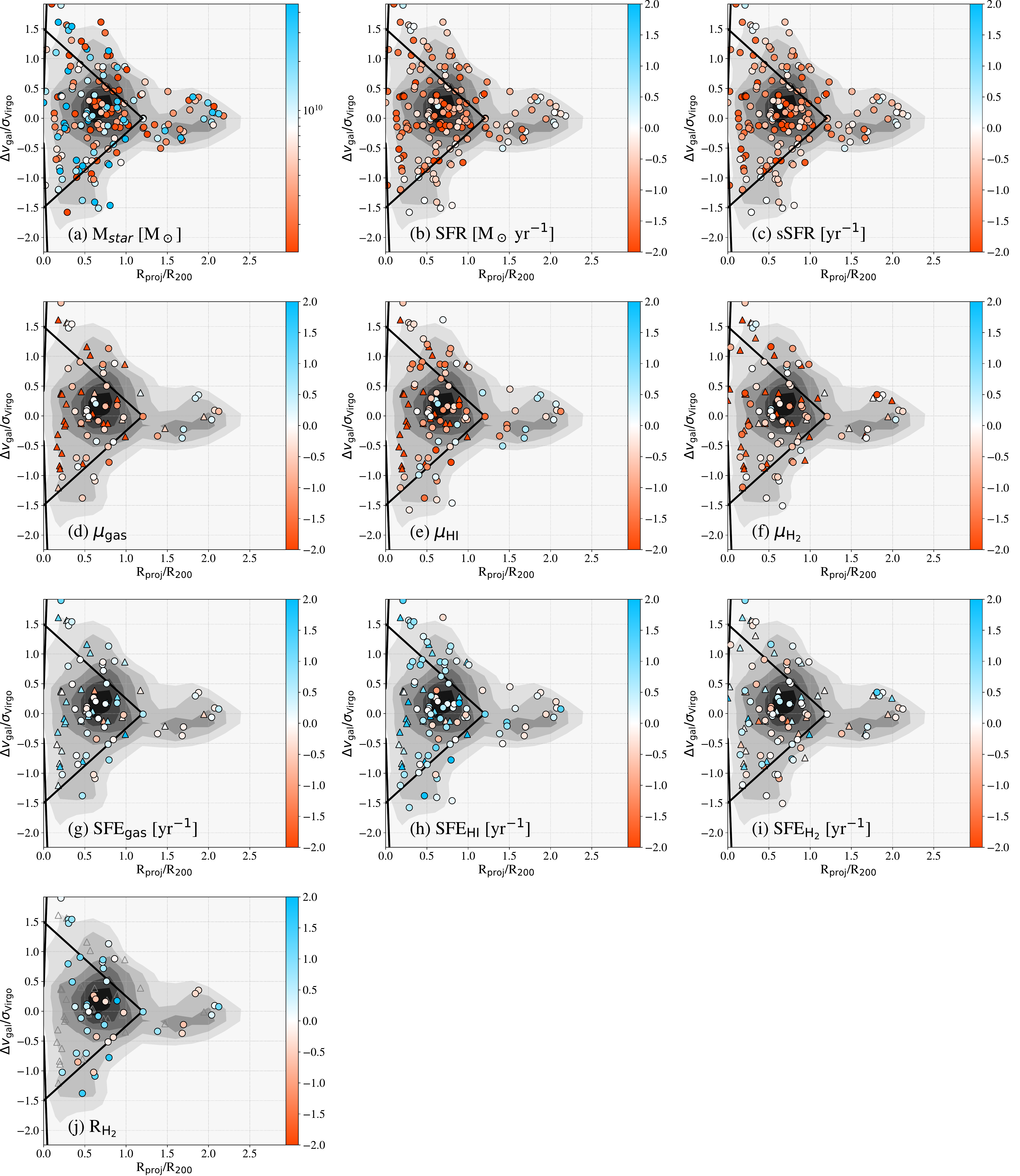}
\end{center}
\caption{
Stellar mass and the key quantities [$\Delta({\rm Field})$] varieties on PSD without the EVCC possible members.
The colors of the symbols indicate the values of the key quantities.
Zero of each key quantity indicates the values of the field galaxies except for the stellar mass (white), i.e., the blue-ish and red-ish colors indicate the values are higher and lower than field galaxies, respectively.
}
\label{fig:psd_wop}
\end{figure*}

In this section, we present supplementary figures of the key quantities distributions on the PSD (Sections~\ref{sec:accretion} and \ref{sec:membership}).
Figures~\ref{fig:psd_inclp} and \ref{fig:psd_wop} are respectively the PSDs of the secure+possible and only secure members, color-coded according to the key quantities without the stellar-mass dependences of field galaxies [$\Delta({\rm Field})$].
The zero value indicates the values for the field galaxies and the blue-ish/red-ish colors indicate that the values are higher/lower than the field galaxies.


\bibliographystyle{aasjournal}
\bibliography{myref_moro}

\begin{thebibliography}{}
\expandafter\ifx\csname natexlab\endcsname\relax\def\natexlab#1{#1}\fi
\providecommand{\url}[1]{\href{#1}{#1}}
\providecommand{\dodoi}[1]{doi:~\href{http://doi.org/#1}{\nolinkurl{#1}}}
\providecommand{\doeprint}[1]{\href{http://ascl.net/#1}{\nolinkurl{http://ascl.net/#1}}}
\providecommand{\doarXiv}[1]{\href{https://arxiv.org/abs/#1}{\nolinkurl{https://arxiv.org/abs/#1}}}

\bibitem[{{Abadi} {et~al.}(1999){Abadi}, {Moore}, \& {Bower}}]{Abadi:1999id}
{Abadi}, M.~G., {Moore}, B., \& {Bower}, R.~G. 1999, \mnras, 308, 947,
  \dodoi{10.1046/j.1365-8711.1999.02715.x}

\bibitem[{{Abazajian} {et~al.}(2009){Abazajian}, {Adelman-McCarthy},
  {Ag{\"u}eros}, {Allam}, {Allende Prieto}, {An}, {Anderson}, {Anderson},
  {Annis}, {Bahcall}, \& et~al.}]{Abazajian:2009cz}
{Abazajian}, K.~N., {Adelman-McCarthy}, J.~K., {Ag{\"u}eros}, M.~A., {et~al.}
  2009, \apjs, 182, 543, \dodoi{10.1088/0067-0049/182/2/543}

\bibitem[{{Abell}(1958)}]{Abell:1958hs}
{Abell}, G.~O. 1958, \apjs, 3, 211, \dodoi{10.1086/190036}

\bibitem[{{Aguerri} {et~al.}(2005){Aguerri}, {Gerhard}, {Arnaboldi},
  {Napolitano}, {Castro-Rodriguez}, \& {Freeman}}]{Aguerri:2005fo}
{Aguerri}, J.~A.~L., {Gerhard}, O.~E., {Arnaboldi}, M., {et~al.} 2005, \aj,
  129, 2585, \dodoi{10.1086/429936}

\bibitem[{{Alatalo} {et~al.}(2014){Alatalo}, {Appleton}, {Lisenfeld},
  {Bitsakis}, {Guillard}, {Charmandaris}, {Cluver}, {Dopita}, {Freeland},
  {Jarrett}, {Kewley}, {Ogle}, {Rasmussen}, {Rich}, {Verdes-Montenegro}, {Xu},
  \& {Yun}}]{Alatalo:2014lf}
{Alatalo}, K., {Appleton}, P.~N., {Lisenfeld}, U., {et~al.} 2014, \apj, 795,
  159, \dodoi{10.1088/0004-637X/795/2/159}

\bibitem[{{Alatalo} {et~al.}(2015){Alatalo}, {Appleton}, {Lisenfeld},
  {Bitsakis}, {Lanz}, {Lacy}, {Charmandaris}, {Cluver}, {Dopita}, {Guillard},
  {Jarrett}, {Kewley}, {Nyland}, {Ogle}, {Rasmussen}, {Rich},
  {Verdes-Montenegro}, {Xu}, \& {Yun}}]{Alatalo:2015fg}
---. 2015, \apj, 812, 117, \dodoi{10.1088/0004-637X/812/2/117}

\bibitem[{{Andersen}(1996)}]{Andersen:1996vm}
{Andersen}, V. 1996, \aj, 111, 1805, \dodoi{10.1086/117918}

\bibitem[{{Andrews} \& {Martini}(2013)}]{Andrews:2013nx}
{Andrews}, B.~H., \& {Martini}, P. 2013, \apj, 765, 140,
  \dodoi{10.1088/0004-637X/765/2/140}

\bibitem[{{Arnaboldi} {et~al.}(2004){Arnaboldi}, {Gerhard}, {Aguerri},
  {Freeman}, {Napolitano}, {Okamura}, \& {Yasuda}}]{Arnaboldi:2004ti}
{Arnaboldi}, M., {Gerhard}, O., {Aguerri}, J. A.~L., {et~al.} 2004, \apjl, 614,
  L33, \dodoi{10.1086/425417}

\bibitem[{{Asplund} {et~al.}(2009){Asplund}, {Grevesse}, {Sauval}, \&
  {Scott}}]{Asplund:2009fh}
{Asplund}, M., {Grevesse}, N., {Sauval}, A.~J., \& {Scott}, P. 2009, \araa, 47,
  481, \dodoi{10.1146/annurev.astro.46.060407.145222}

\bibitem[{{Astropy Collaboration} {et~al.}(2013){Astropy Collaboration},
  {Robitaille}, {Tollerud}, {Greenfield}, {Droettboom}, {Bray}, {Aldcroft},
  {Davis}, {Ginsburg}, {Price-Whelan}, {Kerzendorf}, {Conley}, {Crighton},
  {Barbary}, {Muna}, {Ferguson}, {Grollier}, {Parikh}, {Nair}, {Unther},
  {Deil}, {Woillez}, {Conseil}, {Kramer}, {Turner}, {Singer}, {Fox}, {Weaver},
  {Zabalza}, {Edwards}, {Azalee Bostroem}, {Burke}, {Casey}, {Crawford},
  {Dencheva}, {Ely}, {Jenness}, {Labrie}, {Lim}, {Pierfederici}, {Pontzen},
  {Ptak}, {Refsdal}, {Servillat}, \&
  {Streicher}}]{Astropy-Collaboration:2013uf}
{Astropy Collaboration}, {Robitaille}, T.~P., {Tollerud}, E.~J., {et~al.} 2013,
  \aap, 558, A33, \dodoi{10.1051/0004-6361/201322068}

\bibitem[{{Baldry} {et~al.}(2006){Baldry}, {Balogh}, {Bower}, {Glazebrook},
  {Nichol}, {Bamford}, \& {Budavari}}]{Baldry:2006kd}
{Baldry}, I.~K., {Balogh}, M.~L., {Bower}, R.~G., {et~al.} 2006, \mnras, 373,
  469, \dodoi{10.1111/j.1365-2966.2006.11081.x}

\bibitem[{{Baldry} {et~al.}(2004){Baldry}, {Glazebrook}, {Brinkmann},
  {Ivezi{\'c}}, {Lupton}, {Nichol}, \& {Szalay}}]{Baldry:2004tw}
{Baldry}, I.~K., {Glazebrook}, K., {Brinkmann}, J., {et~al.} 2004, \apj, 600,
  681, \dodoi{10.1086/380092}

\bibitem[{{Balogh} {et~al.}(2004){Balogh}, {Eke}, {Miller}, {Lewis}, {Bower},
  {Couch}, {Nichol}, {Bland -Hawthorn}, {Baldry}, {Baugh}, {Bridges}, {Cannon},
  {Cole}, {Colless}, {Collins}, {Cross}, {Dalton}, {de Propris}, {Driver},
  {Efstathiou}, {Ellis}, {Frenk}, {Glazebrook}, {Gomez}, {Gray}, {Hawkins},
  {Jackson}, {Lahav}, {Lumsden}, {Maddox}, {Madgwick}, {Norberg}, {Peacock},
  {Percival}, {Peterson}, {Sutherland}, \& {Taylor}}]{Balogh:2004vf}
{Balogh}, M., {Eke}, V., {Miller}, C., {et~al.} 2004, \mnras, 348, 1355,
  \dodoi{10.1111/j.1365-2966.2004.07453.x}

\bibitem[{{Balogh} {et~al.}(2009){Balogh}, {McGee}, {Wilman}, {Bower}, {Hau},
  {Morris}, {Mulchaey}, {Oemler}, {Parker}, \& {Gwyn}}]{Balogh:2009cd}
{Balogh}, M.~L., {McGee}, S.~L., {Wilman}, D., {et~al.} 2009, \mnras, 398, 754,
  \dodoi{10.1111/j.1365-2966.2009.15193.x}

\bibitem[{{Balsara} {et~al.}(1994){Balsara}, {Livio}, \&
  {O'Dea}}]{Balsara:1994nn}
{Balsara}, D., {Livio}, M., \& {O'Dea}, C.~P. 1994, \apj, 437, 83,
  \dodoi{10.1086/174977}

\bibitem[{{Bekki}(2014)}]{Bekki:2014ip}
{Bekki}, K. 2014, \mnras, 438, 444, \dodoi{10.1093/mnras/stt2216}

\bibitem[{{Bekki} \& {Couch}(2003)}]{Bekki:2003tu}
{Bekki}, K., \& {Couch}, W.~J. 2003, \apjl, 596, L13, \dodoi{10.1086/379054}

\bibitem[{{Bekki} {et~al.}(2002){Bekki}, {Couch}, \& {Shioya}}]{Bekki:2002qa}
{Bekki}, K., {Couch}, W.~J., \& {Shioya}, Y. 2002, \apj, 577, 651,
  \dodoi{10.1086/342221}

\bibitem[{{Bertram} {et~al.}(2006){Bertram}, {Eckart}, {Krips}, {Staguhn}, \&
  {Hackenberg}}]{Bertram:2006kj}
{Bertram}, T., {Eckart}, A., {Krips}, M., {Staguhn}, J.~G., \& {Hackenberg}, W.
  2006, \aap, 448, 29, \dodoi{10.1051/0004-6361:20042564}

\bibitem[{{Bialas} {et~al.}(2015){Bialas}, {Lisker}, {Olczak}, {Spurzem}, \&
  {Kotulla}}]{Bialas:2015ty}
{Bialas}, D., {Lisker}, T., {Olczak}, C., {Spurzem}, R., \& {Kotulla}, R. 2015,
  \aap, 576, A103, \dodoi{10.1051/0004-6361/201425235}

\bibitem[{{Binggeli} {et~al.}(1993){Binggeli}, {Popescu}, \&
  {Tammann}}]{Binggeli:1993qt}
{Binggeli}, B., {Popescu}, C.~C., \& {Tammann}, G.~A. 1993, \aaps, 98, 275

\bibitem[{{Binggeli} {et~al.}(1985){Binggeli}, {Sandage}, \&
  {Tammann}}]{Binggeli:1985os}
{Binggeli}, B., {Sandage}, A., \& {Tammann}, G.~A. 1985, \aj, 90, 1681,
  \dodoi{10.1086/113874}

\bibitem[{{Binggeli} {et~al.}(1987){Binggeli}, {Tammann}, \&
  {Sandage}}]{Binggeli:1987ur}
{Binggeli}, B., {Tammann}, G.~A., \& {Sandage}, A. 1987, \aj, 94, 251,
  \dodoi{10.1086/114467}

\bibitem[{{Biviano} {et~al.}(2002){Biviano}, {Katgert}, {Thomas}, \&
  {Adami}}]{Biviano:2002df}
{Biviano}, A., {Katgert}, P., {Thomas}, T., \& {Adami}, C. 2002, \aap, 387, 8,
  \dodoi{10.1051/0004-6361:20020340}

\bibitem[{{B{\"o}hringer} {et~al.}(1994){B{\"o}hringer}, {Briel}, {Schwarz},
  {Voges}, {Hartner}, \& {Tr{\"u}mper}}]{Bohringer:1994td}
{B{\"o}hringer}, H., {Briel}, U.~G., {Schwarz}, R.~A., {et~al.} 1994, \nat,
  368, 828, \dodoi{10.1038/368828a0}

\bibitem[{{Bolatto} {et~al.}(2013){Bolatto}, {Wolfire}, \&
  {Leroy}}]{Bolatto:2013vn}
{Bolatto}, A.~D., {Wolfire}, M., \& {Leroy}, A.~K. 2013, \araa, 51, 207,
  \dodoi{10.1146/annurev-astro-082812-140944}

\bibitem[{{Boquien} {et~al.}(2019){Boquien}, {Burgarella}, {Roehlly}, {Buat},
  {Ciesla}, {Corre}, {Inoue}, \& {Salas}}]{Boquien:2019ex}
{Boquien}, M., {Burgarella}, D., {Roehlly}, Y., {et~al.} 2019, \aap, 622, A103,
  \dodoi{10.1051/0004-6361/201834156}

\bibitem[{{Borthakur} {et~al.}(2010){Borthakur}, {Yun}, \&
  {Verdes-Montenegro}}]{Borthakur:2010uk}
{Borthakur}, S., {Yun}, M.~S., \& {Verdes-Montenegro}, L. 2010, \apj, 710, 385,
  \dodoi{10.1088/0004-637X/710/1/385}

\bibitem[{{Boselli}(1994)}]{Boselli:1994gj}
{Boselli}, A. 1994, \aap, 292, 1

\bibitem[{{Boselli} {et~al.}(1995){Boselli}, {Casoli}, \&
  {Lequeux}}]{Boselli:1995sf}
{Boselli}, A., {Casoli}, F., \& {Lequeux}, J. 1995, \aaps, 110, 521

\bibitem[{{Boselli} {et~al.}(2014{\natexlab{a}}){Boselli}, {Cortese}, \&
  {Boquien}}]{Boselli:2014yq}
{Boselli}, A., {Cortese}, L., \& {Boquien}, M. 2014{\natexlab{a}}, \aap, 564,
  A65, \dodoi{10.1051/0004-6361/201322311}

\bibitem[{{Boselli} {et~al.}(2014{\natexlab{b}}){Boselli}, {Cortese},
  {Boquien}, {Boissier}, {Catinella}, {Gavazzi}, {Lagos}, \&
  {Saintonge}}]{Boselli:2014qs}
{Boselli}, A., {Cortese}, L., {Boquien}, M., {et~al.} 2014{\natexlab{b}}, \aap,
  564, A67, \dodoi{10.1051/0004-6361/201322313}

\bibitem[{{Boselli} {et~al.}(1997){Boselli}, {Gavazzi}, {Lequeux}, {Buat},
  {Casoli}, {Dickey}, \& {Donas}}]{Boselli:1997jk}
{Boselli}, A., {Gavazzi}, G., {Lequeux}, J., {et~al.} 1997, \aap, 327, 522

\bibitem[{{Brown} {et~al.}(2017){Brown}, {Catinella}, {Cortese}, {Lagos},
  {Dav{\'e}}, {Kilborn}, {Haynes}, {Giovanelli}, \&
  {Rafieferantsoa}}]{Brown:2017bu}
{Brown}, T., {Catinella}, B., {Cortese}, L., {et~al.} 2017, \mnras, 466, 1275,
  \dodoi{10.1093/mnras/stw2991}

\bibitem[{{Burchett} {et~al.}(2018){Burchett}, {Tripp}, {Wang}, {Willmer},
  {Bowen}, \& {Jenkins}}]{Burchett:2018jv}
{Burchett}, J.~N., {Tripp}, T.~M., {Wang}, Q.~D., {et~al.} 2018, \mnras, 475,
  2067, \dodoi{10.1093/mnras/stx3170}

\bibitem[{{Butcher} \& {Oemler}(1984)}]{Butcher:1984xt}
{Butcher}, H., \& {Oemler}, A., J. 1984, \apj, 285, 426, \dodoi{10.1086/162519}

\bibitem[{{Cairns} {et~al.}(2019){Cairns}, {Stroe}, {De Breuck}, {Mroczkowski},
  \& {Clements}}]{Cairns:2019ju}
{Cairns}, J., {Stroe}, A., {De Breuck}, C., {Mroczkowski}, T., \& {Clements},
  D. 2019, \apj, 882, 132, \dodoi{10.3847/1538-4357/ab3392}

\bibitem[{{Cappellari}(2013)}]{Cappellari:2013rg}
{Cappellari}, M. 2013, \apjl, 778, L2, \dodoi{10.1088/2041-8205/778/1/L2}

\bibitem[{{Casoli} {et~al.}(1991){Casoli}, {Boisse}, {Combes}, \&
  {Dupraz}}]{Casoli:1991jr}
{Casoli}, F., {Boisse}, P., {Combes}, F., \& {Dupraz}, C. 1991, \aap, 249, 359

\bibitem[{{Castignani} {et~al.}(2021){Castignani}, {Combes}, {Jablonka},
  {Finn}, {Rudnick}, {Vulcani}, {Desai}, {Zaritsky}, \&
  {Salom{\'e}}}]{Castignani:2021tr}
{Castignani}, G., {Combes}, F., {Jablonka}, P., {et~al.} 2021, arXiv e-prints,
  arXiv:2101.04389.
\newblock \doarXiv{2101.04389}

\bibitem[{{Catinella} {et~al.}(2018){Catinella}, {Saintonge}, {Janowiecki},
  {Cortese}, {Dav{\'e}}, {Lemonias}, {Cooper}, {Schiminovich}, {Hummels},
  {Fabello}, {Ger{\'e}b}, {Kilborn}, \& {Wang}}]{Catinella:2018ib}
{Catinella}, B., {Saintonge}, A., {Janowiecki}, S., {et~al.} 2018, \mnras, 476,
  875, \dodoi{10.1093/mnras/sty089}

\bibitem[{{Cavaliere} \& {Fusco-Femiano}(1976)}]{Cavaliere:1976mb}
{Cavaliere}, A., \& {Fusco-Femiano}, R. 1976, \aap, 500, 95

\bibitem[{{Cayatte} {et~al.}(1994){Cayatte}, {Kotanyi}, {Balkowski}, \& {van
  Gorkom}}]{Cayatte:1994hr}
{Cayatte}, V., {Kotanyi}, C., {Balkowski}, C., \& {van Gorkom}, J.~H. 1994,
  \aj, 107, 1003, \dodoi{10.1086/116913}

\bibitem[{{Cayatte} {et~al.}(1990){Cayatte}, {van Gorkom}, {Balkowski}, \&
  {Kotanyi}}]{Cayatte:1990qy}
{Cayatte}, V., {van Gorkom}, J.~H., {Balkowski}, C., \& {Kotanyi}, C. 1990,
  \aj, 100, 604, \dodoi{10.1086/115545}

\bibitem[{{Chabrier}(2003)}]{Chabrier:2003oe}
{Chabrier}, G. 2003, \pasp, 115, 763, \dodoi{10.1086/376392}

\bibitem[{{Chamaraux} {et~al.}(1980){Chamaraux}, {Balkowski}, \&
  {Gerard}}]{Chamaraux:1980rs}
{Chamaraux}, P., {Balkowski}, C., \& {Gerard}, E. 1980, \aap, 83, 38

\bibitem[{{Chang} {et~al.}(2013){Chang}, {Macci{\`o}}, \&
  {Kang}}]{Chang:2013qv}
{Chang}, J., {Macci{\`o}}, A.~V., \& {Kang}, X. 2013, \mnras, 431, 3533,
  \dodoi{10.1093/mnras/stt434}

\bibitem[{{Chung} {et~al.}(2009{\natexlab{a}}){Chung}, {van Gorkom}, {Kenney},
  {Crowl}, \& {Vollmer}}]{Chung:2009ys}
{Chung}, A., {van Gorkom}, J.~H., {Kenney}, J. D.~P., {Crowl}, H., \&
  {Vollmer}, B. 2009{\natexlab{a}}, \aj, 138, 1741,
  \dodoi{10.1088/0004-6256/138/6/1741}

\bibitem[{{Chung} {et~al.}(2009{\natexlab{b}}){Chung}, {Rhee}, {Kim}, {Yun},
  {Heyer}, \& {Young}}]{Chung:2009xq}
{Chung}, E.~J., {Rhee}, M.-H., {Kim}, H., {et~al.} 2009{\natexlab{b}}, \apjs,
  184, 199, \dodoi{10.1088/0067-0049/184/2/199}

\bibitem[{{Chung} {et~al.}(2017){Chung}, {Yun}, {Verheijen}, \&
  {Chung}}]{Chung:2017kj}
{Chung}, E.~J., {Yun}, M.~S., {Verheijen}, M. A.~W., \& {Chung}, A. 2017, \apj,
  843, 50, \dodoi{10.3847/1538-4357/aa756b}

\bibitem[{{Coenda} \& {Muriel}(2009)}]{Coenda:2009ff}
{Coenda}, V., \& {Muriel}, H. 2009, \aap, 504, 347,
  \dodoi{10.1051/0004-6361/200811504}

\bibitem[{{Coenda} {et~al.}(2009){Coenda}, {Muriel}, \&
  {Donzelli}}]{Coenda:2009kz}
{Coenda}, V., {Muriel}, H., \& {Donzelli}, C. 2009, \apj, 700, 1382,
  \dodoi{10.1088/0004-637X/700/2/1382}

\bibitem[{{Combes} {et~al.}(1988){Combes}, {Dupraz}, {Casoli}, \&
  {Pagani}}]{Combes:1988jm}
{Combes}, F., {Dupraz}, C., {Casoli}, F., \& {Pagani}, L. 1988, \aap, 203, L9

\bibitem[{{Cooper} {et~al.}(2008){Cooper}, {Tremonti}, {Newman}, \&
  {Zabludoff}}]{Cooper:2008td}
{Cooper}, M.~C., {Tremonti}, C.~A., {Newman}, J.~A., \& {Zabludoff}, A.~I.
  2008, \mnras, 390, 245, \dodoi{10.1111/j.1365-2966.2008.13714.x}

\bibitem[{{Corbelli} {et~al.}(2012){Corbelli}, {Bianchi}, {Cortese},
  {Giovanardi}, {Magrini}, {Pappalardo}, {Boselli}, {Bendo}, {Davies},
  {Grossi}, {Madden}, {Smith}, {Vlahakis}, {Auld}, {Baes}, {De Looze}, {Fritz},
  {Pohlen}, \& {Verstappen}}]{Corbelli:2012wv}
{Corbelli}, E., {Bianchi}, S., {Cortese}, L., {et~al.} 2012, \aap, 542, A32,
  \dodoi{10.1051/0004-6361/201117329}

\bibitem[{{Cortese} {et~al.}(2007){Cortese}, {Marcillac}, {Richard},
  {Bravo-Alfaro}, {Kneib}, {Rieke}, {Covone}, {Egami}, {Rigby}, {Czoske}, \&
  {Davies}}]{Cortese:2007ty}
{Cortese}, L., {Marcillac}, D., {Richard}, J., {et~al.} 2007, \mnras, 376, 157,
  \dodoi{10.1111/j.1365-2966.2006.11369.x}

\bibitem[{{Cortese} {et~al.}(2016){Cortese}, {Bekki}, {Boselli}, {Catinella},
  {Ciesla}, {Hughes}, {Baes}, {Bendo}, {Boquien}, {de Looze}, {Smith},
  {Spinoglio}, \& {Viaene}}]{Cortese:2016yh}
{Cortese}, L., {Bekki}, K., {Boselli}, A., {et~al.} 2016, \mnras, 459, 3574,
  \dodoi{10.1093/mnras/stw801}

\bibitem[{{Couch} \& {Sharples}(1987)}]{Couch:1987yw}
{Couch}, W.~J., \& {Sharples}, R.~M. 1987, \mnras, 229, 423,
  \dodoi{10.1093/mnras/229.3.423}

\bibitem[{{Cramer} {et~al.}(2020){Cramer}, {Kenney}, {Cortes}, {Cortes P.~C.},
  {Vlahakis}, {J{\'a}chym}, {Pompei}, \& {Rubio}}]{Cramer:2020nt}
{Cramer}, W.~J., {Kenney}, J.~D.~P., {Cortes}, J.~R., {et~al.} 2020, \apj, 901,
  95, \dodoi{10.3847/1538-4357/abaf54}

\bibitem[{{Darvish} {et~al.}(2016){Darvish}, {Mobasher}, {Sobral}, {Rettura},
  {Scoville}, {Faisst}, \& {Capak}}]{Darvish:2016gj}
{Darvish}, B., {Mobasher}, B., {Sobral}, D., {et~al.} 2016, \apj, 825, 113,
  \dodoi{10.3847/0004-637X/825/2/113}

\bibitem[{{Davies} \& {Lewis}(1973)}]{Davies:1973jj}
{Davies}, R.~D., \& {Lewis}, B.~M. 1973, \mnras, 165, 231,
  \dodoi{10.1093/mnras/165.2.231}

\bibitem[{{de Vaucouleurs}(1961)}]{de-Vaucouleurs:1961ln}
{de Vaucouleurs}, G. 1961, \apjs, 6, 213, \dodoi{10.1086/190064}

\bibitem[{{Deb} {et~al.}(2020){Deb}, {Verheijen}, {Gullieuszik}, {Poggianti},
  {van Gorkom}, {Ramatsoku}, {Serra}, {Moretti}, {Vulcani}, {Bettoni},
  {Jaff{\'e}}, {Tonnesen}, \& {Fritz}}]{Deb:2020rl}
{Deb}, T., {Verheijen}, M. A.~W., {Gullieuszik}, M., {et~al.} 2020, \mnras,
  494, 5029, \dodoi{10.1093/mnras/staa968}

\bibitem[{{Dressler}(1980)}]{Dressler:1980yl}
{Dressler}, A. 1980, \apj, 236, 351, \dodoi{10.1086/157753}

\bibitem[{{Ebeling} {et~al.}(2014){Ebeling}, {Stephenson}, \&
  {Edge}}]{Ebeling:2014ez}
{Ebeling}, H., {Stephenson}, L.~N., \& {Edge}, A.~C. 2014, \apjl, 781, L40,
  \dodoi{10.1088/2041-8205/781/2/L40}

\bibitem[{{Eke} {et~al.}(2005){Eke}, {Baugh}, {Cole}, {Frenk}, {King}, \&
  {Peacock}}]{Eke:2005pz}
{Eke}, V.~R., {Baugh}, C.~M., {Cole}, S., {et~al.} 2005, \mnras, 362, 1233,
  \dodoi{10.1111/j.1365-2966.2005.09384.x}

\bibitem[{{Eke} {et~al.}(2004){Eke}, {Baugh}, {Cole}, {Frenk}, {Norberg},
  {Peacock}, {Baldry}, {Bland-Hawthorn}, {Bridges}, {Cannon}, {Colless},
  {Collins}, {Couch}, {Dalton}, {de Propris}, {Driver}, {Efstathiou}, {Ellis},
  {Glazebrook}, {Jackson}, {Lahav}, {Lewis}, {Lumsden}, {Maddox}, {Madgwick},
  {Peterson}, {Sutherland}, \& {Taylor}}]{Eke:2004mt}
---. 2004, \mnras, 348, 866, \dodoi{10.1111/j.1365-2966.2004.07408.x}

\bibitem[{{Ellingson} {et~al.}(2001){Ellingson}, {Lin}, {Yee}, \&
  {Carlberg}}]{Ellingson:2001ms}
{Ellingson}, E., {Lin}, H., {Yee}, H.~K.~C., \& {Carlberg}, R.~G. 2001, \apj,
  547, 609, \dodoi{10.1086/318423}

\bibitem[{{Ellison} {et~al.}(2009){Ellison}, {Simard}, {Cowan}, {Baldry},
  {Patton}, \& {McConnachie}}]{Ellison:2009dd}
{Ellison}, S.~L., {Simard}, L., {Cowan}, N.~B., {et~al.} 2009, \mnras, 396,
  1257, \dodoi{10.1111/j.1365-2966.2009.14817.x}

\bibitem[{{Ellison} {et~al.}(2020){Ellison}, {Thorp}, {Lin}, {Pan}, {Bluck},
  {Scudder}, {Teimoorinia}, {S{\'a}nchez}, \& {Sargent}}]{Ellison:2020kf}
{Ellison}, S.~L., {Thorp}, M.~D., {Lin}, L., {et~al.} 2020, \mnras, 493, L39,
  \dodoi{10.1093/mnrasl/slz179}

\bibitem[{{Emerick} {et~al.}(2015){Emerick}, {Bryan}, \&
  {Putman}}]{Emerick:2015dm}
{Emerick}, A., {Bryan}, G., \& {Putman}, M.~E. 2015, \mnras, 453, 4051,
  \dodoi{10.1093/mnras/stv1936}

\bibitem[{{Ftaclas} {et~al.}(1984){Ftaclas}, {Fanelli}, \&
  {Struble}}]{Ftaclas:1984ae}
{Ftaclas}, C., {Fanelli}, M.~N., \& {Struble}, M.~F. 1984, \apj, 282, 19,
  \dodoi{10.1086/162172}

\bibitem[{{Fujita}(2004)}]{Fujita:2004fa}
{Fujita}, Y. 2004, \pasj, 56, 29, \dodoi{10.1093/pasj/56.1.29}

\bibitem[{{Fujita} \& {Nagashima}(1999)}]{Fujita:1999iv}
{Fujita}, Y., \& {Nagashima}, M. 1999, \apj, 516, 619, \dodoi{10.1086/307139}

\bibitem[{{Fumagalli} \& {Gavazzi}(2008)}]{Fumagalli:2008gf}
{Fumagalli}, M., \& {Gavazzi}, G. 2008, \aap, 490, 571,
  \dodoi{10.1051/0004-6361:200810604}

\bibitem[{{Fumagalli} {et~al.}(2009){Fumagalli}, {Krumholz}, {Prochaska},
  {Gavazzi}, \& {Boselli}}]{Fumagalli:2009bt}
{Fumagalli}, M., {Krumholz}, M.~R., {Prochaska}, J.~X., {Gavazzi}, G., \&
  {Boselli}, A. 2009, \apj, 697, 1811, \dodoi{10.1088/0004-637X/697/2/1811}

\bibitem[{{Gavazzi} {et~al.}(2005){Gavazzi}, {Boselli}, {van Driel}, \&
  {O'Neil}}]{Gavazzi:2005kb}
{Gavazzi}, G., {Boselli}, A., {van Driel}, W., \& {O'Neil}, K. 2005, \aap, 429,
  439, \dodoi{10.1051/0004-6361:20041678}

\bibitem[{{Gavazzi} {et~al.}(2006){Gavazzi}, {O'Neil}, {Boselli}, \& {van
  Driel}}]{Gavazzi:2006xu}
{Gavazzi}, G., {O'Neil}, K., {Boselli}, A., \& {van Driel}, W. 2006, \aap, 449,
  929, \dodoi{10.1051/0004-6361:20053844}

\bibitem[{{Geller} \& {Huchra}(1983)}]{Geller:1983mw}
{Geller}, M.~J., \& {Huchra}, J.~P. 1983, \apjs, 52, 61, \dodoi{10.1086/190859}

\bibitem[{{Genzel} {et~al.}(2015){Genzel}, {Tacconi}, {Lutz}, {Saintonge},
  {Berta}, {Magnelli}, {Combes}, {Garc{\'\i}a-Burillo}, {Neri}, {Bolatto},
  {Contini}, {Lilly}, {Boissier}, {Boone}, {Bouch{\'e}}, {Bournaud}, {Burkert},
  {Carollo}, {Colina}, {Cooper}, {Cox}, {Feruglio}, {F{\"o}rster Schreiber},
  {Freundlich}, {Gracia-Carpio}, {Juneau}, {Kovac}, {Lippa}, {Naab}, {Salome},
  {Renzini}, {Sternberg}, {Walter}, {Weiner}, {Weiss}, \&
  {Wuyts}}]{Genzel:2015gn}
{Genzel}, R., {Tacconi}, L.~J., {Lutz}, D., {et~al.} 2015, \apj, 800, 20,
  \dodoi{10.1088/0004-637X/800/1/20}

\bibitem[{{Giovanelli} \& {Haynes}(1985)}]{Giovanelli:1985kt}
{Giovanelli}, R., \& {Haynes}, M.~P. 1985, \apj, 292, 404,
  \dodoi{10.1086/163170}

\bibitem[{{Giovanelli} {et~al.}(2005){Giovanelli}, {Haynes}, {Kent},
  {Perillat}, {Saintonge}, {Brosch}, {Catinella}, {Hoffman}, {Stierwalt},
  {Spekkens}, {Lerner}, {Masters}, {Momjian}, {Rosenberg}, {Springob},
  {Boselli}, {Charmand aris}, {Darling}, {Davies}, {Garcia Lambas}, {Gavazzi},
  {Giovanardi}, {Hardy}, {Hunt}, {Iovino}, {Karachentsev}, {Karachentseva},
  {Koopmann}, {Marinoni}, {Minchin}, {Muller}, {Putman}, {Pantoja}, {Salzer},
  {Scodeggio}, {Skillman}, {Solanes}, {Valotto}, {van Driel}, \& {van
  Zee}}]{Giovanelli:2005ua}
{Giovanelli}, R., {Haynes}, M.~P., {Kent}, B.~R., {et~al.} 2005, \aj, 130,
  2598, \dodoi{10.1086/497431}

\bibitem[{{G{\'o}mez} {et~al.}(2003){G{\'o}mez}, {Nichol}, {Miller}, {Balogh},
  {Goto}, {Zabludoff}, {Romer}, {Bernardi}, {Sheth}, {Hopkins}, {Castander},
  {Connolly}, {Schneider}, {Brinkmann}, {Lamb}, {SubbaRao}, \&
  {York}}]{Gomez:2003hb}
{G{\'o}mez}, P.~L., {Nichol}, R.~C., {Miller}, C.~J., {et~al.} 2003, \apj, 584,
  210, \dodoi{10.1086/345593}

\bibitem[{{Gu} {et~al.}(2013){Gu}, {Gandhi}, {Inada}, {Kawaharada}, {Kodama},
  {Konami}, {Nakazawa}, {Shimasaku}, {Xu}, \& {Makishima}}]{Gu:2013xa}
{Gu}, L., {Gandhi}, P., {Inada}, N., {et~al.} 2013, \apj, 767, 157,
  \dodoi{10.1088/0004-637X/767/2/157}

\bibitem[{{Gu} {et~al.}(2016){Gu}, {Wen}, {Gandhi}, {Inada}, {Kawaharada},
  {Kodama}, {Konami}, {Nakazawa}, {Xu}, \& {Makishima}}]{Gu:2016td}
{Gu}, L., {Wen}, Z., {Gandhi}, P., {et~al.} 2016, \apj, 826, 72,
  \dodoi{10.3847/0004-637X/826/1/72}

\bibitem[{{Gunn} \& {Gott}(1972)}]{Gunn:1972kc}
{Gunn}, J.~E., \& {Gott}, III, J.~R. 1972, \apj, 176, 1, \dodoi{10.1086/151605}

\bibitem[{{Hayashi} {et~al.}(2017){Hayashi}, {Kodama}, {Kohno}, {Yamaguchi},
  {Tadaki}, {Hatsukade}, {Koyama}, {Shimakawa}, {Tamura}, \&
  {Suzuki}}]{Hayashi:2017dn}
{Hayashi}, M., {Kodama}, T., {Kohno}, K., {et~al.} 2017, \apjl, 841, L21,
  \dodoi{10.3847/2041-8213/aa71ad}

\bibitem[{{Haynes} {et~al.}(2007){Haynes}, {Giovanelli}, \&
  {Kent}}]{Haynes:2007wy}
{Haynes}, M.~P., {Giovanelli}, R., \& {Kent}, B.~R. 2007, \apjl, 665, L19,
  \dodoi{10.1086/521188}

\bibitem[{{Haynes} {et~al.}(2018){Haynes}, {Giovanelli}, {Kent}, {Adams},
  {Balonek}, {Craig}, {Fertig}, {Finn}, {Giovanardi}, {Hallenbeck}, {Hess},
  {Hoffman}, {Huang}, {Jones}, {Koopmann}, {Kornreich}, {Leisman}, {Miller},
  {Moorman}, {O'Connor}, {O'Donoghue}, {Papastergis}, {Troischt}, {Stark}, \&
  {Xiao}}]{Haynes:2018rt}
{Haynes}, M.~P., {Giovanelli}, R., {Kent}, B.~R., {et~al.} 2018, \apj, 861, 49,
  \dodoi{10.3847/1538-4357/aac956}

\bibitem[{{Healy} {et~al.}(2020){Healy}, {Blyth}, {Verheijen}, {Hess}, {Serra},
  {van der Hulst}, {Jarrett}, {Yim}, \& {Jozsa}}]{Healy:2020bz}
{Healy}, J., {Blyth}, S.-L., {Verheijen}, M.~A.~W., {et~al.} 2020, arXiv
  e-prints, arXiv:2011.06285.
\newblock \doarXiv{2011.06285}

\bibitem[{{Hern{\'a}ndez-Fern{\'a}ndez}
  {et~al.}(2014){Hern{\'a}ndez-Fern{\'a}ndez}, {Haines}, {Diaferio},
  {Iglesias-P{\'a}ramo}, {Mendes de Oliveira}, \&
  {Vilchez}}]{Hernandez-Fernandez:2014zx}
{Hern{\'a}ndez-Fern{\'a}ndez}, J.~D., {Haines}, C.~P., {Diaferio}, A., {et~al.}
  2014, \mnras, 438, 2186, \dodoi{10.1093/mnras/stt2354}

\bibitem[{{Hickson}(1982)}]{Hickson:1982pe}
{Hickson}, P. 1982, \apj, 255, 382, \dodoi{10.1086/159838}

\bibitem[{{Hogg} {et~al.}(2004){Hogg}, {Blanton}, {Brinchmann}, {Eisenstein},
  {Schlegel}, {Gunn}, {McKay}, {Rix}, {Bahcall}, {Brinkmann}, \&
  {Meiksin}}]{Hogg:2004oj}
{Hogg}, D.~W., {Blanton}, M.~R., {Brinchmann}, J., {et~al.} 2004, \apjl, 601,
  L29, \dodoi{10.1086/381749}

\bibitem[{{Hou} {et~al.}(2013){Hou}, {Parker}, {Balogh}, {McGee}, {Wilman},
  {Connelly}, {Harris}, {Mok}, {Mulchaey}, {Bower}, \&
  {Finoguenov}}]{Hou:2013aq}
{Hou}, A., {Parker}, L.~C., {Balogh}, M.~L., {et~al.} 2013, \mnras, 435, 1715,
  \dodoi{10.1093/mnras/stt1410}

\bibitem[{{Huchra} \& {Geller}(1982)}]{Huchra:1982gk}
{Huchra}, J.~P., \& {Geller}, M.~J. 1982, \apj, 257, 423,
  \dodoi{10.1086/160000}

\bibitem[{{Hunt} {et~al.}(2015){Hunt}, {Garc{\'\i}a-Burillo}, {Casasola},
  {Caselli}, {Combes}, {Henkel}, {Lundgren}, {Maiolino}, {Menten}, {Testi}, \&
  {Weiss}}]{Hunt:2015yd}
{Hunt}, L.~K., {Garc{\'\i}a-Burillo}, S., {Casasola}, V., {et~al.} 2015, \aap,
  583, A114, \dodoi{10.1051/0004-6361/201526553}

\bibitem[{{J{\'a}chym} {et~al.}(2014){J{\'a}chym}, {Combes}, {Cortese}, {Sun},
  \& {Kenney}}]{Jachym:2014oc}
{J{\'a}chym}, P., {Combes}, F., {Cortese}, L., {Sun}, M., \& {Kenney}, J. D.~P.
  2014, \apj, 792, 11, \dodoi{10.1088/0004-637X/792/1/11}

\bibitem[{{J{\'a}chym} {et~al.}(2017){J{\'a}chym}, {Sun}, {Kenney}, {Cortese},
  {Combes}, {Yagi}, {Yoshida}, {Palou{\v{s}}}, \& {Roediger}}]{Jachym:2017vd}
{J{\'a}chym}, P., {Sun}, M., {Kenney}, J. D.~P., {et~al.} 2017, \apj, 839, 114,
  \dodoi{10.3847/1538-4357/aa6af5}

\bibitem[{{J{\'a}chym} {et~al.}(2019){J{\'a}chym}, {Kenney}, {Sun}, {Combes},
  {Cortese}, {Scott}, {Sivanandam}, {Brinks}, {Roediger}, {Palou{\v{s}}}, \&
  {Fumagalli}}]{Jachym:2019hx}
{J{\'a}chym}, P., {Kenney}, J. D.~P., {Sun}, M., {et~al.} 2019, \apj, 883, 145,
  \dodoi{10.3847/1538-4357/ab3e6c}

\bibitem[{{Jaff{\'e}} {et~al.}(2015){Jaff{\'e}}, {Smith}, {Candlish},
  {Poggianti}, {Sheen}, \& {Verheijen}}]{Jaffe:2015pq}
{Jaff{\'e}}, Y.~L., {Smith}, R., {Candlish}, G.~N., {et~al.} 2015, \mnras, 448,
  1715, \dodoi{10.1093/mnras/stv100}

\bibitem[{{Kauffmann} {et~al.}(2004){Kauffmann}, {White}, {Heckman},
  {M{\'e}nard}, {Brinchmann}, {Charlot}, {Tremonti}, \&
  {Brinkmann}}]{Kauffmann:2004rj}
{Kauffmann}, G., {White}, S. D.~M., {Heckman}, T.~M., {et~al.} 2004, \mnras,
  353, 713, \dodoi{10.1111/j.1365-2966.2004.08117.x}

\bibitem[{{Kenney} \& {Young}(1986)}]{Kenney:1986eu}
{Kenney}, J.~D., \& {Young}, J.~S. 1986, \apjl, 301, L13,
  \dodoi{10.1086/184614}

\bibitem[{{Kenney} \& {Young}(1989)}]{Kenney:1989vp}
{Kenney}, J. D.~P., \& {Young}, J.~S. 1989, \apj, 344, 171,
  \dodoi{10.1086/167787}

\bibitem[{{Kim} {et~al.}(2014){Kim}, {Rey}, {Jerjen}, {Lisker}, {Sung}, {Lee},
  {Chung}, {Pak}, {Yi}, \& {Lee}}]{Kim:2014nb}
{Kim}, S., {Rey}, S.-C., {Jerjen}, H., {et~al.} 2014, \apjs, 215, 22,
  \dodoi{10.1088/0067-0049/215/2/22}

\bibitem[{{Kleiner} {et~al.}(2021){Kleiner}, {Serra}, {Maccagni}, {Venhola},
  {Morokuma-Matsui}, {Peletier}, {Iodice}, {Raj}, {de Blok}, {Comrie},
  {J{\'o}zsa}, {Kamphuis}, {Loni}, {Loubser}, {Moln{\'a}r}, {Passmoor},
  {Ramatsoku}, {Sivitilli}, {Smirnov}, {Thorat}, \& {Vitello}}]{Kleiner:2021un}
{Kleiner}, D., {Serra}, P., {Maccagni}, F.~M., {et~al.} 2021, arXiv e-prints,
  arXiv:2101.10347.
\newblock \doarXiv{2101.10347}

\bibitem[{{Kodama} \& {Bower}(2001)}]{Kodama:2001zx}
{Kodama}, T., \& {Bower}, R.~G. 2001, \mnras, 321, 18,
  \dodoi{10.1046/j.1365-8711.2001.03981.x}

\bibitem[{{Kodama} {et~al.}(2001){Kodama}, {Smail}, {Nakata}, {Okamura}, \&
  {Bower}}]{Kodama:2001zg}
{Kodama}, T., {Smail}, I., {Nakata}, F., {Okamura}, S., \& {Bower}, R.~G. 2001,
  \apjl, 562, L9, \dodoi{10.1086/338100}

\bibitem[{{Koopmann} {et~al.}(2006){Koopmann}, {Haynes}, \&
  {Catinella}}]{Koopmann:2006hj}
{Koopmann}, R.~A., {Haynes}, M.~P., \& {Catinella}, B. 2006, \aj, 131, 716,
  \dodoi{10.1086/498713}

\bibitem[{{Koribalski} \& {L{\'o}pez-S{\'a}nchez}(2009)}]{Koribalski:2009zi}
{Koribalski}, B.~S., \& {L{\'o}pez-S{\'a}nchez}, {\'A}.~R. 2009, \mnras, 400,
  1749, \dodoi{10.1111/j.1365-2966.2009.15610.x}

\bibitem[{{Koribalski} {et~al.}(2018){Koribalski}, {Wang}, {Kamphuis},
  {Westmeier}, {Staveley-Smith}, {Oh}, {L{\'o}pez-S{\'a}nchez}, {Wong}, {Ott},
  {de Blok}, \& {Shao}}]{Koribalski:2018la}
{Koribalski}, B.~S., {Wang}, J., {Kamphuis}, P., {et~al.} 2018, \mnras, 478,
  1611, \dodoi{10.1093/mnras/sty479}

\bibitem[{{Koyama} {et~al.}(2017){Koyama}, {Koyama}, {Yamashita},
  {Morokuma-Matsui}, {Matsuhara}, {Nakagawa}, {Hayashi}, {Kodama}, {Shimakawa},
  {Suzuki}, {Tadaki}, {Tanaka}, \& {Yamamoto}}]{Koyama:2017lf}
{Koyama}, S., {Koyama}, Y., {Yamashita}, T., {et~al.} 2017, \apj, 847, 137,
  \dodoi{10.3847/1538-4357/aa8a6c}

\bibitem[{{Koyama} {et~al.}(2008){Koyama}, {Kodama}, {Shimasaku}, {Okamura},
  {Tanaka}, {Lee}, {Im}, {Matsuhara}, {Takagi}, {Wada}, \&
  {Oyabu}}]{Koyama:2008dn}
{Koyama}, Y., {Kodama}, T., {Shimasaku}, K., {et~al.} 2008, \mnras, 391, 1758,
  \dodoi{10.1111/j.1365-2966.2008.13931.x}

\bibitem[{{Kronberger} {et~al.}(2008){Kronberger}, {Kapferer}, {Ferrari},
  {Unterguggenberger}, \& {Schindler}}]{Kronberger:2008li}
{Kronberger}, T., {Kapferer}, W., {Ferrari}, C., {Unterguggenberger}, S., \&
  {Schindler}, S. 2008, \aap, 481, 337, \dodoi{10.1051/0004-6361:20078904}

\bibitem[{{Larson} {et~al.}(1980){Larson}, {Tinsley}, \&
  {Caldwell}}]{Larson:1980ok}
{Larson}, R.~B., {Tinsley}, B.~M., \& {Caldwell}, C.~N. 1980, \apj, 237, 692,
  \dodoi{10.1086/157917}

\bibitem[{{Lee} {et~al.}(2017){Lee}, {Chung}, {Tonnesen}, {Kenney}, {Wong},
  {Vollmer}, {Petitpas}, {Crowl}, \& {van Gorkom}}]{Lee:2017ut}
{Lee}, B., {Chung}, A., {Tonnesen}, S., {et~al.} 2017, \mnras, 466, 1382,
  \dodoi{10.1093/mnras/stw3162}

\bibitem[{{Leon} {et~al.}(1998){Leon}, {Combes}, \& {Menon}}]{Leon:1998qq}
{Leon}, S., {Combes}, F., \& {Menon}, T.~K. 1998, \aap, 330, 37.
\newblock \doarXiv{astro-ph/9709121}

\bibitem[{{Leroy} {et~al.}(2019){Leroy}, {Sandstrom}, {Lang}, {Lewis}, {Salim},
  {Behrens}, {Chastenet}, {Chiang}, {Gallagher}, {Kessler}, \&
  {Utomo}}]{Leroy:2019cu}
{Leroy}, A.~K., {Sandstrom}, K.~M., {Lang}, D., {et~al.} 2019, \apjs, 244, 24,
  \dodoi{10.3847/1538-4365/ab3925}

\bibitem[{{Lewis} {et~al.}(2002){Lewis}, {Balogh}, {De Propris}, {Couch},
  {Bower}, {Offer}, {Bland -Hawthorn}, {Baldry}, {Baugh}, {Bridges}, {Cannon},
  {Cole}, {Colless}, {Collins}, {Cross}, {Dalton}, {Driver}, {Efstathiou},
  {Ellis}, {Frenk}, {Glazebrook}, {Hawkins}, {Jackson}, {Lahav}, {Lumsden},
  {Maddox}, {Madgwick}, {Norberg}, {Peacock}, {Percival}, {Peterson},
  {Sutherland}, \& {Taylor}}]{Lewis:2002pw}
{Lewis}, I., {Balogh}, M., {De Propris}, R., {et~al.} 2002, \mnras, 334, 673,
  \dodoi{10.1046/j.1365-8711.2002.05558.x}

\bibitem[{{Lisenfeld} {et~al.}(2011){Lisenfeld}, {Espada}, {Verdes-Montenegro},
  {Kuno}, {Leon}, {Sabater}, {Sato}, {Sulentic}, {Verley}, \&
  {Yun}}]{Lisenfeld:2011sy}
{Lisenfeld}, U., {Espada}, D., {Verdes-Montenegro}, L., {et~al.} 2011, \aap,
  534, A102, \dodoi{10.1051/0004-6361/201117056}

\bibitem[{{Lu} {et~al.}(2012){Lu}, {Gilbank}, {McGee}, {Balogh}, \&
  {Gallagher}}]{Lu:2012zy}
{Lu}, T., {Gilbank}, D.~G., {McGee}, S.~L., {Balogh}, M.~L., \& {Gallagher}, S.
  2012, \mnras, 420, 126, \dodoi{10.1111/j.1365-2966.2011.20008.x}

\bibitem[{{Mahajan} {et~al.}(2011){Mahajan}, {Mamon}, \&
  {Raychaudhury}}]{Mahajan:2011qg}
{Mahajan}, S., {Mamon}, G.~A., \& {Raychaudhury}, S. 2011, \mnras, 416, 2882,
  \dodoi{10.1111/j.1365-2966.2011.19236.x}

\bibitem[{{Margoniner} {et~al.}(2001){Margoniner}, {de Carvalho}, {Gal}, \&
  {Djorgovski}}]{Margoniner:2001xy}
{Margoniner}, V.~E., {de Carvalho}, R.~R., {Gal}, R.~R., \& {Djorgovski}, S.~G.
  2001, \apjl, 548, L143, \dodoi{10.1086/319099}

\bibitem[{{Martin} {et~al.}(2005){Martin}, {Fanson}, {Schiminovich},
  {Morrissey}, {Friedman}, {Barlow}, {Conrow}, {Grange}, {Jelinsky},
  {Milliard}, {Siegmund}, {Bianchi}, {Byun}, {Donas}, {Forster}, {Heckman},
  {Lee}, {Madore}, {Malina}, {Neff}, {Rich}, {Small}, {Surber}, {Szalay},
  {Welsh}, \& {Wyder}}]{Martin:2005wd}
{Martin}, D.~C., {Fanson}, J., {Schiminovich}, D., {et~al.} 2005, \apjl, 619,
  L1, \dodoi{10.1086/426387}

\bibitem[{{Martinez-Badenes} {et~al.}(2012){Martinez-Badenes}, {Lisenfeld},
  {Espada}, {Verdes-Montenegro}, {Garc{\'\i}a-Burillo}, {Leon}, {Sulentic}, \&
  {Yun}}]{Martinez-Badenes:2012dj}
{Martinez-Badenes}, V., {Lisenfeld}, U., {Espada}, D., {et~al.} 2012, \aap,
  540, A96, \dodoi{10.1051/0004-6361/201117281}

\bibitem[{{Mastropietro} {et~al.}(2005){Mastropietro}, {Moore}, {Mayer},
  {Debattista}, {Piffaretti}, \& {Stadel}}]{Mastropietro:2005gf}
{Mastropietro}, C., {Moore}, B., {Mayer}, L., {et~al.} 2005, \mnras, 364, 607,
  \dodoi{10.1111/j.1365-2966.2005.09579.x}

\bibitem[{{Matsuki} {et~al.}(2017){Matsuki}, {Koyama}, {Nakagawa}, \&
  {Takita}}]{Matsuki:2017nl}
{Matsuki}, Y., {Koyama}, Y., {Nakagawa}, T., \& {Takita}, S. 2017, \mnras, 466,
  2517, \dodoi{10.1093/mnras/stw2929}

\bibitem[{{McGee} {et~al.}(2011){McGee}, {Balogh}, {Wilman}, {Bower},
  {Mulchaey}, {Parker}, \& {Oemler}}]{McGee:2011wd}
{McGee}, S.~L., {Balogh}, M.~L., {Wilman}, D.~J., {et~al.} 2011, \mnras, 413,
  996, \dodoi{10.1111/j.1365-2966.2010.18189.x}

\bibitem[{{McLaughlin}(1999)}]{McLaughlin:1999gx}
{McLaughlin}, D.~E. 1999, \apjl, 512, L9, \dodoi{10.1086/311860}

\bibitem[{{Mei} {et~al.}(2007){Mei}, {Blakeslee}, {C{\^o}t{\'e}}, {Tonry},
  {West}, {Ferrarese}, {Jord{\'a}n}, {Peng}, {Anthony}, \&
  {Merritt}}]{Mei:2007wt}
{Mei}, S., {Blakeslee}, J.~P., {C{\^o}t{\'e}}, P., {et~al.} 2007, \apj, 655,
  144, \dodoi{10.1086/509598}

\bibitem[{{Mok} {et~al.}(2017){Mok}, {Wilson}, {Knapen}, {S{\'a}nchez-Gallego},
  {Brinks}, \& {Rosolowsky}}]{Mok:2017ey}
{Mok}, A., {Wilson}, C.~D., {Knapen}, J.~H., {et~al.} 2017, \mnras, 467, 4282,
  \dodoi{10.1093/mnras/stx345}

\bibitem[{{Moore} {et~al.}(1996){Moore}, {Katz}, {Lake}, {Dressler}, \&
  {Oemler}}]{Moore:1996mv}
{Moore}, B., {Katz}, N., {Lake}, G., {Dressler}, A., \& {Oemler}, A. 1996,
  \nat, 379, 613, \dodoi{10.1038/379613a0}

\bibitem[{{Moore} {et~al.}(1998){Moore}, {Lake}, \& {Katz}}]{Moore:1998ps}
{Moore}, B., {Lake}, G., \& {Katz}, N. 1998, \apj, 495, 139,
  \dodoi{10.1086/305264}

\bibitem[{{Moore} {et~al.}(1999){Moore}, {Lake}, {Quinn}, \&
  {Stadel}}]{Moore:1999hw}
{Moore}, B., {Lake}, G., {Quinn}, T., \& {Stadel}, J. 1999, \mnras, 304, 465,
  \dodoi{10.1046/j.1365-8711.1999.02345.x}

\bibitem[{{Moretti} {et~al.}(2018){Moretti}, {Paladino}, {Poggianti},
  {D'Onofrio}, {Bettoni}, {Gullieuszik}, {Jaff{\'e}}, {Vulcani}, {Fasano},
  {Fritz}, \& {Torstensson}}]{Moretti:2018hp}
{Moretti}, A., {Paladino}, R., {Poggianti}, B.~M., {et~al.} 2018, \mnras, 480,
  2508, \dodoi{10.1093/mnras/sty2021}

\bibitem[{{Moretti} {et~al.}(2020{\natexlab{a}}){Moretti}, {Paladino},
  {Poggianti}, {Serra}, {Roediger}, {Gullieuszik}, {Tomi{\v{c}}i{\'c}},
  {Radovich}, {Vulcani}, {Jaff{\'e}}, {Fritz}, {Bettoni}, {Ramatsoku}, \&
  {Wolter}}]{Moretti:2020we}
---. 2020{\natexlab{a}}, \apj, 889, 9, \dodoi{10.3847/1538-4357/ab616a}

\bibitem[{{Moretti} {et~al.}(2020{\natexlab{b}}){Moretti}, {Paladino},
  {Poggianti}, {Serra}, {Ramatsoku}, {Franchetto}, {Deb}, {Gullieuszik},
  {Tomi{\v{c}}i{\'c}}, {Mingozzi}, {Vulcani}, {Radovich}, {Bettoni}, \&
  {Fritz}}]{Moretti:2020td}
---. 2020{\natexlab{b}}, \apjl, 897, L30, \dodoi{10.3847/2041-8213/ab9f3b}

\bibitem[{{Mori} \& {Burkert}(2000)}]{Mori:2000om}
{Mori}, M., \& {Burkert}, A. 2000, \apj, 538, 559, \dodoi{10.1086/309140}

\bibitem[{{Mouhcine} {et~al.}(2007){Mouhcine}, {Baldry}, \&
  {Bamford}}]{Mouhcine:2007dd}
{Mouhcine}, M., {Baldry}, I.~K., \& {Bamford}, S.~P. 2007, \mnras, 382, 801,
  \dodoi{10.1111/j.1365-2966.2007.12405.x}

\bibitem[{{Mulchaey}(2000)}]{Mulchaey:2000zn}
{Mulchaey}, J.~S. 2000, \araa, 38, 289, \dodoi{10.1146/annurev.astro.38.1.289}

\bibitem[{{Muzzin} {et~al.}(2014){Muzzin}, {van der Burg}, {McGee}, {Balogh},
  {Franx}, {Hoekstra}, {Hudson}, {Noble}, {Taranu}, {Webb}, {Wilson}, \&
  {Yee}}]{Muzzin:2014ev}
{Muzzin}, A., {van der Burg}, R.~F.~J., {McGee}, S.~L., {et~al.} 2014, \apj,
  796, 65, \dodoi{10.1088/0004-637X/796/1/65}

\bibitem[{{Nakanishi} {et~al.}(2006){Nakanishi}, {Kuno}, {Sofue}, {Sato},
  {Nakai}, {Shioya}, {Tosaki}, {Onodera}, {Sorai}, {Egusa}, \&
  {Hirota}}]{Nakanishi:2006zv}
{Nakanishi}, H., {Kuno}, N., {Sofue}, Y., {et~al.} 2006, \apj, 651, 804,
  \dodoi{10.1086/507974}

\bibitem[{{Noble} {et~al.}(2013){Noble}, {Webb}, {Muzzin}, {Wilson}, {Yee}, \&
  {van der Burg}}]{Noble:2013he}
{Noble}, A.~G., {Webb}, T.~M.~A., {Muzzin}, A., {et~al.} 2013, \apj, 768, 118,
  \dodoi{10.1088/0004-637X/768/2/118}

\bibitem[{{Noble} {et~al.}(2016){Noble}, {Webb}, {Yee}, {Muzzin}, {Wilson},
  {van der Burg}, {Balogh}, \& {Shupe}}]{Noble:2016wl}
{Noble}, A.~G., {Webb}, T.~M.~A., {Yee}, H.~K.~C., {et~al.} 2016, \apj, 816,
  48, \dodoi{10.3847/0004-637X/816/2/48}

\bibitem[{{Okamoto} \& {Nagashima}(2003)}]{Okamoto:2003xb}
{Okamoto}, T., \& {Nagashima}, M. 2003, \apj, 587, 500, \dodoi{10.1086/368251}

\bibitem[{{Oman} {et~al.}(2013){Oman}, {Hudson}, \& {Behroozi}}]{Oman:2013tf}
{Oman}, K.~A., {Hudson}, M.~J., \& {Behroozi}, P.~S. 2013, \mnras, 431, 2307,
  \dodoi{10.1093/mnras/stt328}

\bibitem[{{Paccagnella} {et~al.}(2016){Paccagnella}, {Vulcani}, {Poggianti},
  {Moretti}, {Fritz}, {Gullieuszik}, {Couch}, {Bettoni}, {Cava}, {D'Onofrio},
  \& {Fasano}}]{Paccagnella:2016nw}
{Paccagnella}, A., {Vulcani}, B., {Poggianti}, B.~M., {et~al.} 2016, \apjl,
  816, L25, \dodoi{10.3847/2041-8205/816/2/L25}

\bibitem[{{Paccagnella} {et~al.}(2017){Paccagnella}, {Vulcani}, {Poggianti},
  {Fritz}, {Fasano}, {Moretti}, {Jaff{\'e}}, {Biviano}, {Gullieuszik},
  {Bettoni}, {Cava}, {Couch}, \& {D'Onofrio}}]{Paccagnella:2017tx}
---. 2017, \apj, 838, 148, \dodoi{10.3847/1538-4357/aa64d7}

\bibitem[{{Peng} {et~al.}(2015){Peng}, {Maiolino}, \& {Cochrane}}]{Peng:2015ar}
{Peng}, Y., {Maiolino}, R., \& {Cochrane}, R. 2015, \nat, 521, 192,
  \dodoi{10.1038/nature14439}

\bibitem[{{Peng} {et~al.}(2010){Peng}, {Lilly}, {Kova{\v{c}}}, {Bolzonella},
  {Pozzetti}, {Renzini}, {Zamorani}, {Ilbert}, {Knobel}, {Iovino}, {Maier},
  {Cucciati}, {Tasca}, {Carollo}, {Silverman}, {Kampczyk}, {de Ravel},
  {Sanders}, {Scoville}, {Contini}, {Mainieri}, {Scodeggio}, {Kneib}, {Le
  F{\`e}vre}, {Bardelli}, {Bongiorno}, {Caputi}, {Coppa}, {de la Torre},
  {Franzetti}, {Garilli}, {Lamareille}, {Le Borgne}, {Le Brun}, {Mignoli},
  {Perez Montero}, {Pello}, {Ricciardelli}, {Tanaka}, {Tresse}, {Vergani},
  {Welikala}, {Zucca}, {Oesch}, {Abbas}, {Barnes}, {Bordoloi}, {Bottini},
  {Cappi}, {Cassata}, {Cimatti}, {Fumana}, {Hasinger}, {Koekemoer},
  {Leauthaud}, {Maccagni}, {Marinoni}, {McCracken}, {Memeo}, {Meneux}, {Nair},
  {Porciani}, {Presotto}, \& {Scaramella}}]{Peng:2010eq}
{Peng}, Y.-j., {Lilly}, S.~J., {Kova{\v{c}}}, K., {et~al.} 2010, \apj, 721,
  193, \dodoi{10.1088/0004-637X/721/1/193}

\bibitem[{{Perea} {et~al.}(1997){Perea}, {del Olmo}, {Verdes-Montenegro}, \&
  {Yun}}]{Perea:1997wd}
{Perea}, J., {del Olmo}, A., {Verdes-Montenegro}, L., \& {Yun}, M.~S. 1997,
  \apj, 490, 166, \dodoi{10.1086/304876}

\bibitem[{{Pimbblet} {et~al.}(2013){Pimbblet}, {Shabala}, {Haines},
  {Fraser-McKelvie}, \& {Floyd}}]{Pimbblet:2013wp}
{Pimbblet}, K.~A., {Shabala}, S.~S., {Haines}, C.~P., {Fraser-McKelvie}, A., \&
  {Floyd}, D.~J.~E. 2013, \mnras, 429, 1827, \dodoi{10.1093/mnras/sts470}

\bibitem[{{Poggianti} {et~al.}(2016){Poggianti}, {Fasano}, {Omizzolo},
  {Gullieuszik}, {Bettoni}, {Moretti}, {Paccagnella}, {Jaff{\'e}}, {Vulcani},
  {Fritz}, {Couch}, \& {D'Onofrio}}]{Poggianti:2016bc}
{Poggianti}, B.~M., {Fasano}, G., {Omizzolo}, A., {et~al.} 2016, \aj, 151, 78,
  \dodoi{10.3847/0004-6256/151/3/78}

\bibitem[{{Poggianti} {et~al.}(2017){Poggianti}, {Moretti}, {Gullieuszik},
  {Fritz}, {Jaff{\'e}}, {Bettoni}, {Fasano}, {Bellhouse}, {Hau}, {Vulcani},
  {Biviano}, {Omizzolo}, {Paccagnella}, {D'Onofrio}, {Cava}, {Sheen}, {Couch},
  \& {Owers}}]{Poggianti:2017qg}
{Poggianti}, B.~M., {Moretti}, A., {Gullieuszik}, M., {et~al.} 2017, \apj, 844,
  48, \dodoi{10.3847/1538-4357/aa78ed}

\bibitem[{{Quilis} {et~al.}(2000){Quilis}, {Moore}, \& {Bower}}]{Quilis:2000gz}
{Quilis}, V., {Moore}, B., \& {Bower}, R. 2000, Science, 288, 1617,
  \dodoi{10.1126/science.288.5471.1617}

\bibitem[{{Rakos} \& {Schombert}(1995)}]{Rakos:1995xt}
{Rakos}, K.~D., \& {Schombert}, J.~M. 1995, \apj, 439, 47,
  \dodoi{10.1086/175150}

\bibitem[{{Ramatsoku} {et~al.}(2019){Ramatsoku}, {Serra}, {Poggianti},
  {Moretti}, {Gullieuszik}, {Bettoni}, {Deb}, {Fritz}, {van Gorkom},
  {Jaff{\'e}}, {Tonnesen}, {Verheijen}, {Vulcani}, {Hugo}, {J{\'o}zsa},
  {Maccagni}, {Makhathini}, {Ramaila}, {Smirnov}, \&
  {Thorat}}]{Ramatsoku:2019kt}
{Ramatsoku}, M., {Serra}, P., {Poggianti}, B.~M., {et~al.} 2019, \mnras, 487,
  4580, \dodoi{10.1093/mnras/stz1609}

\bibitem[{{Ramatsoku} {et~al.}(2020){Ramatsoku}, {Serra}, {Poggianti},
  {Moretti}, {Gullieuszik}, {Bettoni}, {Deb}, {Franchetto}, {van Gorkom},
  {Jaff{\'e}}, {Tonnesen}, {Verheijen}, {Vulcani}, {Andati}, {de Blok},
  {J{\'o}zsa}, {Kamphuis}, {Kleiner}, {Maccagni}, {Makhathini}, {Moln{\'a}r},
  {Ramaila}, {Smirnov}, \& {Thorat}}]{Ramatsoku:2020ut}
---. 2020, \aap, 640, A22, \dodoi{10.1051/0004-6361/202037759}

\bibitem[{{Rasmussen} {et~al.}(2012){Rasmussen}, {Mulchaey}, {Bai}, {Ponman},
  {Raychaudhury}, \& {Dariush}}]{Rasmussen:2012ej}
{Rasmussen}, J., {Mulchaey}, J.~S., {Bai}, L., {et~al.} 2012, \apj, 757, 122,
  \dodoi{10.1088/0004-637X/757/2/122}

\bibitem[{{Rasmussen} {et~al.}(2008){Rasmussen}, {Ponman}, {Verdes-Montenegro},
  {Yun}, \& {Borthakur}}]{Rasmussen:2008kb}
{Rasmussen}, J., {Ponman}, T.~J., {Verdes-Montenegro}, L., {Yun}, M.~S., \&
  {Borthakur}, S. 2008, \mnras, 388, 1245,
  \dodoi{10.1111/j.1365-2966.2008.13451.x}

\bibitem[{{Rengarajan} \& {Iyengar}(1992)}]{Rengarajan:1992kx}
{Rengarajan}, T.~N., \& {Iyengar}, K.~V.~K. 1992, \mnras, 259, 559,
  \dodoi{10.1093/mnras/259.3.559}

\bibitem[{{Rhee} {et~al.}(2017){Rhee}, {Smith}, {Choi}, {Yi}, {Jaff{\'e}},
  {Candlish}, \& {S{\'a}nchez-J{\'a}nssen}}]{Rhee:2017kl}
{Rhee}, J., {Smith}, R., {Choi}, H., {et~al.} 2017, \apj, 843, 128,
  \dodoi{10.3847/1538-4357/aa6d6c}

\bibitem[{{Roberts} \& {Parker}(2020)}]{Roberts:2020he}
{Roberts}, I.~D., \& {Parker}, L.~C. 2020, \mnras, 495, 554,
  \dodoi{10.1093/mnras/staa1213}

\bibitem[{{Roberts} {et~al.}(2019){Roberts}, {Parker}, {Brown}, {Joshi},
  {Hlavacek-Larrondo}, \& {Wadsley}}]{Roberts:2019bp}
{Roberts}, I.~D., {Parker}, L.~C., {Brown}, T., {et~al.} 2019, \apj, 873, 42,
  \dodoi{10.3847/1538-4357/ab04f7}

\bibitem[{{Robitaille} \& {Bressert}(2012)}]{Robitaille:2012lz}
{Robitaille}, T., \& {Bressert}, E. 2012, {APLpy: Astronomical Plotting Library
  in Python}.
\newblock \doeprint{1208.017}

\bibitem[{{Ruggiero} \& {Lima Neto}(2017)}]{Ruggiero:2017qq}
{Ruggiero}, R., \& {Lima Neto}, G.~B. 2017, \mnras, 468, 4107,
  \dodoi{10.1093/mnras/stx744}

\bibitem[{{Saintonge} {et~al.}(2008){Saintonge}, {Tran}, \&
  {Holden}}]{Saintonge:2008tj}
{Saintonge}, A., {Tran}, K.-V.~H., \& {Holden}, B.~P. 2008, \apjl, 685, L113,
  \dodoi{10.1086/592730}

\bibitem[{{Saintonge} {et~al.}(2012){Saintonge}, {Tacconi}, {Fabello}, {Wang},
  {Catinella}, {Genzel}, {Graci{\'a}-Carpio}, {Kramer}, {Moran}, {Heckman},
  {Schiminovich}, {Schuster}, \& {Wuyts}}]{Saintonge:2012nj}
{Saintonge}, A., {Tacconi}, L.~J., {Fabello}, S., {et~al.} 2012, \apj, 758, 73,
  \dodoi{10.1088/0004-637X/758/2/73}

\bibitem[{{Saintonge} {et~al.}(2017){Saintonge}, {Catinella}, {Tacconi},
  {Kauffmann}, {Genzel}, {Cortese}, {Dav{\'e}}, {Fletcher},
  {Graci{\'a}-Carpio}, {Kramer}, {Heckman}, {Janowiecki}, {Lutz}, {Rosario},
  {Schiminovich}, {Schuster}, {Wang}, {Wuyts}, {Borthakur}, {Lamperti}, \&
  {Roberts-Borsani}}]{Saintonge:2017ve}
{Saintonge}, A., {Catinella}, B., {Tacconi}, L.~J., {et~al.} 2017, \apjs, 233,
  22, \dodoi{10.3847/1538-4365/aa97e0}

\bibitem[{{Salim} {et~al.}(2016){Salim}, {Lee}, {Janowiecki}, {da Cunha},
  {Dickinson}, {Boquien}, {Burgarella}, {Salzer}, \& {Charlot}}]{Salim:2016wi}
{Salim}, S., {Lee}, J.~C., {Janowiecki}, S., {et~al.} 2016, \apjs, 227, 2,
  \dodoi{10.3847/0067-0049/227/1/2}

\bibitem[{{Sandage} {et~al.}(1985){Sandage}, {Binggeli}, \&
  {Tammann}}]{Sandage:1985bw}
{Sandage}, A., {Binggeli}, B., \& {Tammann}, G.~A. 1985, \aj, 90, 1759,
  \dodoi{10.1086/113875}

\bibitem[{{Schr{\"o}der} {et~al.}(2001){Schr{\"o}der}, {Drinkwater}, \&
  {Richter}}]{Schroder:2001rc}
{Schr{\"o}der}, A., {Drinkwater}, M.~J., \& {Richter}, O.~G. 2001, \aap, 376,
  98, \dodoi{10.1051/0004-6361:20010997}

\bibitem[{{Scoville} {et~al.}(2017){Scoville}, {Lee}, {Vanden Bout},
  {Diaz-Santos}, {Sanders}, {Darvish}, {Bongiorno}, {Casey}, {Murchikova},
  {Koda}, {Capak}, {Vlahakis}, {Ilbert}, {Sheth}, {Morokuma-Matsui}, {Ivison},
  {Aussel}, {Laigle}, {McCracken}, {Armus}, {Pope}, {Toft}, \&
  {Masters}}]{Scoville:2017jw}
{Scoville}, N., {Lee}, N., {Vanden Bout}, P., {et~al.} 2017, \apj, 837, 150,
  \dodoi{10.3847/1538-4357/aa61a0}

\bibitem[{{Serra} {et~al.}(2012){Serra}, {Oosterloo}, {Morganti}, {Alatalo},
  {Blitz}, {Bois}, {Bournaud}, {Bureau}, {Cappellari}, {Crocker}, {Davies},
  {Davis}, {de Zeeuw}, {Duc}, {Emsellem}, {Khochfar}, {Krajnovi{\'c}},
  {Kuntschner}, {Lablanche}, {McDermid}, {Naab}, {Sarzi}, {Scott}, {Trager},
  {Weijmans}, \& {Young}}]{Serra:2012oc}
{Serra}, P., {Oosterloo}, T., {Morganti}, R., {et~al.} 2012, \mnras, 422, 1835,
  \dodoi{10.1111/j.1365-2966.2012.20219.x}

\bibitem[{{Sheen} {et~al.}(2012){Sheen}, {Yi}, {Ree}, \& {Lee}}]{Sheen:2012nt}
{Sheen}, Y.-K., {Yi}, S.~K., {Ree}, C.~H., \& {Lee}, J. 2012, \apjs, 202, 8,
  \dodoi{10.1088/0067-0049/202/1/8}

\bibitem[{{Smethurst} {et~al.}(2017){Smethurst}, {Lintott}, {Bamford}, {Hart},
  {Kruk}, {Masters}, {Nichol}, \& {Simmons}}]{Smethurst:2017ws}
{Smethurst}, R.~J., {Lintott}, C.~J., {Bamford}, S.~P., {et~al.} 2017, \mnras,
  469, 3670, \dodoi{10.1093/mnras/stx973}

\bibitem[{{Smith} {et~al.}(2010){Smith}, {Davies}, \& {Nelson}}]{Smith:2010xy}
{Smith}, R., {Davies}, J.~I., \& {Nelson}, A.~H. 2010, \mnras, 405, 1723,
  \dodoi{10.1111/j.1365-2966.2010.16545.x}

\bibitem[{{Smith} {et~al.}(2015){Smith}, {S{\'a}nchez-Janssen}, {Beasley},
  {Cand lish}, {Gibson}, {Puzia}, {Janz}, {Knebe}, {Aguerri}, {Lisker},
  {Hensler}, {Fellhauer}, {Ferrarese}, \& {Yi}}]{Smith:2015dv}
{Smith}, R., {S{\'a}nchez-Janssen}, R., {Beasley}, M.~A., {et~al.} 2015,
  \mnras, 454, 2502, \dodoi{10.1093/mnras/stv2082}

\bibitem[{{Solanes} {et~al.}(2001){Solanes}, {Manrique},
  {Garc{\'\i}a-G{\'o}mez}, {Gonz{\'a}lez-Casado}, {Giovanelli}, \&
  {Haynes}}]{Solanes:2001nq}
{Solanes}, J.~M., {Manrique}, A., {Garc{\'\i}a-G{\'o}mez}, C., {et~al.} 2001,
  \apj, 548, 97, \dodoi{10.1086/318672}

\bibitem[{{Sorai} {et~al.}(2019){Sorai}, {Kuno}, {Muraoka}, {Miyamoto},
  {Kaneko}, {Nakanishi}, {Nakai}, {Yanagitani}, {Tanaka}, {Sato}, {Salak},
  {Umei}, {Morokuma-Matsui}, {Matsumoto}, {Ueno}, {Pan}, {Noma}, {Takeuchi},
  {Yoda}, {Kuroda}, {Yasuda}, {Yajima}, {Oi}, {Shibata}, {Seta}, {Watanabe},
  {Kita}, {Komatsuzaki}, {Kajikawa}, {Yashima}, {Cooray}, {Baji}, {Segawa},
  {Tashiro}, {Takeda}, {Kishida}, {Hatakeyama}, {Tomiyasu}, \&
  {Saita}}]{Sorai:2019hs}
{Sorai}, K., {Kuno}, N., {Muraoka}, K., {et~al.} 2019, \pasj, 71, S14,
  \dodoi{10.1093/pasj/psz115}

\bibitem[{{Speagle} {et~al.}(2014){Speagle}, {Steinhardt}, {Capak}, \&
  {Silverman}}]{Speagle:2014by}
{Speagle}, J.~S., {Steinhardt}, C.~L., {Capak}, P.~L., \& {Silverman}, J.~D.
  2014, \apjs, 214, 15, \dodoi{10.1088/0067-0049/214/2/15}

\bibitem[{{Stark} {et~al.}(1986){Stark}, {Knapp}, {Bally}, {Wilson}, {Penzias},
  \& {Rowe}}]{Stark:1986js}
{Stark}, A.~A., {Knapp}, G.~R., {Bally}, J., {et~al.} 1986, \apj, 310, 660,
  \dodoi{10.1086/164717}

\bibitem[{{Steinhauser} {et~al.}(2016){Steinhauser}, {Schindler}, \&
  {Springel}}]{Steinhauser:2016dk}
{Steinhauser}, D., {Schindler}, S., \& {Springel}, V. 2016, \aap, 591, A51,
  \dodoi{10.1051/0004-6361/201527705}

\bibitem[{{Tal} {et~al.}(2014){Tal}, {Dekel}, {Oesch}, {Muzzin}, {Brammer},
  {van Dokkum}, {Franx}, {Illingworth}, {Leja}, {Magee}, {Marchesini},
  {Momcheva}, {Nelson}, {Patel}, {Quadri}, {Rix}, {Skelton}, {Wake}, \&
  {Whitaker}}]{Tal:2014mx}
{Tal}, T., {Dekel}, A., {Oesch}, P., {et~al.} 2014, \apj, 789, 164,
  \dodoi{10.1088/0004-637X/789/2/164}

\bibitem[{{Tanaka} {et~al.}(2004){Tanaka}, {Goto}, {Okamura}, {Shimasaku}, \&
  {Brinkmann}}]{Tanaka:2004uh}
{Tanaka}, M., {Goto}, T., {Okamura}, S., {Shimasaku}, K., \& {Brinkmann}, J.
  2004, \aj, 128, 2677, \dodoi{10.1086/425529}

\bibitem[{{Tanaka} {et~al.}(2005){Tanaka}, {Kodama}, {Arimoto}, {Okamura},
  {Umetsu}, {Shimasaku}, {Tanaka}, \& {Yamada}}]{Tanaka:2005vn}
{Tanaka}, M., {Kodama}, T., {Arimoto}, N., {et~al.} 2005, \mnras, 362, 268,
  \dodoi{10.1111/j.1365-2966.2005.09300.x}

\bibitem[{{van der Burg} {et~al.}(2018){van der Burg}, {McGee}, {Aussel},
  {Dahle}, {Arnaud}, {Pratt}, \& {Muzzin}}]{van-der-Burg:2018tf}
{van der Burg}, R. F.~J., {McGee}, S., {Aussel}, H., {et~al.} 2018, \aap, 618,
  A140, \dodoi{10.1051/0004-6361/201833572}

\bibitem[{{Verdes-Montenegro} {et~al.}(1998){Verdes-Montenegro}, {Yun},
  {Perea}, {del Olmo}, \& {Ho}}]{Verdes-Montenegro:1998xt}
{Verdes-Montenegro}, L., {Yun}, M.~S., {Perea}, J., {del Olmo}, A., \& {Ho},
  P.~T.~P. 1998, \apj, 497, 89, \dodoi{10.1086/305454}

\bibitem[{{Verdes-Montenegro} {et~al.}(2001){Verdes-Montenegro}, {Yun},
  {Williams}, {Huchtmeier}, {Del Olmo}, \& {Perea}}]{Verdes-Montenegro:2001dh}
{Verdes-Montenegro}, L., {Yun}, M.~S., {Williams}, B.~A., {et~al.} 2001, \aap,
  377, 812, \dodoi{10.1051/0004-6361:20011127}

\bibitem[{{Verdugo} {et~al.}(2015){Verdugo}, {Combes}, {Dasyra}, {Salom{\'e}},
  \& {Braine}}]{Verdugo:2015is}
{Verdugo}, C., {Combes}, F., {Dasyra}, K., {Salom{\'e}}, P., \& {Braine}, J.
  2015, \aap, 582, A6, \dodoi{10.1051/0004-6361/201526551}

\bibitem[{{Villalobos} {et~al.}(2012){Villalobos}, {De Lucia}, {Borgani}, \&
  {Murante}}]{Villalobos:2012mp}
{Villalobos}, {\'A}., {De Lucia}, G., {Borgani}, S., \& {Murante}, G. 2012,
  \mnras, 424, 2401, \dodoi{10.1111/j.1365-2966.2012.20667.x}

\bibitem[{{Vollmer} {et~al.}(2008){Vollmer}, {Braine}, {Pappalardo}, \&
  {Hily-Blant}}]{Vollmer:2008dk}
{Vollmer}, B., {Braine}, J., {Pappalardo}, C., \& {Hily-Blant}, P. 2008, \aap,
  491, 455, \dodoi{10.1051/0004-6361:200810432}

\bibitem[{{Vollmer} {et~al.}(2001){Vollmer}, {Cayatte}, {Balkowski}, \&
  {Duschl}}]{Vollmer:2001ir}
{Vollmer}, B., {Cayatte}, V., {Balkowski}, C., \& {Duschl}, W.~J. 2001, \apj,
  561, 708, \dodoi{10.1086/323368}

\bibitem[{{Vollmer} {et~al.}(2009){Vollmer}, {Soida}, {Chung}, {Chemin},
  {Braine}, {Boselli}, \& {Beck}}]{Vollmer:2009zg}
{Vollmer}, B., {Soida}, M., {Chung}, A., {et~al.} 2009, \aap, 496, 669,
  \dodoi{10.1051/0004-6361/200811140}

\bibitem[{{Vollmer} {et~al.}(2012){Vollmer}, {Soida}, {Braine}, {Abramson},
  {Beck}, {Chung}, {Crowl}, {Kenney}, \& {van Gorkom}}]{Vollmer:2012jz}
{Vollmer}, B., {Soida}, M., {Braine}, J., {et~al.} 2012, \aap, 537, A143,
  \dodoi{10.1051/0004-6361/201117680}

\bibitem[{{Vulcani} {et~al.}(2018{\natexlab{a}}){Vulcani}, {Poggianti},
  {Gullieuszik}, {Moretti}, {Tonnesen}, {Jaff{\'e}}, {Fritz}, {Fasano}, \&
  {Bettoni}}]{Vulcani:2018oy}
{Vulcani}, B., {Poggianti}, B.~M., {Gullieuszik}, M., {et~al.}
  2018{\natexlab{a}}, \apjl, 866, L25, \dodoi{10.3847/2041-8213/aae68b}

\bibitem[{{Vulcani} {et~al.}(2018{\natexlab{b}}){Vulcani}, {Poggianti},
  {Jaff{\'e}}, {Moretti}, {Fritz}, {Gullieuszik}, {Bettoni}, {Fasano},
  {Tonnesen}, \& {McGee}}]{Vulcani:2018ku}
{Vulcani}, B., {Poggianti}, B.~M., {Jaff{\'e}}, Y.~L., {et~al.}
  2018{\natexlab{b}}, \mnras, 480, 3152, \dodoi{10.1093/mnras/sty2095}

\bibitem[{{Wang} {et~al.}(2020){Wang}, {Xu}, {Lee}, {Du}, {Overzier}, \&
  {Shao}}]{Wang:2020av}
{Wang}, J., {Xu}, W., {Lee}, B., {et~al.} 2020, arXiv e-prints,
  arXiv:2009.08159.
\newblock \doarXiv{2009.08159}

\bibitem[{{Wang} {et~al.}(2018){Wang}, {Elbaz}, {Daddi}, {Liu}, {Kodama},
  {Tanaka}, {Schreiber}, {Zanella}, {Valentino}, {Sargent}, {Kohno}, {Xiao},
  {Pannella}, {Ciesla}, {Gobat}, \& {Koyama}}]{Wang:2018rz}
{Wang}, T., {Elbaz}, D., {Daddi}, E., {et~al.} 2018, \apjl, 867, L29,
  \dodoi{10.3847/2041-8213/aaeb2c}

\bibitem[{{Warren} {et~al.}(2004){Warren}, {Jerjen}, \&
  {Koribalski}}]{Warren:2004hx}
{Warren}, B.~E., {Jerjen}, H., \& {Koribalski}, B.~S. 2004, \aj, 128, 1152,
  \dodoi{10.1086/422923}

\bibitem[{{Waugh} {et~al.}(2002){Waugh}, {Drinkwater}, {Webster},
  {Staveley-Smith}, {Kilborn}, {Barnes}, {Bhathal}, {de Blok}, {Boyce},
  {Disney}, {Ekers}, {Freeman}, {Gibson}, {Henning}, {Jerjen}, {Knezek},
  {Koribalski}, {Marquarding}, {Minchin}, {Price}, {Putman}, {Ryder}, {Sadler},
  {Stootman}, \& {Zwaan}}]{Waugh:2002rr}
{Waugh}, M., {Drinkwater}, M.~J., {Webster}, R.~L., {et~al.} 2002, \mnras, 337,
  641, \dodoi{10.1046/j.1365-8711.2002.05942.x}

\bibitem[{{Webb} {et~al.}(2013){Webb}, {O'Donnell}, {Yee}, {Gilbank}, {Coppin},
  {Ellingson}, {Faloon}, {Geach}, {Gladders}, {Noble}, {Muzzin}, {Wilson}, \&
  {Yan}}]{Webb:2013gr}
{Webb}, T.~M.~A., {O'Donnell}, D., {Yee}, H.~K.~C., {et~al.} 2013, \aj, 146,
  84, \dodoi{10.1088/0004-6256/146/4/84}

\bibitem[{{Weinmann} {et~al.}(2006){Weinmann}, {van den Bosch}, {Yang}, \&
  {Mo}}]{Weinmann:2006ve}
{Weinmann}, S.~M., {van den Bosch}, F.~C., {Yang}, X., \& {Mo}, H.~J. 2006,
  \mnras, 366, 2, \dodoi{10.1111/j.1365-2966.2005.09865.x}

\bibitem[{{Wild}(1952)}]{Wild:1952bt}
{Wild}, J.~P. 1952, \apj, 115, 206, \dodoi{10.1086/145533}

\bibitem[{{Wilman} {et~al.}(2005){Wilman}, {Balogh}, {Bower}, {Mulchaey},
  {Oemler}, {Carlberg}, {Morris}, \& {Whitaker}}]{Wilman:2005vm}
{Wilman}, D.~J., {Balogh}, M.~L., {Bower}, R.~G., {et~al.} 2005, \mnras, 358,
  71, \dodoi{10.1111/j.1365-2966.2005.08744.x}

\bibitem[{{Wilman} {et~al.}(2009){Wilman}, {Oemler}, {Mulchaey}, {McGee},
  {Balogh}, \& {Bower}}]{Wilman:2009to}
{Wilman}, D.~J., {Oemler}, A., J., {Mulchaey}, J.~S., {et~al.} 2009, \apj, 692,
  298, \dodoi{10.1088/0004-637X/692/1/298}

\bibitem[{{Wright} {et~al.}(2010){Wright}, {Eisenhardt}, {Mainzer}, {Ressler},
  {Cutri}, {Jarrett}, {Kirkpatrick}, {Padgett}, {McMillan}, {Skrutskie},
  {Stanford}, {Cohen}, {Walker}, {Mather}, {Leisawitz}, {Gautier}, {McLean},
  {Benford}, {Lonsdale}, {Blain}, {Mendez}, {Irace}, {Duval}, {Liu}, {Royer},
  {Heinrichsen}, {Howard}, {Shannon}, {Kendall}, {Walsh}, {Larsen}, {Cardon},
  {Schick}, {Schwalm}, {Abid}, {Fabinsky}, {Naes}, \& {Tsai}}]{Wright:2010oi}
{Wright}, E.~L., {Eisenhardt}, P. R.~M., {Mainzer}, A.~K., {et~al.} 2010, \aj,
  140, 1868, \dodoi{10.1088/0004-6256/140/6/1868}

\bibitem[{{Yoon} {et~al.}(2017){Yoon}, {Chung}, {Smith}, \&
  {Jaff{\'e}}}]{Yoon:2017jl}
{Yoon}, H., {Chung}, A., {Smith}, R., \& {Jaff{\'e}}, Y.~L. 2017, \apj, 838,
  81, \dodoi{10.3847/1538-4357/aa6579}

\bibitem[{{Yoon} \& {Putman}(2017)}]{Yoon:2017fi}
{Yoon}, J.~H., \& {Putman}, M.~E. 2017, \apj, 839, 117,
  \dodoi{10.3847/1538-4357/aa697b}

\bibitem[{{Yoon} {et~al.}(2012){Yoon}, {Putman}, {Thom}, {Chen}, \&
  {Bryan}}]{Yoon:2012pj}
{Yoon}, J.~H., {Putman}, M.~E., {Thom}, C., {Chen}, H.-W., \& {Bryan}, G.~L.
  2012, \apj, 754, 84, \dodoi{10.1088/0004-637X/754/2/84}

\bibitem[{{Yoshida} {et~al.}(2004){Yoshida}, {Ohyama}, {Iye}, {Aoki},
  {Kashikawa}, {Sasaki}, {Shimasaku}, {Yagi}, {Okamura}, {Doi}, {Furusawa},
  {Hamabe}, {Kimura}, {Komiyama}, {Miyazaki}, {Miyazaki}, {Nakata}, {Ouchi},
  {Sekiguchi}, \& {Yasuda}}]{Yoshida:2004vc}
{Yoshida}, M., {Ohyama}, Y., {Iye}, M., {et~al.} 2004, \aj, 127, 90,
  \dodoi{10.1086/380221}

\bibitem[{{Young} {et~al.}(1995){Young}, {Xie}, {Tacconi}, {Knezek}, {Viscuso},
  {Tacconi-Garman}, {Scoville}, {Schneider}, {Schloerb}, {Lord}, {Lesser},
  {Kenney}, {Huang}, {Devereux}, {Claussen}, {Case}, {Carpenter}, {Berry}, \&
  {Allen}}]{Young:1995jq}
{Young}, J.~S., {Xie}, S., {Tacconi}, L., {et~al.} 1995, \apjs, 98, 219,
  \dodoi{10.1086/192159}

\bibitem[{{Young} {et~al.}(2011){Young}, {Bureau}, {Davis}, {Combes},
  {McDermid}, {Alatalo}, {Blitz}, {Bois}, {Bournaud}, {Cappellari}, {Davies},
  {de Zeeuw}, {Emsellem}, {Khochfar}, {Krajnovi{\'c}}, {Kuntschner},
  {Lablanche}, {Morganti}, {Naab}, {Oosterloo}, {Sarzi}, {Scott}, {Serra}, \&
  {Weijmans}}]{Young:2011sq}
{Young}, L.~M., {Bureau}, M., {Davis}, T.~A., {et~al.} 2011, \mnras, 414, 940,
  \dodoi{10.1111/j.1365-2966.2011.18561.x}

\bibitem[{{Zabel} {et~al.}(2019){Zabel}, {Davis}, {Smith}, {Maddox}, {Bendo},
  {Peletier}, {Iodice}, {Venhola}, {Baes}, {Davies}, {de Looze}, {Gomez},
  {Grossi}, {Kenney}, {Serra}, {van de Voort}, {Vlahakis}, \&
  {Young}}]{Zabel:2019ne}
{Zabel}, N., {Davis}, T.~A., {Smith}, M. W.~L., {et~al.} 2019, \mnras, 483,
  2251, \dodoi{10.1093/mnras/sty3234}

\bibitem[{{Zabludoff} \& {Mulchaey}(1998)}]{Zabludoff:1998ew}
{Zabludoff}, A.~I., \& {Mulchaey}, J.~S. 1998, \apj, 496, 39,
  \dodoi{10.1086/305355}

\bibitem[{{Zwicky}(1938)}]{Zwicky:1938fz}
{Zwicky}, F. 1938, \pasp, 50, 218, \dodoi{10.1086/124935}

\end{thebibliography}


\end{document}